\documentclass[preprint,aps,amsmath,nofootinbib]{revtex4}
\usepackage{graphicx,array,dcolumn}
\usepackage{calc,tabularx, epsfig}
\usepackage{hyperref}

\usepackage{amsmath}
\usepackage{amssymb}
\usepackage{multirow}
\usepackage{xspace}
\usepackage{slashed}
\allowdisplaybreaks[1]
\newlength{\figurewidth}
\newcommand{\beq}{\begin{equation}}
\newcommand{\eeq}{\end{equation}}
\newcommand{\bea}{\begin{eqnarray}}
\newcommand{\eea}{\end{eqnarray}}
\newcommand{\ba}{\begin{array}}
\newcommand{\ea}{\end{array}}
\newcommand{\bg}{\bar{g}}
\newcommand{\mn}{{\mu\nu}}

\newcommand{\pt}{\partial}
\newcommand{\pd}{(2\pi)^d}

\newcommand{\al}{\alpha}
\newcommand{\bt}{\beta}
\newcommand{\g}{\gamma}
\newcommand{\ep}{\epsilon}
\newcommand{\ta}{\theta}
\newcommand{\lam}{\lambda}

\newcommand{\G}{\Gamma}

\newcommand{\de}{\delta}
\newcommand{\D}{\Delta}

\newcommand{\om}{\omega}
\newcommand{\sg}{\sigma}
\newcommand{\kp}{\kappa}
\newcommand{\bnb}{\bar{\nabla}}
\makeatother
\begin{document}
%
\title{Exorcising Ghosts in Induced Gravity}
\setlength{\figurewidth}{\columnwidth}
%
\author{Gaurav Narain}
\email{gaunarain@itp.ac.cn}
\affiliation{
Kavli Institute for Theoretical Physics China (KITPC), \\
Key Laboratory of Theoretical Physics,
Institute of Theoretical Physics (ITP), 
Chinese Academy of Sciences (CAS), Beijing 100190, P.R. China.}
%
%
\begin{abstract}
Unitarity of scale-invariant coupled theory of higher-derivative gravity 
and matter is investigated. A scalar field coupled with 
dirac fermion is taken as matter sector.
Following the idea of induced gravity Einstein-Hilbert term 
is generated via dynamical symmetry breaking of scale-invariance.
The renormalisation group flows are computed and one-loop RG improved effective potential 
of scalar is calculated. Scalar field develops a new minimum 
via Coleman-Weinberg procedure inducing 
Newton's constant and masses in the matter sector. 
The spin-2 problematic ghost and the spin-0 mode of the metric 
fluctuation gets a mass in the broken phase of 
theory. The energy-dependence of VeV in the RG improved 
scenario implies a running for the induced parameters. 
This sets up platform to ask whether it is possible to 
evade the spin-2 ghost by keeping its mass always 
above the running energy scale? In broken phase 
this question is satisfactorily answered for a large domain of 
coupling parameter space where the ghost is evaded. 
The spin-0 mode can be made physically realisable or not depending 
upon the choice of initial parameters. Induced Newton's constant is seen to 
vanishes in ultraviolet. By properly choosing parameters it is possible to make the 
matter fields physically unrealisable. 
\end{abstract}

\maketitle
%
%

\section{Introduction}
\label{intro}

Finding a well-defined and mathematically consistent theory of quantum gravity 
is one of the most important problems of theoretical physics. Moreover, finding 
experimental evidence validating or falsifying one is equally hard. Presently there 
are several models of quantum gravity which are aimed at studying the quantum 
nature of space-time and investigating physics at ultra-high energies. Recently 
a minimalistic model in the framework of four dimension quantum field theory (QFT) 
in lorentizian space-time was investigated, which was shown to be renormalizable 
to all loops \cite{Stelle1977c, Stelle77}, and was recently shown 
to be unitary \cite{NarainA1, NarainA2} (see also references therein).
This then offers a sufficiently good and simple model of quantum field theory 
of gravity whose arena can be used to investigate physics at ultra-high energies. 

Here in this paper, motivated by the results of \cite{Stelle77, Salam1978, 
Julve1978, NarainA1, NarainA2, NarainA3, NarainA4} 
we study the scale-invariant higher-derivative gravitational 
system coupled with matter fields. These constitute interesting systems. The 
scale-invariant purely gravitational sector consist of only dimensionless couplings.
This makes the theory perturbatively renormalizable to all loops 
in four space-time dimensions by power-counting \cite{Fradkin1981, Fradkin1982}
(for classical picture of these theories see \cite{Stelle1977c, Gaume2015}).
Coupling this with scale-invariant matter sector doesn't change the picture. The resulting 
theory is still perturbatively UV renormalizable in four dimensions due to lack 
of any dimensionful parameter. Classically the matter sector however has
local conformal invariance, which is broken under quantum corrections
due to conformal anomalies (local and non-local) \cite{Capper1974, Deser1976}. 
The interesting thing to note here is that in this quantum theory the 
counter terms generated still possess scale invariant structure (due to 
lack of any dimensionful parameter in theory) \cite{Duff1977}.
This therefore preserves the renormalizability of theory \cite{Stelle77}, 
even though trace anomalies are present. 

Scale-invariant gravitational systems coupled with matter 
have been investigated in past. Some of the first studies were
conducted in \cite{Julve1978,Fradkin1981, Fradkin1982,Barth1983,Avramidi1985}, 
where the renormalisation group running of various couplings was computed 
and fixed point structure was analysed. Further investigation 
for more complicated systems were done in 
\cite{Odintsov1989,Buchbinder1989,Shapiro1989,Elizalde1994,Elizalde1995_1,Elizalde1994_2}
(see also the book \cite{Buchbinder1992}).
Matter coupling with conformal quantum gravity along with 
Gauss-Bonnet term were investigated in 
\cite{deBerredoPeixoto2003,deBerredoPeixoto2004}.
Recently it has gained some momentum and these models 
have been reinvestigated \cite{Strumia1,Einhorn2014,Jones1,Jones2,Jones2015}.
The purpose in these papers were to see if it is possible to generate 
a scale dynamically starting from scale-invariant system. In 
\cite{Strumia1} the authors called their model `Agravity', where 
Planck scale is dynamically generated from the 
vacuum-expectation-value(VeV) of potential in Einstein frame (not in Jordan frame).
By this they achieve a negligible cosmological constant, 
generates Planck's mass, and addresses the naturalness \cite{Strumia1, Salvio2016} and 
inflation \cite{Kannike2015}, but unitarity issues was not explored
\footnote{
In \cite{Salvio2015} a quantum mechanical treatment of 
$4$-derivative theories was suggested, which when suitably extended 
can tackle more complicated field theoretic systems. This can 
perhaps address issues of ghost and unitarity in a more 
robust manner.}. 
In \cite{Einhorn2014,Jones2015,Jones1,Jones2}
the authors studies the issue of dynamical generation of 
scale via dimensional transmutation in the presence of 
background curvature. This also induces Einstein-Hilbert 
gravity and generates Newton's constant, but unitarity 
problem was not addressed. An interesting idea 
has been suggested in \cite{Holdom2015,Holdom2016}
by assuming an analogy with QCD, where the authors addresses 
the problems of ghost and tachyon using the wisdom 
acquired from non-perturbative sector of QCD, as
is argued that the gravitational theory enters a non-perturbative regime
below the Planck scale. 

The ideas of induced gravity goes long back. It was first proposed in
\cite{Zeldovich1967,Sakharov1967}, where the quantum matter 
fluctuations at high energy generate gravitational dynamics at 
low energy inducing cosmological and Newton's gravitational constant. 
Another proposal suggested in \cite{Fujii1974, Chudnovsky1976,Zee1978} induces Einstein 
gravity spontaneously via symmetry breaking along 
the lines of Higgs mechanism. Later in \cite{Zee1980,Adler19801, 
Adler19802, Adler19803, Adler1982} the idea of generation of Einstein 
gravity via dynamical symmetry breaking was considered, following 
the methodology of Coleman-Weinberg \cite{Coleman1973}. 
In \cite{Adler1982}, metric fluctuations were also incorporated 
in the generation of induced Newton's constant. Around the same 
time induced gravity from weyl-theory was also studied 
\cite{Nepomechie1983, Zee1983, Buchbinder1986}.
Phase-transitions leading to generation of 
Einstein-Hilbert gravity due to loop-effects from conformal 
factor coupled with scalar field were studied in 
\cite{Shapiro1994}. In \cite{Floreanini1993,Floreanini1994}
the renormalization group improved effective-potential 
of the dilaton leads to running of VeV inducing masses (along with 
Einstein-Hilbert gravity). Furthermore the authors make a proposal  
along lines of \cite{Salam1978,Julve1978} to tackle ghost and tachyons. 
Cosmological consequences of these models were explored 
in \cite{Cooper1982,Finelli2007}.

In this paper we therefore study scale-invariant gravitational 
and matter coupled systems in $(4-\ep)$ dimensional regularisation 
scheme. The beta-functions are computed and compared with the 
past literature. The one-loop RG improved effective potential for the 
scalar field is computed by incorporating the quantum fluctuations 
of both matter and gravity \cite{Ford1992}. The scale 
invariance is broken dynamically when the scalar-field $\phi$ acquires 
a VeV via Coleman-Weinberg mechanism \cite{Coleman1973}. 
This in turn induces gravitational Newton's constant, cosmological constant
and masses in the matter sector. 
We work in lorentzian signature and take the sign of $C_{\mn\rho\sg}^2$
(Weyl tensor square)
to be negative, keeping the sign of $R^2$ term (where $R$ is Ricci-scalar) 
to be always positive (this is done to avoid tachyonic instabilities). 
These choice of signs further allows necessary convergence in 
feynman $+i\ep$-prescription by suppressing fields modes with large action
in the lorentzian path-integral.
The sign of the $R\phi^2$ term is taken negative, so to 
generate the right sign of Newton's constant and to 
avoid tachyonic instabilities in the broken phase. 
The negative sign of $C_{\mn\rho\sg}^2$ term is taken in order 
to satisfy the unitary criterion as stated in \cite{NarainA1, NarainA2}.
In this case we no longer have asymptotic freedom
as has been observed in euclidean case in 
\cite{Fradkin1981,Fradkin1982, Avramidi1985, Strumia1, Einhorn2014, 
Jones2015, Jones1, Jones2}.
The VeV generated in RG improved effective potential has 
running, and therefore induces running in 
Newton's constant and masses of matter fields. 
Due to the generation of Einstein-Hilbert term in the action, 
the propagator of metric fluctuations in the broken phase gets modified, and 
the modes acquires mass. In this broken phase we investigate 
the problem of spin-2 ghost by probing the running of its mass 
along the lines of \cite{NarainA1, NarainA2}
\footnote{
These studies are conducted in the perturbative framework around the gaussian 
fixed point and is therefore different from studies done in 
the asymptotic safety scenario \cite{Percacci2007}, 
where the RG running of couplings were computed 
using functional-renormalization group in euclidean signature 
\cite{Codello2006,Codello2008,Groh2011,Ohta2013,Ohta2015,Ohta2016},
with spectral positivity \cite{Niedermaier2009} and ghost
\cite{Benedetti20091,Benedetti20092} issues analysed around a 
non-gaussian fixed point. 
Present work is also different from the recent studies done 
in the direction of finite and ghost free non-local quantum 
field theories of gravity \cite{Modesto2011,Biswas2011,Modesto2014,Tomboulis2015}, 
in the sense as it doesn't incorporates non-local features. 
}. 
 
The outline of paper is following.
In section \ref{RGrun} the divergent part of the effective action is 
computed in $(4-\ep)$ dimensional regularisation scheme. The 
beta-function are obtained from it. In section \ref{effpot}
one-loop renormalisation group improved effective potential 
for the scalar field is computed by incorporating quantum 
corrections from gravitational and matter degrees of freedom. 
In section \ref{symbr} the breaking of scale-invariance 
is studied via Coleman-Weinberg mechanism, which in turn 
induces gravitational Newton's constant and masses in the 
broken phase. RG equation for VeV is derived,
which induces a running in the generated Newton's constant 
and masses. In section \ref{unitpres} a prescription to avoid
spin-2 massive ghost is given, where a procedure to pick the 
right set of initial conditions is stated. In section \ref{numanal}
numerical analysis is done to give evidence showing that 
there exist a large domain of coupling parameter space where 
spin-2 massive ghost can be made physically unrealisable. 
Finally in section \ref{conc} conclusions are presented. 

\section{RG running}
\label{RGrun}

In this section we compute the renormalisation group running of the various coupling 
parameters present in our scale-invariant theory. We start first with the formalism and 
compute the various diagrams that contain UV divergences. These are then used to 
write the beta-function of the various coupling parameters. We start by considering the 
path-integral of the coupled system $(\hbar=c=1)$
\begin{equation}
\label{eq:pathint}
Z
= \int \, {\cal D} \g_\mn
{\cal D} \phi {\cal D} \bar{\psi} {\cal D}\psi 
\exp \biggl[
i\biggl(
S_{\rm GR} + S_{\rm matter} 
\biggr)
\biggr] \, ,
\end{equation}
where $S_{\rm GR}$ and $S_{\rm matter}$ are given by 
\bea
\label{eq:Act}
S_{\rm GR} &=& \frac{1}{16\pi} \int {\rm d}^dx \sqrt{-\g} \biggl[
- \frac{1}{f^2}\left(R_\mn R^\mn - \frac{1}{3}R^2 \right) + \frac{\om}{6f^2} R^2 
\biggr] \, ,
\notag \\
S_{\rm matter} &=& \int {\rm d}^dx \sqrt{-\g} \biggl[
\frac{1}{2} \pt_\mu \phi \pt^\mu \phi - \frac{\lam}{4} \phi^4 - \frac{\xi}{2} R \phi^2 
+ \bar{\psi} \left( i \gamma^\mu D_\mu - y_t \phi \right) \psi \biggr] \, ,
\eea
where the coupling parameters $f^2$, $\om$, $\lam$, $\xi$ and $y_t$ are all 
dimensionless, and the geometric quantities (curvature and covariant-derivative) 
depends on metric $\g_\mn$.
In the fermionic part of action, the Dirac gamma matrices are defined via tetrads and 
inverse tetrads $(\g^\mu = e_a{}^\mu \g^a, \g_\mu = e^a{}_\mu \g_a)$, 
and $D_\mu$ is the spin-connection covariant derivative.
\beq
\label{eq:spinCcovD}
D_\mu = \pt_\mu - \frac{i}{2} \sg^{cd} \om_{\mu cd} \, .
\eeq
Here greek indices denote the space-time index, while the 
latin indices denote the internal lorentz index, and 
$\sg_{cd} = i/4 [\g_c, \g_d]$. The internal indices are 
raised and lowered using internal metric $\eta_{ab}$. For torsionless 
manifold the spin-connection can be expressed in terms of the
christoffel connection $\G_\mu{}^\al{}_\nu$ (which can be re-expressed 
again in terms of tetrads) as,
\beq
\label{eq:spinwcon}
\om_\mu{}^{ad} = \frac{1}{2} \left[
e^{a\rho}(\pt_\mu e^d{}_\rho - \pt_\rho e^d{}_\mu)
- e^{d\rho} (\pt_\mu e^a{}_\rho - \pt_\rho e^a{}_\mu)
+ e^b{}_\mu e^{a\rho}e^{d\nu} (\pt_\nu e_{b\rho} - \pt_\rho e_{b\nu})
\right] \, .
\eeq
The dimensionless nature of coupling $f^2$ and $\om/f^2$
results in a fully dimensionless scale-invariant coupled action.

We study the diffeomorphism invariant action of the coupled system 
using background field method \cite{DeWitt3, Abbott} in $(4-\ep)$
dimensional regularisation scheme. It is advantageous, as by construction 
it preserves background gauge invariance. The field is decomposed into 
background and fluctuation. Keeping the background fixed the path-integral 
is then reduced to an integral over the fluctuations. The gravitational 
metric field is decomposed into background and fluctuation, while the 
tetrads (and its inverse) are expressed in powers of this fluctuation field. 
The matter fields are similarly decomposed. The gauge invariance of the 
full metric field is then transformed into the invariance over the fluctuation field.
To prevent over-counting of gauge-orbits in the path-integral measure,
a constraint is applied on this fluctuation field, which results in appearance
of auxiliary fields called ghosts. The effective action generated after 
integrating over the fluctuation and auxiliary fields still enjoys invariance 
over the background fields. 

The quantum metric is written as
\beq
\label{eq:qmet}
\g_\mn=\bg_\mn+h_\mn \, ,
\eeq
where $\bg_\mn$ is some arbitrary (but fixed) background and $h_\mn$ is the 
metric fluctuation. The full action can be expanded in powers of $h_\mn$.
The path-integral measure over $\g_\mn$ is then replaced with measure 
over $h_\mn$. Integrating over the fluctuation field implies that in some 
sense they will appear only as internal legs and never as external legs. 
The background gauge invariant effective action formalism allows to 
choose a particular background for the ease of computation. In particular 
writing $\bg_\mn = \eta_\mn+H_\mn$ (while still keeping $H_\mn$ generic)
allows one to use the machinery of the flat space-time QFT, thereby 
giving a particle interpretation to the internal ($h_\mn$)
and external ($H_\mn$) legs, in the sense that the former behaves as 
virtual particle, while the later is the corresponding external particle. 
In this manner one can compute the effective action for the 
external leg $H_\mn$. Alternatively one can expand the 
full action around flat space-time directly, calling the fluctuation 
to be $h_\mn^{\prime}$ though. This is a highly non-linear gauge theory with infinite 
number of interactions terms (however their couplings are related to 
each other by diffeomorphism invariance). Then following the usual 
strategy of background field method and writing $h_\mn^{\prime}=H_\mn+h_\mn$,
it is quickly seen that $H_\mn$ is the external leg corresponding to the 
virtual particle given by $h_\mn$. Integrating over quantum fluctuations 
$h_\mn$ then gives the effective action in terms of $H_\mn$ field. 

One can then set-up Feynman perturbation theory by expanding the 
original action in powers of $h_\mn$ and $H_\mn$. Similarly writing 
the scalar and fermion fields as
\beq
\label{eq:linearSP}
\phi = \varphi + \chi \, , 
\hspace{4mm}
\bar{\psi} = \bar{\ta} + \bar{\eta} \, , 
\hspace{4mm}
\psi = \ta + \eta \, ,
\eeq
one can expand the action in powers of fluctuations $\chi$, $\eta$ and
$\bar{\eta}$. The piece which is quadratic in only fluctuations ($h_\mn$, $\chi$, $\eta$ and
$\bar{\eta}$) gives the propagator while all the other terms gives the 
interactions vertices. In one-loop approximation the terms which are 
exclusively quadratic in fluctuations are retained, all other terms which 
involve higher powers of fluctuations ($h_\mn$, $\chi$, $\eta$ and
$\bar{\eta}$) contribute in higher-loops and will be ignored here. 
In-fact for computing the running of all matter couplings (except $\xi$)
it is sufficient to consider the situation with $H_\mn=0$. However, 
in the case for computing running of $\xi$, terms up-to (and at-least) linear in 
$H_\mn$ should be retained. Similarly if one is interested in studying 
behaviour or $R^2$ and $R_\mn R^\mn$ then one should at-least 
retain terms up-to quadratic in $H_\mn$. 

\subsection{Gauge Fixing and Ghosts}
\label{gfghost}

The path-integration over the gravitational field is ill defined.
This is a general feature of gauge theories where the gauge 
invariance (diffeomorphism invariance for gravity) relates 
two field configuration by gauge transformation. Such field 
configuration will contribute equally to the path-integral. However 
this will lead to over-counting. To prevent such over-counting,
gauge-invariance needs to be broken by constraining the gauge 
field. This procedure of systematically applying the constraint 
leads to ghost, which are elegantly taken care of by the 
Faddeev-Popov prescription \cite{Faddeev}.

However in this style of breaking the invariance one may wonder
whether the gauge (or diffeomorphism) invariance emerges in the 
effective action. To make sure that the effective action 
obtained after integrating out the fluctuation field is gauge 
invariant, background field method is followed. 
It is a method (and procedure) which guarantees that the 
effective action constructed using it will be background 
gauge invariant. Below we describe the procedure for gauge-fixing 
in the background field method.

The diffeomorphism invariance of the full action in eq (\ref{eq:Act}) 
implies that for arbitrary vector field $\ep^\rho$, the action
should be invariant under the following transformation 
of the metric field variable,
\beq
\label{eq:gaugetrgamma}
\de_D \g_\mn = {\cal L}_\ep \g_\mn = \ep^\rho \pt_\rho \g_\mn 
+ \g_{\mu\rho} \pt_\nu \ep^\rho + \g_{\nu\rho}\pt_\mu \ep^\rho \, ,
\eeq
where ${\cal L}_\ep \g_\mn$ is the Lie derivative of the quantum metric 
$\g_\mn$ along the vector field $\ep^\rho$. Decomposing the quantum 
metric $\g_\mn$ into background ($\bg_\mn$) and fluctuation ($h_\mn$)
allows one to figure out the transformation of the fluctuation field while 
keeping the background fixed. This will imply the following 
transformation of $h_\mn$.
\beq
\label{eq:gaugetrh}
\de_D h_\mn = \bar{\nabla}_\mu \ep_\nu + \bar{\nabla}_\nu \ep_\mu
+ \ep^\rho \bar{\nabla}_\rho h_\mn + h_{\mu\rho} \bar{\nabla}_\nu \ep^\rho
+ h_{\nu\rho} \bar{\nabla}_\mu \ep^\rho \, ,
\eeq
where $\bar{\nabla}$ is the covariant derivative whose connection is
constructed using the background metric. This is the full transformation
of the metric fluctuation field. Ignoring terms which are linear in 
$h_\mn$ allows one to investigate only the one-loop effects. These ignored 
terms are however mandatory when dealing with higher-loop effects. 
The invariance of the action is broken by choosing an 
appropriate gauge-fixing condition implemented via Faddeev-Popov 
procedure.

The gauge fixing action chosen for fixing the invariance under the 
transformation of the metric fluctuation field is given by,
\beq
\label{eq:gravity_GF}
S_{GF} = \frac{1}{32 \pi \alpha} 
\int {\rm d}^dx \sqrt{-\bg} \biggl(\bnb^{\rho} h_{\rho\mu}
- \frac{1+\rho}{d} \bnb_\mu h \biggr) \, Y^\mn \, 
\biggl(\bnb^{\sigma}h_{\sigma\nu} 
-\frac{1+\rho}{d} \bnb_\nu h\biggr) \, ,
\eeq
where $\al$ and $\rho$ are gauge parameters, while $Y_\mn$
is either a constant or a differential operator depending 
upon the gravitational theory under consideration. For the theory
considered here in eq. (\ref{eq:Act}), we consider higher-derivative type
gauge fixing with $Y_\mn = (-\bg_\mn \bar{\Box}
+ \bt \bnb_\mu \bnb_\nu)$, where $\bar{\Box} = \bnb_\mu \bnb^\mu$.
Choosing $\rho=-1$ and $\bt=0$ correspond to Landau gauge
condition. Taking $\al\to0$ imposes the gauge condition sharply. 

The ghost action is obtained following the Faddeev-Popov 
procedure \cite{Faddeev}. In general if the gauge-fixing 
condition on the gravitational field $h_\mn$ is written as $F_\mu=0$
(which here is $F_{\mu} = \bnb^{\rho} h_{\rho\mu}
- \frac{1+\rho}{d} \bnb_\mu h$), we introduce it in the path-integral 
by multiplying the later with unity in the following form,
\bea
\label{eq:pathunity}
1= && 
\int {\cal D} F^{\ep}_{\mu}  \left(\det Y \right)^{\frac{1}{2}}
\exp \biggl[ \frac{i}{32 \pi \al} \int {\rm d}^dx \sqrt{-\bg} F^{\ep}_{\mu}
Y^\mn F^{\ep}_{\nu} \biggr] \, ,
\eea
where $F_{\mu}^{\ep}$ is the gauge transformed $F_{\mu}$.
As $Y^\mn$ contains derivative operator, therefore its determinant 
is non-trivial. The original path-integral (without gauge-fixing) being invariant 
under transformation eq. (\ref{eq:gaugetrh}) of the field $h_\mn$
implies that a change of integration variable from $h_\mn$ to
$h_\mn^\ep$ doesn't give rise to any jacobian in the path-integral 
measure. However replacing the measure over $F^{\ep}_{\mu}$
with measure over $\ep^\rho$ introduces a non-trivial jacobian 
in the path-integral. This is obtained as follows,
\beq
\label{eq:gaugejacob1}
dF_{\mu}^{\ep} = \frac{\pt F_{\mu}}{\pt \ep^\rho} d\ep^\rho 
\hspace{5mm}
\Rightarrow
\hspace{5mm}
{\cal D} F_{\mu}^{\ep} = \det \biggl(
\frac{\pt F_{\mu}}{\pt \ep^\rho}
\biggr) {\cal D} \ep^\rho \, .
\eeq
In the background field formalism this jacobian consist of 
background covariant derivative, background and fluctuation 
fields, and is independent of the transformation parameter $\ep^\rho$.
This implies that it can be taken out of the functional integral over
$\ep^\rho$. Changing the integration variable from $h_\mn^\ep$ 
to $h_\mn$, and ignoring the infinite constant generated 
by integrating over $\ep^\rho$, gives us the gauge fixed path 
integral including the determinant. 

The functional determinant appearing in eq. (\ref{eq:gaugejacob1})
can be exponentiated by making use 
of appropriate auxiliary fields. Writing the functional 
determinant $(\det Y)^{1/2}$ as $(\det \, Y) \times (\det \, Y)^{-1/2}$, allows
to combine the former with the Faddeev-Popov determinant in
eq. (\ref{eq:gaugejacob1}), which is then exponentiated 
by making use of anti-commuting auxiliary fields, while the 
later $(\det \, Y)^{-1/2}$ is exponentiated 
by making use of commuting auxiliary fields. The former 
auxiliary fields are known as Faddeev-Popov ghosts, while 
those in later case are known as 
Nielsen-Kallosh ghosts \cite{Kallosh, Nielsen}.
The path integral of the full ghost sector is given by,
\bea
\label{eq:ghostpath}
&&
\int {\cal D}\bar{C}_\mu {\cal D} C_\nu  {\cal D} \ta_\al
\,\,
\exp \biggl[
-i \int {\rm d}^dx \sqrt{-\bg} 
\biggl\{ 
\bar{C}_\mu \left(Y^\mn \, \frac{\pt F_{\nu}}{\pt \ep^\rho} \right) C^\rho 
+\frac{1}{2}\ta_\al Y^{\al\bt} \ta_\bt 
\biggr\}
\biggr] \, ,
\eea
where $\bar{C}_\mu$ and $C_\nu$ are Faddeev-Popov ghost fields arising 
from the gauge fixing in the gravitational sector, and $\ta_\mu$ is the commuting ghost 
arising due to fact that $Y_\mn$ contains derivatives.

In the case when $F_{\mu}$ is given as in eq. (\ref{eq:gravity_GF}), 
the Faddeev-Popov ghost action is given by,
\beq
\label{eq:ghostact}
S^{FP}_{\rm gh} = - \int {\rm d}^dx \sqrt{-\bg}
\bar{C}_\mu X^\mu_{\rho} C^\rho \, ,
\eeq
where,
\bea
\label{eq:ghostet}
X^\mu_{\rho} &=& 
(\bg^\mn \bar{\Box} + \bt \bnb^\mu \bnb^\nu)
\biggl[
\bnb_\rho \bnb_\nu + \bg_{\nu\rho} \bar{\Box}
- \frac{2(1+\rho)}{d} \bnb_\nu \bnb_\rho 
\notag \\
&&
+ \bnb_\rho h_{\sg\nu} \bnb^\sg
+ \bnb^\sg\bnb_\rho h_{\sg\nu} 
+ \bnb^\sg h_{\nu\rho} \bnb_\sg + h_{\nu\rho}\bar{\Box}
+ \bnb^\sg h_{\sg\rho} \bnb_\nu 
+ h_{\sg\rho} \bnb^\sg \bnb_\nu
\notag \\
&&
- \frac{1+\rho}{d} \biggl(
\bnb_\rho h\bnb_\nu + \bnb_\nu\bnb_\rho h 
+ 2 \bnb_\nu h_{\sg\rho} \bnb^\sg
+ 2 h_{\sg\rho}\bnb_\nu \bnb^\sg
\biggr)
\biggr] \, .
\eea
Here the last two lines contains terms linear in $h_\mn$. These are 
not relevant in doing one-loop computations, but at higher-loops they are 
important. In the following we will ignore ghost contributions completely 
as they are not relevant in the computation of the running of matter 
couplings, while the running of gravitational couplings are taken from 
past literature \cite{NarainA1, NarainA2, Avramidi1985, Strumia1, Avramidi2000}.

\subsection{Gravitational Field Propagator}
\label{eq:Gravprop}

The propagator for the gravitational field is obtained by expanding the gravitational 
action around the flat space-time up-to second order in the fluctuation field $h_\mn$.
By decomposing the fluctuation field in terms of various components and writing them 
using the projection operators ($P_2^{\mn \rho\sg}$, $P_1^{\mn \rho\sg}$, $P_s^{\mn \rho\sg}$,
$P_{sw}^{\mn \rho\sg}$, $P_{ws}^{\mn \rho\sg}$ and $P_w^{\mn \rho\sg}$)
as given in appendix \ref{app_proj}, we note that this 
second variation can be expressed in a neat form in momentum space
in the following manner,
\beq
\label{eq:2ndVar_GRproj}
\de^2 S_{\rm GR} = 
\frac{1}{32 \pi} \int \frac{{\rm d}^dq}{\pd}
h_{\mn} 
\biggl[
-\frac{q^4}{2f^2}  P_2^{\mn \rho\sg}
+ \frac{q^4 \omega}{f^2} 
P_s^{\mn \rho\sg}
\biggr] h_{\rho\sg} \, .
\eeq
Moreover the gauge-fixing action can be similarly expressed by using the 
projection operators,
\bea
\label{eq:SgfGR_proj}
&&
S_{\rm GF} = \frac{1}{32 \pi \al} \int \frac{{\rm d}^dq}{\pd}
h_{\mn} \, q^4 \, \biggl[
-\frac{1}{2} P_1^{\mn \rho\sg} + \frac{1-\bt}{d^2} \biggl\{ 
(1+\rho)^2 (d-1) P_s^{\mn \rho\sg} 
\notag \\
&&
+(d-1-\rho)^2 P_w^{\mn \rho\sg} 
- \sqrt{d-1} (1+\rho)(d-1-\rho)
\left(P_{sw}^{\mn \rho\sg}  + P_{ws}^{\mn \rho\sg}  \right)
\biggr\}
\biggr]
h_{\rho\sg} \, .
\eea
By writing the gauge-fixing action in terms of the projection operators
allows us to see clearly which modes of the field are affected by the gauge-fixing.
For example the spin-2 mode is not affected at all by the gauge-fixing condition.
Interestingly it should be noted that there is another gauge-invariant mode of the 
field which arises due to the action of spin-s projection operator on the $h_\mn$ field
(see appendix \ref{app_proj}). 
However under harmonic type gauge-fixing condition this mode doesn't 
remain completely unaffected. Only for some particular gauge 
choices this mode is not affected by the gauge-fixing condition. Landau gauge 
being one such choice $\rho=-1, \bt=0, \al=0$. In this gauge choice only the 
purely longitudinal modes are gauge fixed. 
In this gauge the propagator for the metric fluctuation field is the 
following,
\beq
\label{eq:GR_prop}
D^{\mn \rho\sg} = (\D_G^{-1})^{\mn\rho\sg} 
=  (16 \pi) \frac{f^2}{q^4} \left(
- 2 P_2^{\mn \rho\sg}
+ \frac{1}{\omega} P_s^{\mn\rho\sg} \right) 
=  \sum_{i} Y_{i}(q^2) P_i^{\mn \rho\sg}\, , 
\eeq
where $Y_i$ are the propagators for the various spin-components:
\beq
\label{eq:Yprop}
Y_2 = -(16 \pi) \frac{2f^2}{q^4} \, 
\hspace{5mm}
Y_s = (16 \pi) \frac{f^2}{\om q^4} \, .
\eeq
Here $\D_G^{\mn\al\bt}$ is the inverse propagator for the 
$h_\mn$ field including the gauge fixing and is symmetric 
in $\mn$ and $\al\bt$. As the propagator is $1/q^4$,
it doesn't allow to be decomposed further via partial fractions. 
Here the first term in eq. (\ref{eq:GR_prop}) 
arises due the presence of $C_{\mn\rho\sg}^2$ part of action, 
while the later comes from the $R^2$ part. 
In this form it is not clear how the unitarity will be 
satisfied. 

\subsection{Formalism}
\label{formal}

We employ the background field formalism and decompose the metric and 
matter fields as in eq. (\ref{eq:qmet} and \ref{eq:linearSP}) respectively, 
where we choose the background space-time to be flat. 
In order to do the one-loop computation
the action is expanded up-to second powers of the all fluctuation field ($h_\mn$,
$\chi$, $\eta$ and $\bar{\eta}$). This will result in various vertices and propagators 
that are required for the one-loop analysis. The second variation of the matter 
action is given by the following,
\bea
\label{eq:2ndvarActMat}
&&
\de^2 S_{\rm matter} = \frac{1}{2} \int {\rm d}^dx 
\biggl[
\left(\frac{1}{4} h^2 - \frac{1}{2} h_\mn h^\mn \right)
\left\{ \frac{1}{2} \pt_\al \varphi \pt^\al \varphi - \frac{\lam}{4} \varphi^4 \right\}
- \frac{1}{2} h h^\mn \pt_\mu \varphi \pt_\nu \varphi  
\notag \\
&&
+ h^{\mu\al} h_\al{}^\nu \pt_\mu \varphi \pt_\nu \varphi
- \frac{1}{2} \xi \varphi^2  \biggl(h\pt_\mu\pt_\nu h^\mn - h\Box h 
-4 h^\mn \pt_\mu \pt_\rho h^\rho{}_\nu + 2 h^\mn \Box h_\mn  + 2 h^\mn \pt_\mu\pt_\nu h 
\notag \\
&&
- 2 \pt_\rho h^{\rho\sg} \pt^\mu h_{\sg\mu}
+2 \pt_\rho h^{\rho\sg} \pt_\sg h + \frac{3}{2} \pt_\mu h^{\rho\sg} \pt^\mu h_{\rho\sg}
- \frac{1}{2} \pt_\mu h \pt^\mu h - \pt^\sg h^{\rho\mu} \pt_\rho h_{\sg\mu} 
\biggr)
\notag \\
&&
- \lam \varphi^3 h \chi 
+ h \pt_\mu \varphi \pt^\mu \chi 
- 2 h^\mn \pt_\mu \varphi \pt_\nu \chi - 2 \xi \chi \varphi (\pt_\mu\pt_\nu h^\mn - \Box h) 
+ \pt_\mu \chi \pt^\mu \chi - 3 \lam \varphi^2 \chi^2 \biggr]
\notag \\
&&
+ \int {\rm d}^dx \biggl[
\left\{
\left(\frac{1}{8} h^2 - \frac{1}{4} h_{\al\bt} h^{\al\bt} \right) \de_\rho{}^\mu
- \frac{1}{4} h h_\rho{}^\mu + \frac{3}{8} h_\rho{}^\al h_\al{}^\mu
\right\} \bar{\ta} i \g^\rho \pt_\mu \ta 
\notag \\
&&
+ \frac{i}{4} \bar{\ta} \g^\rho [\g^\al, \g^\bt] \ta \biggl\{
-\frac{1}{4} h_\al{}^\sg \pt_\rho h_{\bt\sg} 
+ \frac{1}{2} \pt_\al(h_{\bt\sg} h^\sg{}_\rho) 
+ \frac{1}{2} h_\al{}^\sg \pt_\sg h_{\bt\rho}
- \frac{1}{2} h \pt_\al h_{\bt\rho} \biggr\} 
\notag \\
&&
- y_t \varphi \left(\frac{1}{8} h^2 - \frac{1}{4} h_{\al\bt} h^{\al\bt} \right)
+ \bar{\eta} \biggl\{
\frac{i}{2} \g^\rho (h\de_\rho{}^\mu - h_\rho{}^\mu) \pt_\mu 
- \frac{i}{4} \g^\rho [\g^\al,\g^\bt] \pt_\al h_{\bt\rho}
- \frac{1}{2} y_t \varphi h
\biggr\} \ta
\notag \\
&&
+ \bar{\ta} \biggl\{
\frac{i}{2} \g^\rho (h\de_\rho{}^\mu - h_\rho{}^\mu) \pt_\mu 
- \frac{i}{4} \g^\rho [\g^\al,\g^\bt] \pt_\al h_{\bt\rho}
- \frac{1}{2} y_t \varphi h
\biggr\} \eta
+ \bar{\eta} (i\g^\rho \pt_\rho - y_t \varphi) \eta
\notag \\
&&
- y_t \left(\chi \bar{\eta} \ta + \chi \bar{\ta} \eta + \frac{1}{2} h \chi \bar{\ta} \ta \right)
\biggr] \, .
\eea
The various vertices and matter propagators are written in detail in 
appendix \ref{propver}. Having obtained the second variation giving 
propagator and the vertices, we set forth by considering the path-integral over the 
fluctuation fields. In this case the zeroth order term will be independent 
of the fluctuation fields and can be taken out of the path-integral. The linear term
can be removed by doing field redefinition. In general, terms proportional 
to equation of motion can be removed by doing field redefinition. Such a 
redefinition will give rise to a trivial jacobian from the functional measure. 
The quadratic piece can now be investigated easily by putting together 
all the field fluctuations in the form of a multiplet 
$\Phi^T=(h_\mn, \chi, \eta^T, \bar{\eta})$. Using this the path-integral 
can be written in a more compact form as,
\beq
\label{eq:Z_exp}
Z[\textbf{J}] = \exp\biggl[ i \biggl(S_{\rm GR}(\bg) + S_{\rm matter}(\varphi, \bar{\ta}, \ta) 
\biggr)\biggl]
\int \, {\cal D} \Phi
\, \exp \biggl[
i \int \, {\rm d}^dx \left(\frac{1}{2} \Phi^T \cdot \textbf{M} \cdot \Phi
+ \Phi^T \cdot \textbf{J} \right)
\biggr] \, ,
\eeq
where $\textbf{J}=\{ t_\mn, t, \rho, \bar{\rho}^T \}$ is the source multiplet
which couples with the fluctuation field multiplet
$\Phi = \left( h_{\mu\nu}, \chi, \eta^T, \bar{\eta} \right)$. 
The super matrix $\textbf{M}$ is given by 
\beq
\label{eq:Mmat}
\textbf{M} = \left[
\begin{array}{c c c c}
\Delta_G^{\mu\nu\rho\sigma} + V^{\mu\nu\rho\sigma} + U^{\mu\nu\rho\sigma} &
V_{h\phi}^{\mn} & (V_{h\psi}^\mn)_b & (V^T_{h\bar{\psi}})^\mn_b \\
V_{\phi h}^{\rho\sg} & \D_s - V_s & (V_{\phi\psi})_b & (V^T_{\phi\bar{\psi}})_b \\
(V^T_{\psi h})^{\rho\sg}_a & (V^T_{\psi \phi})_a & 0 & (\D_F^T)_{ab} + (V^T_{\bar{\psi}\psi})_{ab} \\
(V_{\bar{\psi}h}^{\rho\sg})_a & (V_{\bar{\psi}\phi})_a & (\D_F)_{ab} - (V_{\bar{\psi}\psi})_{ab} & 0 
\end{array}
\right] \, .
\eeq
From the generating functional $Z$, one can define the one-particle-irreducible (1PI)
generating functional $\G= W[\textbf{J}] - \int \, {\rm d}^dx 
\langle\Phi^T \rangle \cdot \textbf{J}$, where $W[\textbf{J}] =
-i \ln Z[\textbf{J}]$ and $\langle \Phi^T \rangle$ is the 
expectation value of $\Phi^T$ field. The 1PI generating functional 
is also the effective action containing the quantum corrections. In the 
one-loop approximation (which we are considering here), one can 
perform the functional integral over the super-field $\Phi$ thereby 
giving an expression for the one-loop effective action to be,
\beq
\label{eq:EA_1loop}
\G^{\rm 1-loop} [\Phi] = S_{\rm GR} (\bg) + S_{\rm matter}(\varphi, \bar{\ta}, \ta) 
+ \frac{i}{2} {\rm STr} \ln \textbf{M} \, ,
\eeq
where the first two terms correspond to tree-level diagrams while the last
term contains one-loop quantum corrections. The appearance of generalised 
trace `${\rm STr}$' means that
\beq
\label{eq:Str_def}
{\rm STr} \left(
\begin{array}{c c}
a & \al \\
\bt & b
\end{array}
\right) = {\rm Tr} (a) - {\rm Tr} (b) \, .
\eeq
In the following we will be computing the divergent pieces present in the 
${\rm STr} \ln \textbf{M}$. There are various ways to compute the 
one-loop quantum corrections. The most common methodology to do is 
via Feynman diagrams after computing vertices and propagator. Here 
we will follow a slightly different strategy of computation via evaluation 
of functional determinant. We start by writing $\textbf{M} = \D + \textbf{V}$,
where the former $\D$ contains the various propagator while the later
$\mathbb{V}$ contains various vertices. They are given by,
\beq
\label{eq:DandV}
\D = \left[
\begin{array}{c c c c}
\Delta_G^{\mu\nu\rho\sigma} &
0 & 0 & 0 \\
0 & \D_s  & 0 & 0 \\
0 & 0 & 0 & (\D_F^T)_{ab}  \\
0 & 0 & (\D_F)_{ab}  & 0 
\end{array}
\right] ,
\textbf{V} =
\left[
\begin{array}{c c c c}
 V^{\mu\nu\rho\sigma} + U^{\mu\nu\rho\sigma} &
V_{h\phi}^{\mn} & (V_{h\psi}^\mn)_b & (V^T_{h\bar{\psi}})^\mn_b \\
V_{\phi h}^{\rho\sg} &  - V_s & (V_{\phi\psi})_b & (V^T_{\phi\bar{\psi}})_b \\
(V^T_{\psi h})^{\rho\sg}_a & (V^T_{\psi \phi})_a & 0 &  (V^T_{\bar{\psi}\psi})_{ab} \\
(V_{\bar{\psi}h}^{\rho\sg})_a & (V_{\bar{\psi}\phi})_a &  - (V_{\bar{\psi}\psi})_{ab} & 0 
\end{array}
\right] .
\eeq
Pulling out $\D$ from the expression for $\textbf{M}$ allows to expand the 
residual expression $(\textbf{I} + \D^{-1}\cdot \textbf{V})$ (where 
$\textbf{I}$ is a generalised identity in super-field space)
under the logarithm in a perturbative manner as follows,
\beq
\label{eq:log_exp}
{\rm STr} \ln \textbf{M} 
= {\rm STr} \ln \D \cdot \bigl(\textbf{I}
+ \mathbb{V} \bigr)
= {\rm STr} \ln \D
+ {\rm STr} \biggl[
\mathbb{V}
- \frac{1}{2} \mathbb{V}^2 + \frac{1}{3} \mathbb{V}^3 - \frac{1}{4} \mathbb{V}^4 
+ \cdots
\biggr] \, .
\eeq
Here $\mathbb{V} = \D^{-1} \cdot \textbf{V}$ is given by 
\beq
\label{eq:mathbbV}
\mathbb{V} = \left[
\begin{array}{c c c c}
(\D_G^{-1})^{\mn\rho\sg} (V+U)_{\rho\sg\al\bt} &
(\D_G^{-1})^{\mn\rho\sg} (V_{h\phi})_{\rho\sg} &
(\D_G^{-1})^{\mn\rho\sg} (V_{h\psi})_{c\rho\sg} &
(\D_G^{-1})^{\mn\rho\sg} (V^T_{h\bar{\psi}})_{c\rho\sg} \\
\D_s^{-1} (V_{\phi h})_{\al\bt} & 
-\D_s^{-1} V_s & 
\D_s^{-1} (V_{\phi\psi})_c &
\D_s^{-1} (V^T_{\phi\bar{\psi}})_c \\
(\D_F^{-1})_{ab} (V_{\bar{\psi}h})_{b\al\bt} &
(\D_F^{-1})_{ab} (V_{\bar{\psi}\phi})_b &
- (\D_F^{-1})_{ab}(V_{\bar{\psi}\psi})_{bc} & 
0 \\
(\D_F^{-1})^T_{ab} (V^T_{\psi h})_{b\al\bt} &
(\D_F^{-1})^T_{ab} (V^T_{\psi \phi})_b &
0 & 
(\D_F^{-1})^T_{ab} (V^T_{\bar{\psi}\psi})_{bc} 
\end{array}
\right] \, .
\eeq
It should be mentioned here that so far we took background metric to be 
flat with $H_\mn=0$. This is enough to compute the counter-terms 
involving quantum gravity corrections to all matter couplings 
including their anomalous dimensions. If we had included 
terms with $H_\mn\neq0$, then it is also possible to compute 
the counter-term proportional to $R\varphi^2$. But here for 
simplicity we keep $H_\mn=0$, and the counter-term 
proportional to $R\varphi^2$ will be computed using methodology
of heat-kernels (HK) later. Heat-kernel method is quick, as the HK 
coefficients have already been computed in past 
\cite{DeWitt1965,Vassilevich2003,Yajima1988,Avramidi2000}.
Besides, it also gives an alternative check on the computation 
done using feynman diagrams.
For flat background the term ${\rm STr} \ln \D$ is irrelevant, but it is not 
so if the background is non-flat for which case this gives 
purely curvature dependent divergent contributions. 
Such contributions have been computed elsewhere in 
literature \cite{Avramidi1985, Avramidi2000} 
and here we will just take their results. 

In the following we will be computing the various graphs that are giving 
quantum gravity corrections to the running of the matter couplings
and the fields anomalous dimensions. 

\subsection{Graphs}
\label{graphs}

Here we will be computing the various graphs that contain 
divergent contributions. These are basically the terms in the 
series expansion given in eq. (\ref{eq:log_exp}), which will be 
evaluated one by one. The first term of the series contain 
tadpole graphs (those having single vertex), the second term 
of series has bubble graphs (those having two vertices), the third 
term of the series are triangle graphs containing three vertices 
while the fourth term of series are square graphs with four vertices. 
The series has infinite number of graphs, but the divergent ones are only 
present in the first four terms of the series expansion, and below we will 
be computing them.

\subsubsection{Tadpole}
\label{tad}

\begin{figure}
\centerline{
\vspace{0pt}
\centering
\includegraphics[width=6in]{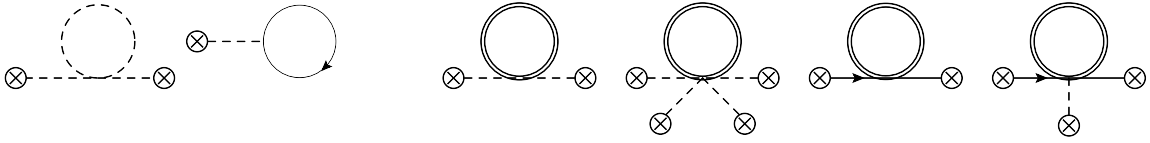}
}
 \caption[]{
Various diagrams containing divergences at the tadpole level. Here the dashed line represent 
scalar field, solid line with arrow represent fermion field while double-line represent 
$h_\mn$-field. The first two graphs are purely matter ones, while the other 
four graphs contain quantum gravity corrections. 
}
\label{fig:tad}
\end{figure}

These set of graphs arises from the first term of the series in eq. (\ref{eq:log_exp})
which is ${\rm STr} (\mathbb{V}) = {\rm STr} (\D^{-1}\cdot \textbf{V})$. Here the 
super-trace takes care of trace over bosonic and fermionic part, and includes 
the trace not only over field space but also over lorentz indices. This will imply 
the following,
\bea
\label{eq:tadTrexp}
&&
{\rm STr} \mathbb{V} = {\rm Tr} \left[
\mathbb{V}_{11} + \mathbb{V}_{22} - \mathbb{V}_{33} - \mathbb{V}_{44}
\right]
\notag \\
&&
= {\rm Tr} \left[
(\D_G^{-1})^{\mn\rho\sg} (V+U)_{\rho\sg\al\bt} 
-\D_s^{-1} V_s
+(\D_F^{-1})_{ab}(V_{\bar{\psi}\psi})_{bc}
-(\D_F^{-1})^T_{ab} (V^T_{\bar{\psi}\psi})_{bc} 
\right] \, .
\eea
Here the first term contains graphs having an internal graviton line, 
while the next three terms contains the usual diagrams which are 
present without gravity. The former gives quantum gravity contribution. The set
of graphs present in the tadpole order are shown in figure \ref{fig:tad}.

Each of these diagrams can be evaluated using the vertices given in 
appendix \ref{vert}. Here we will write their contribution. 
However, the last three terms in eq. (\ref{eq:tadTrexp}) vanish in scale-invariant theory.  
The gravitational ones are complicated and lengthy as the 
vertices are cumbersome. Below we write this 
\bea
\label{eq:scalarGRTad}
&&
{\rm Tr} \left\{ (\D_G^{-1})^{\mn\rho\sg} V_{\rho\sg\al\bt} \right\}
= \frac{1}{4}  \int {\rm d}^dx \left[
\frac{d-4}{2d} (\pt \varphi)^2 - \frac{\lam}{4} \varphi^4
\right] \int \frac{{\rm d}^dp}{\pd} 
\bigl\{ -(d-2)(d+1) Y_{2} 
\notag \\
&&
+ (d-3) Y_{s} \bigr\}
+ \frac{(d-2)\xi}{8} \int {\rm d}^dx \varphi^2 \int \frac{{\rm d}^dp}{\pd} p^2 
\left\{(d+1) Y_{2} -2Y_{s} \right\} \, .
\eea
For the other one there is more algebra as it involves 
Dirac-matrices. Here we will write the expression after performing the 
lorentz and Dirac matrix algebra. This is given by, 
\bea
\label{eq:fermiGRTad}
&&
{\rm Tr} \left\{ (\D_G^{-1})^{\mn\rho\sg} U_{\rho\sg\al\bt} \right\}
= \frac{(d-2)(d+1)}{4} \int {\rm d}^dx \biggl[
\frac{3-2d}{2d} \bar{\ta} i \g^\al \pt_\al \ta 
+ y_t \varphi \bar{\ta} \ta
\biggr] \int \frac{{\rm d}^dp}{\pd} Y_{2}
\notag \\
&&
+ \int {\rm d}^dx \biggl[
\frac{d^2-5d+5}{4d} \bar{\ta} i \g^\al \pt_\al \ta 
- \frac{d-3}{4} y_t \varphi \bar{\ta} \ta
\biggr] \int \frac{{\rm d}^dp}{\pd} Y_{s} \, .
\eea
The momentum integrals can be evaluated in the $(4-\ep)$ dimensional 
regularisation scheme and the divergent piece can be singled out easily. 
The divergent piece of all the above tadpole contribution is,
\bea
\label{eq:Taddiv}
\G^{\rm Tad}_{\rm div} &=& -\frac{\mu^\ep}{16\pi^2\ep} \frac{M^2}{Z} \biggl[
\frac{\lam}{8} \left(10+ \frac{1}{2\om} \right)  \int {\rm d}^dx \varphi^4
- \left(\frac{25}{8} + \frac{1}{16\om} \right) 
\int {\rm d}^dx \bar{\ta} (i\g^\mu \pt_\mu) \ta 
\notag \\
&&
+ 5y_t \left(1+\frac{1}{4\om} \right) 
\int {\rm d}^dx \bar{\ta} \varphi \ta 
\biggr] \, ,
\eea
where $M^2/Z = 16 \pi f^2$ is introduced for convenience. 

\subsubsection{Bubble}
\label{bubble}

\begin{figure}
\centerline{
\vspace{0pt}
\centering
\includegraphics[width=3in]{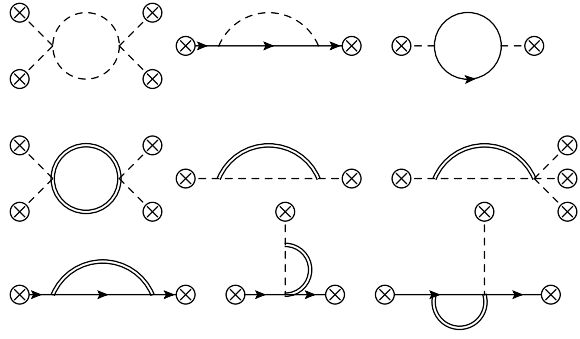}
}
 \caption[]{
Various diagrams containing divergences with two vertices. 
Here the dashed line represent scalar field, solid line with 
arrow represent fermion field while double-line represent $h_\mn$-field. 
Here the graphs on the first line are purely matter ones. The second 
and third line contain graphs having quantum gravity corrections. 
}
\label{fig:bub}
\end{figure}

These set of graphs arise from second term in eq. (\ref{eq:log_exp})
which is $-1/2 {\rm STr} (\D^{-1}\cdot \textbf{V})^2$. Here again the 
super trace is evaluated as before. This will imply,
\beq
\label{eq:BubTrexp}
{\rm STr} \mathbb{V}^2 = {\rm Tr} \left[ \left(\mathbb{V}^2\right)_{11}
+ \left(\mathbb{V}^2\right)_{22} - \left(\mathbb{V}^2\right)_{33}
- \left(\mathbb{V}^2\right)_{44} \right] \, .
\eeq
Here each of the term will contain several diagrams, but only few 
contain the divergences that are relevant for our purpose. These 
diagrams contain two vertices. They can be classified in three categories:
(a) those with two internal graviton lines, (b) those with one internal
graviton and one internal matter line and (c) those with two internal matter lines.
The set of diagrams are shown in figure \ref{fig:bub}.

%
Each of these diagrams can be evaluated using the vertices given in 
appendix \ref{vert}. The super-trace given in eq. (\ref{eq:BubTrexp})
contains lot of diagrams, but not all contain UV divergence. Here we 
will mention only the ones having the UV divergences. These come from,
\bea
\label{eq:divBubcont}
&&
{\rm Tr} \bigl[
\D_s^{-1} V_s \D_s^{-1} V_s 
+ 2 \D_s^{-1} (V_{\phi\psi})_c (\D_F^{-1})_{cd} (V_{\bar{\psi}\phi})_d
+ 2 \D_s^{-1} (V_{\phi\psi}^T)_c (\D_F^{-1})^T_{cd} (V^T_{\psi\phi})_d
\notag \\
&&
- (D_F^{-1})_{ab} (V_{\bar{\psi}\psi})_{bd} (\D_F^{-1})_{de} (V_{\bar{\psi}\psi})_{ec}
- (D_F^{-1})^T_{ab} (V^T_{\bar{\psi}\psi})_{bd} (\D_F^{-1})^T_{de} (V^T_{\bar{\psi}\psi})_{ec}
\notag \\
&&
+ \left(\D_G^{-1}\right)^{\mn\rho\sg} V_{\rho\sg\al\bt} 
\left(\D_G^{-1}\right)^{\al\bt\ta\tau} V_{\ta\tau\mu'\nu'}
+ 2 \left(\D_G^{-1}\right)^{\mn\rho\sg} (V_{h\phi})_{\rho\sg} 
\D_s^{-1} (V_{\phi h})_{\mu'\nu'}
\notag \\
&&
+ 2 \left(\D_G^{-1}\right)^{\mn\rho\sg} (V_{h\psi})_{\rho\sg c}
(\D_F^{-1})_{cd} (V_{\bar{\psi}h})_{d\mu'\nu'}
\notag \\
&&
+ 2 \left(\D_G^{-1}\right)^{\mn\rho\sg} (V^T_{h\bar{\psi}})_{c\rho\sg}
(\D_F^{-1})^T_{cd} (V^T_{\psi h})_{d\mu'\nu'}
\bigr] \, .
\eea
From the various terms written in eq. (\ref{eq:divBubcont}) the first 
two lines contain diagrams which are purely matter ones and 
correspond to the diagrams shown in first row of figure \ref{fig:bub}, while
the last two lines contain diagrams having quantum contributions 
and correspond to the diagrams shown in last two rows of the 
figure \ref{fig:bub}. The trace is over the Lorentz and space-time 
indices. After performing the algebra over Dirac matrices and 
doing the contraction of the tensors we get the simplified expression 
involving momentum integrals. The divergent contributions of the purely 
matter diagrams is,
\beq
\label{eq:BubDivNGR}
\frac{i}{16\pi^2} \frac{1}{\ep} \biggl[
9 \lam^2 \int {\rm d}^4x \varphi^4 
+ 2 y_t^2 \int {\rm d}^4x 
\biggl\{
\bar{\ta} i \slashed{\pt} \ta + 2 \pt_\mu\varphi \pt^\mu \varphi 
\biggr\}
\biggr] \, .
\eeq
The diagrams containing the internal graviton legs are bit complicated,
as it involves lengthy Dirac matrix algebra and tensor manipulations. 
For doing these we have used various tricks to extract the divergent piece 
and also used MATHEMATICA packages (xAct \cite{xAct}, xTras \cite{xTras} 
and FEYNCALC \cite{Mertig1990}). 
Below for simplicity we will mention only the divergent piece 
of these diagrams to evade unnecessary complications, while some 
of the details will be mentioned in the appendix. The diagrams 
with internal graviton line has the following contributions
\bea
\label{eq:BubDivGR}
&&
\frac{i}{16\pi^2} \frac{1}{\ep} \biggl[
\biggl\{
\frac{\xi^2}{4} \left(\frac{M^2}{Z}\right)^2 \left(5+\frac{1}{\om^2} \right)
+ \frac{6 \xi \lam M^2}{Z\om} \biggr\}
\int {\rm d}^4x \varphi^4
+ \frac{3}{8} \frac{M^2}{Z\om} \int {\rm d}^4x \pt_\mu \varphi \pt^\mu \varphi 
\notag \\
&&
+ \frac{M^2}{Z\om} \biggl\{
\frac{9}{4} \int {\rm d}^4x \bar{\ta} i\slashed{\pt} \ta
+3(2\xi-1) y_t \int {\rm d}^4x \varphi \bar{\ta} \ta 
\biggr\}
\biggr] \, .
\eea
Here the first row contains contributions to the scalar sector, while the 
second row contain contributions to the fermion sector. The former 
correspond to diagrams of the second row in figure \ref{fig:bub}, while 
the later correspond to diagrams in the third row of figure \ref{fig:bub}
respectively.  

Putting together the full contribution of the bubble kind of diagrams, we get
contribution to the one-loop effective action of the diagrams having two
vertices. This is given by,
\bea
\label{eq:G1LPbub}
&&
\G^{\rm Bub}_{\rm div} = - \frac{1}{16\pi^2} \frac{1}{\ep} 
\biggl[
\left\{
2y_t^2 + \frac{3M^2}{16 Z\om} 
\right\} \int {\rm d}^4x \pt_\mu \varphi \pt^\mu \varphi 
+ \biggl\{
\frac{9\lam^2}{2}  + \frac{3\xi\lam M^2}{Z\om}
\notag \\
&&
+ \frac{1}{2} \left(\frac{\xi M^2}{2Z}\right)^2 
\left(5+ \frac{1}{\om^2} \right)
\biggr\} \int {\rm d}^4x \varphi^4
+ \left(y_t^2 + \frac{9M^2}{8Z\om} \right) 
\int {\rm d}^4x \bar{\ta} i \slashed{\pt} \ta
\notag \\
&&
+ \frac{3(2\xi-1) y_t}{2} \frac{M^2}{Z\om} \int {\rm d}^4 x \varphi\bar{\ta} \ta
\biggr] \, .
\eea
%

\subsubsection{Triangular Graphs}
\label{triangle}

These diagram are generated from the third order terms in the
series of eq. (\ref{eq:log_exp}), $1/3 {\rm STr} \mathbb{V}^3$ where
\beq
\label{eq:thirdExp}
{\rm STr} \mathbb{V}^3 = {\rm Tr}
\left[ \left(\mathbb{V}^3\right)_{11} +  \left(\mathbb{V}^3\right)_{22}
- \left(\mathbb{V}^3\right)_{33} -  \left(\mathbb{V}^3\right)_{44} \right] \, .
\eeq
These diagrams have three vertices and consist of graphs which are 
either purely matter oriented or ones which include a mixture of matter and 
gravity. These graphs give correction to vertex: either to $\varphi^4$ or 
to yukawa vertex $\varphi \bar{\ta}\ta$. On expansion we will see that 
there are many diagrams but we will consider those which carry divergent 
pieces and give correction to running couplings. 
%
\begin{figure}
\centerline{
\vspace{0pt}
\centering
\includegraphics[width=5in]{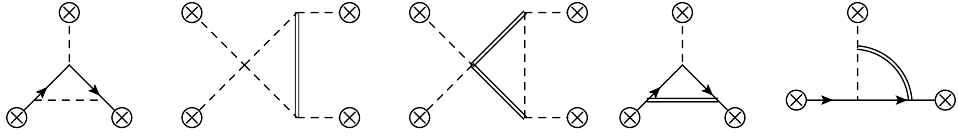}
}
 \caption[]{
Various diagrams containing divergences with three vertices. 
Here the first graph is purely matter oriented and gives correction 
to yukawa coupling. The next two diagrams are giving correction 
to $\varphi^4$ coupling. They are only present in the quantum 
gravity context. The last two diagrams are giving quantum 
gravity correction to the yukawa coupling.  
}
\label{fig:triangle}
\end{figure}
%
Here we will only mention the trace pieces which will be carrying the divergences,
however in principle there will be many more diagrams. The relevant terms in the 
trace which will be of relevance can be guessed by looking at the set of third 
order diagrams in figure \ref{fig:triangle}. These are given by,
\bea
\label{eq:TRtriangle}
&&
3 {\rm Tr} \bigl[
\D_s^{-1} (V^T_{\phi\bar{\psi}})_a (\D_F^{-1})^T_{ab} (V^T_{\bar{\psi}\psi})_{bc}
(\D_F^{-1})^T_{cd} (V^T_{\psi \phi})_d
-\D_s^{-1} (V_{\phi\psi})_a (D_F^{-1})_{ab} (V_{\bar{\psi}\psi})_{bc}
(\D_F^{-1})_{cd} (V_{\bar{\psi}\phi})_d
\notag \\
&&
+\left(\D_G^{-1}\right)^{\mn\rho\sg} 
\bigl\{
(V_{h\phi})_{\rho\sg} \D_s^{-1} 
(V_{\phi h})^{\al\bt} \left(\D_G^{-1}\right)_{\al\bt\ta\tau} V^{\ta\tau\g\de}
- (V_{h\phi})_{\rho\sg} \D_s^{-1} 
V_s \D_s^{-1} (V_{\phi h})^{\al\bt} 
\notag \\
&&
+ (V_{h\phi})_{\rho\sg} \D_s^{-1}
(V_{\phi\psi})_a (\D_F^{-1})_{ab} (V_{\bar{\psi}h})_{b\al\bt}
+ (V_{h\phi})_{\rho\sg} \D_s^{-1} (V^T_{\phi\bar{\psi}})_a (\D_F^{-1})^T_{ab} (V^T_{\psi h})_{b\mn}
\notag \\
&&
+ (V_{h\psi})_{a\rho\sg} (\D_F^{-1})_{ab} (V_{\bar{\psi} \phi})_b \D_s^{-1} (V_{\phi h})_{\al\bt}
+ (V^T_{h\bar{\psi}})_{a\rho\sg} (D_F^{-1})^T_{ab} (V_{\psi\phi}^T)_b \D_s^{-1} (V_{\phi h})_{\al\bt}
\notag \\
&&
- (V_{h\psi})_{a\rho\sg} (\D_F^{-1})_{ab} (V_{\bar{\psi}\psi})_{bc} (\D_F^{-1})_{cd} (V_{h\psi})_{d\al\bt}
\notag \\
&&
+ (V_{h\bar{\psi}}^T)_{a\rho\sg}(\D_F^{-1})^T_{ab} (V^T_{\bar{\psi}\psi})_{bc} 
(\D_F^{-1})^T_{cd} (V^T_{\psi h})_{d\al\bt}
\bigr\}
\bigr] \, .
\eea
Here the first line correspond to set of diagrams of purely matter type, while 
the rest of terms contain quantum gravity corrections. The second line 
contribute to running of $\varphi^4$ coupling while the rest of the terms 
correspond to the running of yukawa coupling. Interestingly while doing the 
computation involving fermions it is noticed that not all of terms are non-zero. 
The divergent contributions of these diagrams and corresponding their
contribution to the effective action is given by,
\bea
\label{eq:triangleDiv}
\G^{\rm Triangle}_{\rm div} = -\frac{1}{16\pi^2} \frac{1}{\ep} \int {\rm d}^4x \biggl[
\left(2y_t^3 - \frac{3M^2 \xi}{Z\om} y_t \right) \varphi \bar{\ta}\ta
+ \left(\frac{9M^2}{Z\om} \lam \xi^2 + \frac{3M^4}{2Z^2 \om^2} \xi^3 \right)
\varphi^4 
\biggr] \, .
\eea
%

\subsubsection{Square Graphs}
\label{square}

These set of contribution arise at fourth order of the perturbative 
expansion given in the series eq. (\ref{eq:log_exp}) and comes 
from $-1/4 {\rm STr} \mathbb{V}^4$. Here again the super-trace is given by,
\beq
\label{eq:fourthexp}
{\rm STr} \mathbb{V}^4 = {\rm Tr} \left[
\left(\mathbb{V}^4\right)_{11} + \left(\mathbb{V}^4\right)_{22} 
- \left(\mathbb{V}^4\right)_{33} - \left(\mathbb{V}^4\right)_{44}
\right] \, .
\eeq
These diagrams consist of 
four vertices and all of them contribute to the running of 
$\varphi^4$ coupling. There are only two diagrams at this order. 
One is purely matter type, contains a fermion loop with four 
external scalar legs, while the other one contains quantum gravity 
correction. 
%
\begin{figure}
\centerline{
\vspace{0pt}
\centering
\includegraphics[width=2.5in]{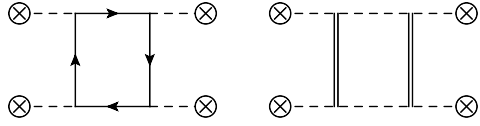}
}
 \caption[]{
Various diagrams containing divergences with four vertices. 
Here there are only two diagrams. The first one is purely matter 
while the second one contain quantum gravity correction. 
Both give correction to the $\varphi^4$ coupling.   
}
\label{fig:square}
\end{figure}
%
The trace can be expanded as before and consists of large number of graphs 
but the ones containing the divergences are only two. These are the following,
\bea
\label{eq:forthTr}
&&
2 {\rm Tr} \bigl[
- (\D_F^{-1})_{ab} (V_{\bar{\psi}\psi})_{bc} (\D_F^{-1})_{cd}
(V_{\bar{\psi}\psi})_{de} (\D_F^{-1})_{ef} (V_{\bar{\psi}\psi})_{fg}
(\D_F^{-1})_{gk} (V_{\bar{\psi}\psi})_{kl}
\notag \\
&&
+ \left(\D_G^{-1}\right)^{\mn\rho\sg} (V_{h\phi})_{\rho\sg} \D_s^{-1} 
(V_{\phi h})^{\al\bt} \left(\D_G^{-1}\right)_{\al\bt\ta\tau}
(V_{h\phi})^{\ta\tau} \D_s^{-1} (V_{\phi h})_{\g\de}
\bigr] \, .
\eea
Here the former term is purely matter and contains a fermion loop, while the second 
term contain virtual gravitons. Considering as before just the divergent part
and their corresponding contribution to the effective action, we find that,
\beq
\label{eq:EAfourth}
\G^{\rm Square}_{\rm div} = - \frac{1}{\ep} \frac{1}{16\pi^2}  
\biggl(
-2 y_t^4 + \frac{9M^4}{2 Z^2 \om^2} \xi^4 
\biggr) \int {\rm d}^4x \varphi^4 \, .
\eeq
%

\subsubsection{$R\varphi^2$ divergence}
\label{Rphi2}

Here we compute the divergence proportional to $R\varphi^2$. There 
are two ways to compute it. The first is via computation of 
feynman graphs and second via heat kernel. Conceptually both are 
same and give same results, however the later is quicker. In each 
case we break the metric $\g_\mn$ appearing in path-integral
is written as in eq. (\ref{eq:qmet}). In the former the background is further 
written as $\eta_\mn + H_\mn$ (see also paragraph following eq. (\ref{eq:qmet})). 
Here $H_\mn$ will act as external 
graviton for the corresponding internal leg $h_\mn$. The action 
is then expanded under this decomposition. This way we get additional 
vertices. The vertices in the previous section are the ones for 
which $H_\mn=0$. If $H_\mn\neq0$ then we get contributions 
which contains dependence on external graviton leg, and if there 
are derivatives acting on $H_\mn$, then those will give terms 
proportional to background curvature. This was employed in \cite{Strumia1}.

Alternatively, one can take the background metric $\bg_\mn$ 
to be maximally symmetric and compute the contributions 
proportional to background curvature using Heat-Kernel. 
This will project directly the contribution proportional to $\bar{R}\varphi^2$.
We use the heat-kernel methodology to compute the 
one-loop divergence proportional to $\bar{R}\varphi^2$.
The matter fields are decomposed as in eq. (\ref{eq:linearSP}), 
but this time we take background matter fields to be constant.
The fluctuation metric $h_\mn$ is further decomposed 
into various components as
\beq
\label{eq:TTdecompA}
h_\mn = h^{TT}_\mn + \bnb_\mu \xi_\nu +\bnb_\nu \xi_\mu
+ \bnb_\mu\bnb_\nu \sg - \frac{1}{4} \bg_\mn \bar{\Box} \sg 
+ \frac{1}{4} \bg_\mn h \, ,
\eeq
where 
\begin{gather}
\label{eq:fdcond}
h^{TT}_\mu{}^\mu=0 \, ,
\hspace{3mm}
\bnb^\mu h^{TT}_\mn =0 \, ,
\hspace{3mm}
\bnb^\mu \xi_\mu =0 \, .
\end{gather}
This decomposition of $h_\mn$ introduces jacobians in the 
path-integral, which can be cancelled by redefining the 
fields as
\beq
\label{eq:Fredef}
\hat{\xi}_\mu = \sqrt{-\bar{\Box} - \frac{\bar{R}}{4}} \xi_\mu \, ,
\hspace{3mm}
\hat{\sg} = \sqrt{-\bar{\Box}\left(-\bar{\Box} - \frac{\bar{R}}{3} \right)} \sg \, .
\eeq
Under this decomposition the Hessian for the fluctuation fields is obtained 
on a maximally symmetric background. This will be same as in 
eq. (\ref{eq:Z_exp}), except now $\textbf{M}$ will be different. The multiplet 
$\Phi$ also gets modified $\Phi^T = \left(h_\mn^{TT}, \hat{\xi}_\mu, \hat{\sg},
h, \chi, \eta^T, \bar{\eta} \right)$. As the background matter fields are 
constant, therefore the fermion sector is completely decoupled with mixing of
fermion fluctuation field with metric and scalar fluctuations. This is not 
so when background matter is not constant. On the maximally symmetric 
background metric with constant background matter the matrix $\textbf{M}$ 
is given by (in the Landau gauge $\rho=-1$)
\bea
\label{eq:2ndvarG}
&&
\int {\rm d}^dx \sqrt{-\bg} \Phi^T \textbf{M} \Phi 
= \int {\rm d}^dx \sqrt{-\bg} 
\biggl[h^{TT}_\mn \D_2 (h^{TT})^\mn
+ \hat{\xi}_\mu \D_1 \hat{\xi}^\mu
\notag \\
&&
+ \left(
\begin{array}{c c c}
\hat{\sg} & h & \chi
\end{array}
\right)
\left(
\begin{array}{c c c}
S_{\hat{\sg}\hat{\sg}} & S_{\hat{\sg}h} & S_{\hat{\sg}\phi} \\
S_{h\hat{\sg}} & S_{hh} & S_{h\phi} \\
S_{\phi \hat{\sg}} & S_{\phi h} & S_{\phi\phi} 
\end{array}
\right)
\left(
\begin{array}{c}
\hat{\sg} \\
h \\
\chi
\end{array}
\right)
+ \frac{1}{2} \bar{\eta} \D_{1/2} \eta 
- \frac{1}{2} \eta^T \D_{1/2}^T \bar{\eta}^T
\biggr] \, ,
\eea
where 
\bea
\label{eq:hess1L}
&&
\D_2 = \biggl\{
- \frac{Z}{M^2} \bar{\Box}^2 + \biggl( - \xi \varphi^2 
+ \frac{(3+2\om) Z \bar{R}}{3M^2} \biggr) \frac{\bar{\Box}}{4}
+ \biggl(
\frac{\lam \varphi^4}{8} 
+ \frac{\xi \bar{R} \varphi^2}{6}
+ \frac{(1+\om)Z \bar{R}^2}{36M^2} 
\biggr)
\biggr\} \, ,
\notag \\
&&
\D_1 = \biggl\{ \frac{\bar{\Box}^2}{\al} + \frac{\bar{R}\bar{\Box}}{4\al} 
+ \frac{\lam \varphi^4 +\xi \bar{R} \varphi^2}{4} \biggr\} \, ,
\notag \\
&&
S_{\hat{\sg}\hat{\sg}} = \biggl[\left(\frac{3}{\al}+ \frac{Z\om}{M^2} \right) \frac{3\bar{\Box}^2}{16}
+ \left(\frac{6\al\xi \varphi^2 + 21 \bar{R}}{64\al}\right)\bar{\Box}
+ \frac{3 \lam \varphi^4}{32} + \frac{3\xi \bar{R} \varphi^2}{32} 
+ \frac{3\bar{R}^2}{64\al} \biggr] \, ,
\notag \\
&&
S_{h\hat{\sg}} = S_{\hat{\sg}h} = \sqrt{-\bar{\Box}\left(-\bar{\Box} - \frac{\bar{R}}{3} \right)}
\left[\left(\frac{1}{\al} - \frac{Z\om}{M^2} \right) \frac{3\bar{\Box}}{16} 
+ \frac{3\bar{R} - 6\al \xi \varphi^2}{64\al} \right] \, ,
\notag \\
&& 
S_{\hat{\sg}\phi} = S_{\phi \hat{\sg}} 
= - \frac{3 \xi \varphi}{4} \sqrt{-\bar{\Box}\left(-\bar{\Box} - \frac{\bar{R}}{3} \right)} \, ,
\hspace{3mm}
S_{h\phi} = S_{\phi h} =\frac{(3\xi \bar{\Box} - \xi \bar{R} - 2\lam \varphi^2)\varphi}{4} \, ,
\notag \\
&&
S_{hh} = \biggl[
\left(\frac{1}{\al} + \frac{3Z\om}{M^2} \right)\frac{\bar{\Box}^2}{16}
+ \left\{ \left(\frac{1}{\al} + \frac{4Z\om}{M^2} \right) \frac{\bar{R}}{64}
+ \frac{3}{32} \xi \varphi^2 \right\} \bar{\Box} - \frac{\lam \varphi^4}{32}
\biggr] \, ,
\notag \\
&&
S_{\phi\phi} = - \bar{\Box} - \xi \bar{R} - 3 \lam \varphi^2 \, ,
\notag \\
&&
\D_{1/2}=(i \g^\mu \bnb_\mu - m - y_t \varphi) \, ,
\hspace{5mm}
\D^T_{1/2}=(-i \g^{T\mu} \bnb_\mu - m - y_t \varphi) \, .
\eea
We will not be considering the contribution for the 
ghost here, as they will not contribute at 
one-loop to the term proportional to $\bar{R}\varphi^2$. 
The one-loop effective action is given by, 
\bea
\label{eq:divEAR}
&&
\G^{(1)} = \frac{(d+1)(d-2)i}{4} {\rm Tr} \ln \D_2
+ \frac{(d-1)i}{2} {\rm Tr} \ln \D_1
\notag \\
&&
+ \frac{i}{2} {\rm Tr} \ln \left[\det \left(
\begin{array}{c c c}
S_{\hat{\sg}\hat{\sg}} & S_{\hat{\sg}h} & S_{\hat{\sg}\phi} \\
S_{h\hat{\sg}} & S_{hh} & S_{h\phi} \\
S_{\phi \hat{\sg}} & S_{\phi h} & S_{\phi\phi} 
\end{array}
\right) \right] 
- i {\rm Tr} \ln \D_{1/2} \, .
\eea
These functional traces can be tackled using heat-kernel 
\cite{DeWitt1965,Vassilevich2003,Yajima1988,Avramidi2000}. 
One can compute the divergent part of the effective action from this. 
Since the background matter fields are not constant therefore this 
will however not be able to give the anomalous dimensions of the 
matter fields. However, the anomalous dimension has already been computed earlier using 
feynman diagram therefore it will not be considered again here. 
Here we will just look the divergent contribution proportional to 
$\bar{R}\varphi^2$, which is given by, 
\beq
\label{eq:Rphi2div}
\G^{R\varphi^2}_{\rm div} =  - \frac{1}{16\pi^2 \ep}  
\biggl[ \frac{\lam}{2} (1+6\xi) + \frac{1}{3} y_t^2 
+ \frac{M^2}{Z\om} \frac{\xi}{48} (7 + 120 \xi + 144 \xi^2)
- \frac{5 M^2}{6Z} \xi \om 
\biggr] \int {\rm d}^4x \sqrt{-\bg} \bar{R} \varphi^2 \, .
\eeq
Here the first two terms arise due to matter loop while the rest 
of terms contain quantum gravity corrections. This is in 
agreement with \cite{Strumia1}.

\subsection{Effective action and Beta Functions}
\label{betaEA}

Once we have computed all the relevant graphs at various order of the perturbation 
theory and their divergent contributions, it is easy to put them together to write the 
divergent part of effective action and collect all the pieces together. The divergent 
part of the full effective action is given by,
\bea
\label{eq:EAfulldiv}
&&
\G^{(1)}_{\rm div} = - \frac{1}{16\pi^2 \ep}  \int {\rm d}^4x
\biggl[
\biggl( \frac{3 M^2}{16 Z \om} + 2 y_t^2 \biggr) (\pt \varphi)^2
+ \biggl\{
\frac{9\lam^2}{2} - 2 y_t^4 + \frac{5M^2}{8Z} 
\left(2\lam + \frac{M^2}{Z} \xi^2 \right)
\notag \\
&&
+ \frac{M^2}{Z\om} \frac{\lam}{16} (1+48\xi + 144 \xi^2) 
+ \frac{M^4}{8Z^2 \om^2} \xi^2 (1+6\xi)^2
\biggr\} \varphi^4
+ \biggl\{
\frac{\lam}{2} (1+6\xi) + \frac{1}{3} y_t^2 
\notag \\
&&
+ \frac{M^2}{Z\om} \frac{\xi}{48} (7 + 120 \xi + 144 \xi^2)
- \frac{5 M^2}{6Z} \xi \om 
\biggr\} R \varphi^2 
+ \biggl\{
y_t^2 - \frac{25M^2}{8Z} + \frac{17M^2}{16 Z \om} 
\biggr\} \bar{\ta} i \slashed{\pt} \ta 
\notag \\
&&
+ \biggl\{
2y_t^3 + \frac{5M^2}{Z} y_t - \frac{5M^2}{4Z\om} y_t 
\biggr\} \varphi \bar{\ta} \ta
\biggr] \, .
\eea
Once the divergent part of the effective action is written it is 
easy to compute the beta-function of the various couplings by 
incorporating the effect of the wave-function renormalisation. 
These are given by,
\bea
\label{eq:betas_Mat1}
\eta_\varphi &=& 
\frac{{\rm d} \ln Z_\varphi}{{\rm d} t} =
\frac{1}{16\pi^2} \biggl[
\frac{3M^2}{8 Z \om} + 4 y_t^2
\biggr] \, , 
\\
\label{eq:betas_Mat2}
\eta_\psi &=& 
\frac{{\rm d} \ln Z_\psi}{{\rm d} t} =
\frac{1}{16\pi^2} \biggl[
y_t^2 - \frac{25M^2}{8 Z } +\frac{17M^2}{Z \om}
\biggr] \, ,
\\
\label{eq:betas_Mat3}
\frac{{\rm d} \lam}{{\rm d} t} &=& \frac{1}{16\pi^2} \biggl[
18\lam^2 + 8 \lam y_t^2 - 8y_t^4 + \frac{\lam M^2}{Z} \biggl(
5 + \frac{(6\xi+1)^2}{\om} \biggr) + \frac{M^4\xi^2}{2Z^2} \biggl(
5 + \frac{(6\xi+1)^2}{\om^2} \biggr) \biggr] \, ,
\\
\label{eq:betas_Mat4}
\frac{{\rm d} \xi}{{\rm d} t} &=& \frac{1}{16\pi^2} \biggl[
\left(\xi+\frac{1}{6}\right)
\left\{4 y_t^2 +6\lam + \frac{2M^2\xi}{Z\om} (3\xi+2)\right\} - \frac{5M^2}{3Z}\xi \om
\biggr] \, ,
\\
\label{eq:betas_Mat5}
\frac{{\rm d} y_t}{{\rm d} t} &=& \frac{y_t}{16\pi^2} \biggl[
5y_t^2 + \frac{15M^2}{8Z} \biggr] \, ,
\eea
where $t=\ln(\mu/\mu_0)$ ($\mu_0$ is a reference scale) 
and ${\rm d}/ {\rm d}t = \mu {\rm d}/{\rm d} \mu$. 
Here $Z_\varphi$ and $Z_\psi$ is the wave-functional renormalisation of 
scalar and fermion field respectively, while $\eta_\varphi$ and $\eta_\psi$
is the corresponding anomalous dimension. 
The beta-functions obtained here agree fully with \cite{Strumia1,Salvio2016}, 
while there is partial agreement with 
\cite{Buchbinder1992, Elizalde1995, Gorbar2002,Gorbar2003,Gorbar2003yp, Yoon1996}. 
For completeness we also include the running of gravitational couplings 
which has been taken from past literature \cite{Avramidi1985, Avramidi2000}. 
These are given by,
\bea
\label{eq:betas_GR1}
&&
\frac{{\rm d}}{{\rm d} t} \left(\frac{Z}{M^2}\right)
= -\frac{1}{16\pi^2} \biggl[
\frac{133}{10} + \frac{N_s + 6 N_f}{60} 
\biggr] \, , \\
&&
\label{eq:betas_GR2}
\frac{{\rm d}}{{\rm d} t} \left(\frac{Z\om}{M^2}\right)
= \frac{1}{16\pi^2} \biggl[
\frac{5}{3} \om^2 + 5\om + \frac{5}{6} + 3N_s \left(\xi+\frac{1}{6} \right)^2 \biggr] .
\eea
We will be doing the RG analysis of the couplings and exploring the issue of 
unitarity later.  

\section{Effective Potential}
\label{effpot}

Here we compute the effective potential for the background scalar field $\varphi$
which gets contributions not only from matter fields but also from gravitational 
sector.

To compute the effective potential for scalar, the background scalar field 
is taken to be constant. This is sufficient to compute the effective potential. The quantum 
gravitational fluctuations are considered around a flat background. 
The fermion fields are likewise decomposed into a constant background 
(which for simplicity is taken to be zero $\bar{\ta}=\ta=0$) plus 
fluctuations. This simplifies the computation very much. 
As the ghost action doesn't depend on the background 
scalar field $\varphi$, therefore there is no contribution by the 
ghost to the effective potential, and hence will be ignored in the following. 
Once the full second variation of the action is performed, 
we have the hessian needed to compute the one-loop effective potential. 
This can be obtained directly from eq. (\ref{eq:2ndvarG}) by putting background 
$\bar{R}=0$ and replacing background covariant derivative with partial derivative.
Being on flat background allows the freedom to work 
directly in momentum space. 

Moreover, in flat space-time the transverse-traceless decomposition 
of $h_\mn$ given in eq. (\ref{eq:TTdecompA}) can be rewritten in 
an alternative form. In this new decomposition the field components 
$\sg$ and $h$ of $h_\mn$ are replaced by $s$ and $w$. These new 
fields $s$ and $w$ are related to old ones in the following manner
\beq
\label{eq:sgh_sw}
s=\frac{h-\Box \sg}{d} \, , \hspace{5mm}
w = \frac{h + (d-1) \Box \sg}{d} \, .
\eeq
The advantage of doing this transformation is to bring 
out the scalar mode which remains invariant under 
diffeomorphism transformation stated in eq. (\ref{eq:gaugetrh}).
The field $s$ is therefore gauge invariant, while the field $w$ is longitudinal. 
So the decomposition of $h_\mn$ has two gauge-invariant 
components $h^{TT}_\mn$ and $s$, with two longitudinal 
components $\hat{\xi}_\mu$ and $w$. Furthermore, on flat space-time 
one can also use the set of orthogonal projectors to project 
$h_\mn$ on various components $h_\mn^{TT}$, $\hat{\xi}_\mu$, 
$s$ and $w$ (see appendix \ref{app_proj}). In terms of new 
field variables, the hessian mentioned in eq. (\ref{eq:2ndvarG})
can be rewritten (for $R=0$) in a more transparent manner
to see clearly the gauge-dependent and gauge-independent part. The 
hessian for $h^{TT}_\mn$ and $\hat{\xi}_\mu$ remans same, 
while the mixing matrix of $\hat{\sg}$, $h$ and $\chi$ gets 
rotated due to field transformation stated in eq. (\ref{eq:sgh_sw}). 
The new mixing between the field variables $s$, $w$ and $\chi$
has a simplified structure. Moreover, this transformation of 
field variable is unaccompanied by 
any non-trivial jacobian in the path-integral. 
The one-loop effective potential is therefore obtained from
a simplified hessian,
\bea
\label{eq:EPTr}
&&
\G^{(1)} = \frac{(d+1)(d-2)i}{4} {\rm Tr} \ln \biggl\{
- \frac{Z}{M^2} \Box^2 -  \frac{\xi}{2} \varphi^2 \Box 
+ \frac{\lam}{4} \varphi^4
\biggr\}
+ \frac{(d-1)i}{2} {\rm Tr} \ln \biggl\{
\frac{2}{\al} \Box^2 + \frac{\lam}{2} \varphi^4 
\biggr\}
\notag \\
&&
+ \frac{i}{2} {\rm Tr} \ln \left[\det \left(
\begin{array}{c c c}
S_{ss} & S_{sw} & S_{s\phi} \\
S_{ws} & S_{ww} & S_{w\phi} \\
S_{\phi s} & S_{\phi w} & S_{\phi\phi} 
\end{array}
\right) \right] 
-i {\rm Tr} \ln (i \g^\mu \pt_\mu - y_t \varphi) \, ,
\eea
where the entries of the scalar mixing matrix are,
\bea
\label{eq:scamat}
&&
S_{ss} = (d-1) \biggl[
\frac{(d-2)Z \om}{M^2} \Box^2 + \frac{(d-2)}{2d M^2} \xi \varphi^2 \Box
- \frac{(d-3) \lam}{8} \varphi^4 \biggr]\, ,
\notag \\
&&
S_{sw} = S_{ws} = -\frac{(d-1)\lam}{8} \varphi^4 \, ,
\hspace{3mm}
S_{s\phi} = S_{\phi s} = - (d-1)\varphi \bigl[
2\xi \Box -  \lam \varphi^2
\bigr] \, ,
\notag \\
&&
S_{ww} = 
\frac{2}{\al} \Box^2 + \frac{\lam}{8} \varphi^4 \, ,
\hspace{3mm} 
S_{w\phi} = S_{\phi w} = -\lam \varphi^3 \, , 
\hspace{5mm} 
S_{\phi\phi} = -2 \Box - 6 \lam \varphi^2 \, .
\eea
From the entries of mixing matrix we clearly notice that 
$S_{ss}$, $S_{sw}$, $S_{s\phi}$ doesn't depend on gauge parameter. 
The only gauge dependence is in $S_{ww}$. 

For a generic case with an arbitrary field variable, the one-loop hessian can be 
written in the form $(-\Box - m^2)$ (where is $m^2$ contain background field 
contributions and couplings). In this case the effective potential is given by 
the general formula
\beq
\label{eq:veff_gen}
V^{(1)}_{\rm eff} = \frac{d_s (m^2)^2}{64\pi^2} \left(
\ln \frac{m^2}{\mu^2} - \frac{3}{2} \right) \, ,
\eeq
where $d_s$ is the factor coming due to the degrees of freedom 
of the field. The term $3/2$ in the bracket can be absorbed by 
rescaling $\mu^2$ as $\bar{\mu}^2 = \mu^2 e^{3/2}$. 
By exploiting this generic formula one can write the contribution to the 
effective potential from the various field modes of the metric fluctuation 
field, the scalar and the fermion field. In the case for spin-2, the differential operator 
responsible for the contribution can be factored and has the 
form $(-\Box - A_1 \varphi^2)(-\Box - A_2 \varphi^2)$ where $A_1$ 
and $A_2$ are given by,
\bea
\label{eq:A1A2}
&&
A_1 = -2 \sqrt{\pi} \big[ 
\sqrt{f^2 (4 \pi f^2 \xi^2 + \lam)}
-2 \sqrt{\pi} f^2 \xi
\bigr] \, ,
\notag \\
&&
A_2 = 2 \sqrt{\pi} \bigl[
\sqrt{f^2 (4 \pi f^2 \xi^2 + \lam)} 
+ 2 \sqrt{\pi} f^2 \xi\bigr] \, .
\eea
respectively. Here both $A_1$ and $A_2$ are dimensionless. 
It should be noted that for positive $\lam$, $A_1$ is negative while 
$A_2$ are positive. If the sign of $\xi$ is reversed, the roles 
of $A_1$ and $A_2$ gets interchanged. A negative $A_1$ is tachyonic in nature.
This is a source of instability in the effective potential and will give imaginary 
contribution to the effective potential. Plugging $A_1 \varphi^2$ and 
$A_2 \varphi^2$ for the $m^2$ in the expression for the effective potential 
in eq. (\ref{eq:veff_gen}) and summing the two, we 
get the contribution of the spin-2 mode to the effective potential. 

This imaginary piece though is an infrared effect. 
It is an indication that background 
chosen for doing the computation is not stable, and is a generic feature of gravity 
coupled with scalar field in flat space-time at zero temperature 
\cite{Odintsov1989,Buchbinder1992}. 
This is same as the instability seen in the gas of graviton at finite 
temperature, an indication that flat space-time is unstable. 
This issue has been throughly investigated in past 
in \cite{Gross1982, Srijit2014}. This kind of tachyonic mode 
will create issues with unitarity, but this one is different 
from the unitarity issue caused by the ghost of higher-derivative 
gravity, in the sense that the former is an IR problem 
and has no affect on the UV physics, while the later 
does affect the UV physics also. 
Since we are interested in sorting out the problem 
of unitarity caused by ghosts of higher-derivative, therefore we 
study only this by focusing on the real part of 
the effective potential, as the imaginary piece is relevant in 
IR and deals with tachyonic instability only. This is an important 
realisation as it decouples the two problems: (a) problem of 
unitarity caused by higher-derivative ghosts, and (b) problem 
of unitarity caused by tachyons. This paper deals with the former problem. 

The contribution of the spin-0 mode is a bit complicated
as it involves the scalar mixing matrix. We need to compute the determinant 
of this mixing matrix and then compute the effective potential of the 
operator so obtained from this matrix determinant. The operator
obtained after matrix determinant is following,
\beq
\label{eq:sca_op_veff}
-\Box^3 -3 \varphi^2 \left[ \frac{M^2 \xi}{Z \om} \left(\xi + \frac{1}{6} \right)
+ \lam \right] \Box^2
+ \frac{M^2}{16Z \om} \lam (24\xi+1) \varphi^4 \Box 
- \frac{9 M^2}{16 Z \om} \lam \varphi^6 \, .
\eeq
This operator is a cubic polynomial in $-\Box$ and will therefore have three roots. 
The nature of roots can be analysed using 
the discriminant $\D$ of the equation formed by putting the cubic 
polynomial in (\ref{eq:sca_op_veff}) to zero. 
We write
\bea
\label{eq:disc}
&&
d_0 =1, 
d_1 = -3 \varphi^2 \left[ \frac{M^2 \xi}{Z \om} \left(\xi + \frac{1}{6} \right)
+ \lam \right],
d_2 = - \frac{M^2}{16Z \om} \lam (24\xi+1) \varphi^4,
d_3 = - \frac{9 M^2}{16 Z \om} \lam \varphi^6,
\notag \\
&&
\D = 18 d_0 d_1 d_2 d_3 - 4 d_1^3 d_3 + d_1^2 d_2^2
- 4 d_0 d_2^3 - 27 d_0^2 d_3^2 \, .
\eea
If $\D>0$ then all roots are real, if $\D=0$ then there is a multiple 
root, and $\D<0$ then roots are complex. The operator in 
eq. (\ref{eq:sca_op_veff}) can be factorised as $(-\Box - B_1\varphi^2)(-\Box -B_2\varphi^2)
(-\Box - B_3\varphi^2)$ where $B_1$, $B_2$ and $B_3$ are dimensionless.
As the product of roots $B_1 B_2 B_3$ is positive 
and $B_1B_2 + B_2B_3 +B_3B_1$ is negative, therefore 
this will imply that when $\D>0$ then two roots will be negative.
If $\D=0$, then there is a multiple root with negative sign. 
In the case when $\D<0$ there is a pair of complex conjugate 
root with negative real part and a positive real root. 
The cases with $\D\geq0$ has roots carrying negative sign,
while for $\D<0$ the complex conjugate pair has a negative 
real part. In all these cases the roots can be written as 
\beq
\label{eq:rootpara}
B_1=a, B_2=-r e^\ta, B_3=-r e^{-\ta} \, .
\eeq
In the case when $\D>0$, $\ta$ is positive and real, 
in $\D=0$ case $\ta=0$, while in $\D<0$ case 
$\ta$ is imaginary. The factor of $-1$ in the parametrisation of
roots can also be exponentiated as $e^{i\pi}$. This factor 
is the source of tachyonic instability and will give rise to an 
imaginary contribution in the effective potential. This 
is similar to the instability caused in spin-2 case and is 
an indication that flat space-time is not a true vacuum
\cite{Gross1982, Srijit2014}. The contribution to the 
effective potential from the scalar sector 
is now easily computed using the generalised 
expression given in eq. (\ref{eq:veff_gen}). This is done 
by replacing $m^2$ in eq. (\ref{eq:veff_gen}) with $B_1\varphi^2$, 
$B_2\varphi^2$ and $B_3\varphi^2$, and summing all together.
Using the parametrisation for the roots written in eq. (\ref{eq:rootpara})
and employing the properties of exponential functions, one can write the 
effective potential in simple terms\footnote{
The discriminant $\D$ depends on RG time $t$, and during the 
RG evolution can change sign, thereby implying that during 
the RG evolution $\ta$ can switch from real to imaginary and 
viceversa.}.

The contribution of the fermions needs to be done in a different manner. 
It arises from $-i {\rm Tr} \ln (i \g^\mu \pt_\mu - y_t \varphi)$. 
This can be written in an alternative form by squaring the operator and by 
making use of the gamma-matrix properties. This then become 
$-i/2 {\rm Tr} \ln (-\Box - y_t^2 \varphi^2)$. Then using the generalised 
expression in eq. (\ref{eq:veff_gen}) one gets the contribution for the fermions. 
The full effective potential involves the tree-level contributions plus the 
quantum corrections. The one-loop RG improved full effective action is then given by,
\bea
\label{eq:VeffFull}
V^{(1)}_{\rm eff} &=& \frac{\lam(t)}{4} Z_{\phi}^4(t) \varphi^4
+ \frac{Z_{\phi}^4(t) \varphi^4}{64 \pi^2} \biggl[
5 \sum_{i=1}^{2} \lvert A_i\rvert^2  \ln \frac{\lvert A_i\rvert \varphi^2}{\bar{\mu}^2}
+ B_1^2 \ln \frac{B_1 \varphi^2}{\bar{\mu}^2}
+ 2 r^2 \cosh (2\ta) \ln\frac{r\varphi^2}{\bar{\mu}^2}
\notag \\
&&
+ 2 r^2 \ta \sinh (2\ta)
- y_t^4  \ln \frac{y_t^2 \varphi^2}{\bar{\mu}^2}
+ \frac{iZ_{\phi}^4(t)\varphi^4}{64\pi} \left\{5 \lvert A_1\rvert^2 
+ 2 r^2 \cosh (2\ta) \right\}
\biggr] \, ,
\eea
where $A_i$'s, $B_1$, $r$ and $\ta$ are dimensionless and RG-time $t$ dependent. 
When $\ta \to i\ta$, $\cosh(2\ta) \to \cos(2\ta)$ and $\sinh(2\ta) \to i \sin(2\ta)$, 
thereby preventing the switching between real and imaginary part.   

In the following we will study the real part of the effective potential. 
We ignore the imaginary part, as the imaginary part arises from the 
tachyonic modes of theory and is relevant in IR. We are interested in investigating 
unitarity issues caused by higher-derivatives ghosts.
It should be noticed that the effective potential still posses the 
$\mathbf{Z}_2$ symmetry, as $\varphi^2=0$ is an extrema. 
But due to radiative corrections the real part of quantum 
corrected effective potential might develop a vacuum expectation 
value (VeV) away from zero.  

\section{Symmetry breaking}
\label{symbr}

Due to RG corrections a VeV is generated in the effective potential, which 
then becomes the new vacuum. The original $\varphi^2=0$ vacuum becomes 
unstable under RG corrections and the field migrates to the new vacuum
which occur at $\varphi^2=\kp^2$. It is given by,
\beq
\label{eq:Veffmin}
\left. \frac{{\rm d}}{{\rm d}\varphi^2} Re(V_{\rm eff}) \right|_{\varphi^2 = \kp^2} = 0 \, .
\eeq
At the tree level our original action 
of the theory is scale-invariant and there is no mass-parameter to begin with. 
However the mass parameter enters the system via RG running thereby breaking 
scale-invariance. The effective potential so generated not only breaks scale 
invariance but also breaks the $\mathbf{Z}_2$ symmetry. The
generation of VeV consequently gives mass to scalar and fermion fields. 
It also generates an effective Newton's constant, beside generating newer interactions. 
The generated mass and Newton's coupling can be expressed in terms of 
VeV $\kp^2$ and all the other couplings as
\beq
\label{eq:symbr_massG}
m_s^2 = \frac{3}{2} \lam \kp^2 \, , \hspace{5mm}
m_f = y_t \kp \, , \hspace{5mm}
G^{-1} = 8 \pi \xi \kp^2 \, .
\eeq
The generation of mass and Newton's constant makes the propagators 
for various fields massive. In particular the graviton propagator after the 
symmetry breaking is following, 
\beq
\label{eq:GRprop_sb}
D^{\mn,\al\bt} = 16 \pi G\cdot \Biggl[
\frac{(2P_2 - P_s)^{\mu\nu, \alpha\beta}}{q^2 + i \, \epsilon}
+ \frac{(P_s)^{\mu\nu, \alpha\beta}}{q^2 - M^2/\omega + i \epsilon}
- \frac{2 \, (P_2)^{\mu\nu, \alpha\beta}}{q^2 - M^2+ i \epsilon}
\Biggr] \, ,
\eeq
where now $G$ is the induced Newton's constant and is defined 
using eq. (\ref{eq:symbr_massG}). The masses $M^2$ and $M^2/\om$ 
are given by
\beq
\label{eq:M2exp}
M^2 = 8 \pi f^2 \cdot \xi \kp^2 \, , 
\hspace{5mm}
\frac{M^2}{\om} = 8 \pi \frac{f^2}{\om} \xi \kp^2 \, .
\eeq
The interesting thing about the generation of Newton's constant is that now as the propagator 
becomes massive, so there is a spin-2 massive ghost that appears 
in the system, which in the original theory was massless. 
In the original theory we cannot do the partial-fraction trick 
in the $h_\mn$ propagator, which is possible in the broken phase
due the induced Newton's constant $G$.
Not only the spin-2 ghost becomes massive but also the 
scalar mode acquires mass through symmetry breaking. 
We call this massive scalar mode `Riccion'. 
It should be pointed out that if we had taken $\xi$ to be negative,
then there will be tachyons in the broken phase. So the 
presence of higher-derivatives terms and requiring no 
tachyons to be generated in broken phase fixes the sign of $\xi$. 
This also generates right sign for induced Newton's constant. 
The sign of various couplings in the broken phase is then in 
accordance with the sign of parameters taken in \cite{NarainA1,NarainA2}. 

At this point we compare the propagator of metric fluctuation 
written in eq. (\ref{eq:GR_prop}) with the one written in 
eq. (\ref{eq:GRprop_sb}). The former is before symmetry breaking 
while the later is after symmetry breaking. The former has 
no mass, while later contains masses. Although the appearances of 
the two are different one should however be careful while counting the
propagating degrees of freedom in the two. In the later case (broken phase)
it is easy to count: two massless graviton modes, five massive-tensor 
ghost modes and one massive scalar mode, thereby making 
eight propagating degrees of freedom. In the former case 
(unbroken phase) one should count carefully. 
The pure $C_{\mn\rho\sg}^2$-gravity
has six massless propagating degrees of freedom 
\cite{Lee1982,Riegert1984}.
For pure $R^2$ gravity, the theory two massless propagating 
degrees of freedom as the linearised field equation   
$(\pt_\mu\pt_\nu - \eta_\mn \Box) \Box h =0$ shows to 
have fourth-order time derivatives. Thereby totalling the 
degrees of freedom in unbroken phase to be eight. 
This implies that the propagating degrees of freedom 
in both phases is same, except in broken phase 
some of the modes acquire mass due to symmetry-breaking.

The generation of mass for the spin-2 ghost 
and scalar-mode gives us a hope to investigate unitarity by using the 
criterion stated in \cite{Salam1978, NarainA1, NarainA2, NarainA3, NarainA4}. 
In the RG improved effective potential the VeV has $t$-dependence. 
This arises because at each energy scale the effective potential 
has a VeV. This translates into $t$-dependence of VeV. 
The RG running of VeV depends on the running of the other couplings in a 
complicated manner. This running of VeV then translates into running of 
generated Newton's constant. 
The running of the VeV $\kp^2$ can be computed using the expression of the 
real part of effective potential given in eq. (\ref{eq:VeffFull}). When $\varphi^2=\kp^2$, then 
we are at the minima. The minima condition written in eq. (\ref{eq:Veffmin}) then 
translates into following,
\bea
\label{eq:mincond}
&&
\left. \frac{{\rm d}}{{\rm d}\varphi^2} Re(V_{\rm eff}) \right|_{\varphi^2 = \kp^2}
= Z_{\phi}^4(t) \kp^2 \biggl[
\frac{\lam(t) + \rho_1(t)}{2} + \rho_2(t) 
+ \rho_1(t) \ln \frac{\kp^2(t)}{\bar{\mu}^2} \biggr]=0 \, .
\eea
where,
\bea
\label{eq:rho1rho2}
&&
\rho_1(t) = \frac{1}{32\pi^2} \biggl\{
5\sum_{i=1}^2 A_i^2 + B_1^2 + 2 r^2 \cosh(2\ta) - y_t^4
\biggr\}
\, , 
\notag \\
&&
\rho_2(t) = \frac{1}{32\pi^2} \biggl[
5\sum_{i=1}^2 A_i^2 \ln \lvert A_i\rvert +  B_1^2 \ln B_1
+ 2r^2 \cosh(2\ta) \ln r + 2r^2 \ta \sinh(2\ta)
- y_t^4 \ln y_t^2 \biggr] \, .
\eea
As $\kp^2 \neq 0$ and $Z_{\phi} \neq0$ therefore these 
overall factors goes away and the residual condition 
simplifies to the expression in the square bracket written in 
eq. (\ref{eq:mincond}). As $\kp^2(t)/\bar{\mu}^2$ is dimensionless, we call it $K(t)$. 
One can then directly solve for $K(t)$ using eq. (\ref{eq:mincond})
in terms of all couplings of theory. This also gives the flow of $K$ 
which is generated due to the flow of various couplings present in the theory. 
We however take a $t$-derivative of the expression in the 
square-bracket of eq. (\ref{eq:mincond}) to 
compute the beta-function of the $K(t)$. This is needed in checking 
and locating extrema of $K(t)$. Such extrema are crucial 
points as will be seen later. 
\beq
\label{eq:Kbeta}
\frac{{\rm d} K(t)}{{\rm d}t} = -\frac{K(t)}{\rho_1(t)} \biggl[
\frac{\lam^\prime(t) + \rho_1^\prime(t)}{2} + \rho_2^\prime(t)
+ \rho_1^\prime(t) \ln K(t)
\biggr] \, .
\eeq
This is a linear first order differential equation for the $\ln K(t)$. 
Plugging the running of various couplings from eq. (\ref{eq:betas_Mat3},
\ref{eq:betas_Mat4}, \ref{eq:betas_Mat5}, \ref{eq:betas_GR1}, 
\ref{eq:betas_GR2}) in RHS of eq. (\ref{eq:Kbeta}) we get the beta function of $K(t)$. 
This will be a very complicated function of various couplings. 
Using the running of $K(t)$ we can compute the running of the 
effective Newton's constant by exploiting the expression for 
induced $G$ mentioned in eq. (\ref{eq:symbr_massG}). This is 
given by,
\beq
\label{eq:betaGeff}
\frac{{\rm d} G(t)}{{\rm d}t} = -G(t) \biggl[
\frac{1}{\xi(t)} \frac{{\rm d} \xi(t)}{{\rm d}t} 
+ \frac{1}{K(t)}\frac{{\rm d} K(t)}{{\rm d}t}
+ 2
\biggr] \, . 
\eeq
In order to investigate the issues of unitarity caused by 
higher-derivative we consider the following combination $M^2/\mu^2$. 
We first note that this is
\beq
\label{eq:unit1}
\frac{M^2}{\mu^2} = 8 \pi e^{3/2} f^2(t) \cdot \xi(t) \cdot K(t) \, .
\eeq
Taking $t$-derivative of this yields,
\beq
\label{eq:uniBeta}
\frac{{\rm d}}{{\rm d}t} \ln \frac{M^2}{\mu^2} 
= \frac{1}{f^2} \frac{{\rm d} f^2}{{\rm d}t} 
+ \frac{1}{\xi(t)} \frac{{\rm d} \xi(t)}{{\rm d}t} 
+ \frac{1}{K(t)}\frac{{\rm d} K(t)}{{\rm d}t} \, .
\eeq
Similarly the expression for induced $M^2/\om \mu^2$ is,
\beq
\label{eq:unitw1}
\frac{M^2}{\om\mu^2} = 8 \pi e^{3/2} \frac{f^2(t)}{\om(t)} \cdot \xi(t) \cdot K(t) \, ,
\eeq
and the RG flow of this combination is given by,
\beq
\label{eq:uniwBeta}
\frac{{\rm d}}{{\rm d}t} \ln \frac{M^2}{\om\mu^2} 
= \frac{1}{f^2} \frac{{\rm d} f^2}{{\rm d}t} 
- \frac{1}{\om(t)} \frac{{\rm d} \om(t)}{{\rm d}t}
+ \frac{1}{\xi(t)} \frac{{\rm d} \xi(t)}{{\rm d}t} 
+ \frac{1}{K(t)}\frac{{\rm d} K(t)}{{\rm d}t} \, .
\eeq
The generation of VeV also induce masses for the 
scalar and the fermion fields, which is mentioned in 
eq. (\ref{eq:symbr_massG}). Due to the running of 
VeV, these masses inherits a running. Then to investigate 
whether these fields are physically realisable or not, we consider 
the flow of combinations $m_s^2/\mu^2$
and $m_f^2/\mu^2$. The running of these combinations 
is given by,
\bea
\label{eq:matMS_run}
\frac{{\rm d}}{{\rm d}t} \ln \frac{m_s^2}{\mu^2} 
= \frac{1}{\lam(t)} \frac{{\rm d} \lam(t)}{{\rm d}t} 
+ \frac{1}{K(t)}\frac{{\rm d} K(t)}{{\rm d}t} \, ,
\hspace{2mm}
\frac{{\rm d}}{{\rm d}t} \ln \frac{m_f^2}{\mu^2} 
= \frac{2}{y_t(t)} \frac{{\rm d} y_t(t)}{{\rm d}t} 
+ \frac{1}{K(t)}\frac{{\rm d} K(t)}{{\rm d}t} \, .
\eea
%

\section{Unitarity prescription}
\label{unitpres}

In this section we dictate the algorithm to choose the set of initial conditions 
for which the theory will have a unitary flow. We start by analysing the RG 
equations given in eq. (\ref{eq:betas_Mat3},
\ref{eq:betas_Mat4}, \ref{eq:betas_Mat5}, \ref{eq:betas_GR1}, 
\ref{eq:betas_GR2}). The first thing we do is to extract the running 
$\om$ using the eq. (\ref{eq:betas_GR1} and \ref{eq:betas_GR2}). 
This is given by, 
\beq
\label{eq:omBeta}
\frac{{\rm d} \om(t)}{{\rm d}t}
= \frac{f^2}{\pi} \biggl[
\frac{5}{3} \om^2 + \bigg\{
\frac{183}{10} + \frac{N_s+6N_f}{60}
\biggr\} \om + \frac{5}{6} + 3N_s \left(\xi + \frac{1}{6} \right)^2
\biggr] \, .
\eeq
From this running we notice that as the RHS is always positive
therefore $\om$ is a monotonically increasing function of $t$.
In \cite{NarainA1, NarainA2, NarainA3, NarainA4} it was shown 
that in order to avoid tachyonic instability, we should demand 
that $\om\geq0$. Here in the present scale-invariant theory we
should demand the same. This is done in order to prevent the 
occurrence of tachyons in the broken phase. 
For every value of $\xi$, $\om$ will have two fixed points.
\beq
\label{eq:wFP}
\om_{1,2} = -\frac{1}{40} \bigl[
221 \mp \sqrt{47961-960\xi-2880 \xi^2} \bigr] \, .
\eeq
$\om_1$ is repulsive while $\om_2$ is attractive.
For $\xi$ small both these fixed points lie in the 
unphysical tachyonic regime. For large $\xi$ the fixed points are 
complex conjugate with negative real part. Since $\om$ is monotonically increasing 
with $t$, therefore one can alternatively study the RG flows 
of various parameters in terms of $\om$. Prevention of 
tachyonic instability restricts $\om$ to lie between zero and 
infinity. This then serves as a good candidate in terms of which the 
the RG flows can be analysed. In \cite{NarainA1, NarainA2, NarainA3, NarainA4}
the RG flows were studied in terms of $\om$. 

The crucial problem in overcoming the issue of unitarity is to 
choose the right set of initial conditions so that throughout the 
RG evolution the flow remains unitary in the sense that the 
ghost mass remains always above the energy scale, and the
effective potential doesn't develop any further instability 
(other than the ones already present). 
To prevent the occurrence of this instability requires 
that the coupling $\lam$ remains positive throughout the RG flow
(as negative $\lam$ will result in tachyonic instability for scalar field $\phi$). 
This particularly depends on the choice of initial condition
for yukawa coupling. If the yukawa coupling is above a certain 
threshold then $\lam$ becomes negative too soon during the RG evolution, 
making the effective potential unstable. In standard model of particle 
physics this is an important instability problem where the electroweak 
vacuum becomes metastable \cite{Branchina2013} (see references therein). 
In present case of scale invariant gravity, 
we have freedom to explore the set of initial conditions which will give 
unitary evolution. So we just consider those domains where this 
instability can be avoided. 

In \cite{NarainA1, NarainA2, NarainA3, NarainA4}
it was observed that the RG evolution of $M^2/\mu^2$ is such 
that its flow has a unique minima. This was a crucial feature which 
allowed us to seek those RG trajectories for which this minima is 
above unity. These RG trajectories are the ones for which the flow 
is unitary (massive tensor mode is not physically realisable). 
In the present case of gravity being induced from scale-invariant 
theory we seek a similar behaviour of induced $M^2/\mu^2$, where now
$M^2$ is given by eq. (\ref{eq:M2exp}), and the flow of 
$M^2/\mu^2$ is given in eq. (\ref{eq:uniBeta}). 
The flow of $M^2/\mu^2$ is much complicated in the present case 
and it is difficult to give a rigorous analytic proof that there exist a minima in the 
RG evolution of $M^2/\mu^2$. From various numerical 
investigations we realised that a minima does exist in the evolution of
induced $M^2/\mu^2$. We choose this minima to be our reference 
point and choose the initial conditions at this point for all other couplings. 
The appearance of a minima in the flow induced $M^2/\mu^2$ implies that at this 
minima the beta-function of induced $M^2/\mu^2$ will vanish,
\beq
\label{eq:mincond1}
\left. \frac{1}{f^2} \frac{{\rm d} f^2}{{\rm d}t} \right|_{t=t_*}
+ \left. \frac{1}{\xi(t)} \frac{{\rm d} \xi(t)}{{\rm d}t} \right|_{t=t_*}
+ \left. \frac{1}{K(t)}\frac{{\rm d} K(t)}{{\rm d}t} \right|_{t=t_*} = 0 \, .
\eeq
Plugging the RG-flows of various coupling in this, will result in a condition 
satisfied by all the couplings of theory at this minima. This will act as a 
constraint in choosing some of the initial parameters. We first choose the value
of $M^2/\mu^2$ at this point, we call it $\rho_*$. We require $\rho_*>1$.
\beq
\label{eq:mincond_rho}
\left. \frac{M^2}{\mu^2} \right|_{t=t_*}
= 8 \pi e^{3/2} f^2_* \cdot \xi_* \cdot K_* = \rho_* >1 \, ,
\eeq
where $f^2_*$, $\xi_*$ and $K_*$ are initial values of 
$f^2$, $\xi$ and $K$ respectively. 
The imposition of this constraint makes sure that the mass of the 
spin-2 ghost mode remains above the running energy scale. 
This will imply that one of the three unknowns $f^2_*$,
$\xi_*$ and $K_*$ can be expressed in terms of other two. 
We choose to write $K_*$ in terms of $f^2_*$ and $\xi_*$.
At this point we also choose $f^2_*\ll1$. Now the left unknowns 
are $\lam_*$, $(y_t)_*$, $\om_*$ and $\xi_*$. In order to choose the 
matter couplings $\lam_*$ we use our knowledge of 
non-gravitational system. In such system the running $\lam$ always 
hits the Landau pole if the initial value of yukawa coupling is below 
certain threshold, beyond which $\lam$ becomes negative leading to 
instability. We accordingly choose $\lam_* \lesssim 0.1$.

At this point we analyse the beta-function for the coupling $\xi$. In this 
theory we have the freedom to choose $\xi$ to be very large ($\gtrsim10$).
This is primarily because in the perturbation theory 
the coupling strength of vertex containing $n$-gravitons and two scalars
is $\sim\xi (\sqrt{f^2})^n$ and $\sim\xi (\sqrt{f^2/\om})^n$. 
Since $f^2\ll1$, so this give us freedom to choose $\xi$ to be very large
while still being in the realm of perturbation theory
\footnote{In the case on Einstein-Hilbert gravity with only Newton's constant $G$,
the coupling strength is $\xi (\sqrt{G})^n$ for a vertex containing 
$n$-gravitons and two scalars.}. For $\xi$ large the beta-functions of various coupling 
acquires a simplified form. Although the beta functions become simplified 
but still they are complicated enough that it requires the analysis to be done numerically. 
We tend to explore numerically this regime of parameters. 

We choose to work in regime where $-\D/\varphi^6 = \ep \ll 1$, 
where $\D$ is the discriminant mentioned in eq. (\ref{eq:disc}). 
In this regime there will 
one positive root for $-\Box$ and a complex conjugate pair with negative 
real part. Under the RG flow $\D/\varphi^6$ will also run. 
We choose the initial parameters in such a way so that at the initial 
point $\ep\ll1$. By reversing this argument we say that we start with 
$\ep \ll 1$ and solve for the initial parameters under this constraint. 
This fixes the initial value problem completely. With the chosen $f^2_* \ll 1$
($\lesssim 10^{-6}$), $\rho_*>1$, $\lam_* \lesssim 0.1$ and $\xi_* \gg 1$ ($\gtrsim 10^2$), 
we use the constraint dictated by $\ep\ll1$ to solve for $\om_*$. 
From the four different solution for $\om_*$, we choose the one which real and positive
(to avoid tachyons)\footnote{It is noticed that if $\ep<0$, then all solutions 
for $\om_*$ will be negative and will lie in unphysical tachyonic regime. 
This knowledge also demands to take $\ep>0$.}. 
Knowledge of $\xi_*$ gives the initial value of $K_*$ by using the relation given 
in eq. (\ref{eq:mincond_rho}). We then plug these into the minima constraint 
given in eq. (\ref{eq:mincond1}). This constraint contains the yukawa coupling
in quadratic form, and therefore on solving gives two equal and opposite 
values for $y_{t*}$. One can choose either of the sign of yukawa coupling 
for the initial condition. The flow of all the other couplings doesn't depend on this 
sign. Once the initial parameters are known we can solve the RG flows and 
compute the flow of induced $M^2/\mu^2$ to see if it remains above unity 
throughout the RG evolution.

\section{Numerical Analysis}
\label{numanal}

We tried several possible values of various parameters 
in order to see how the flows are for various initial conditions,
and did the analysis case by case systematically.

\subsection{Fixed $\lam_*$}
\label{fixedlam}

We first considered case with fixed $\lam_*$, while we took 
different values for $f^2_*$, and for each $f^2_*$ we explored a 
range of $\xi_*$. Throughout this we took $\rho_*=1.5$ (there is nothing 
special about this number, as long as long as $\rho_*>1$). We considered
three different values for $f^2_*=10^{-6}$, $10^{-7}$ and $10^{-8}$. 
We have freedom over the choice of $-\D_*/\varphi^6=\ep$. It is seen 
that with $f^2_*$ fixed, when $\ep$ is made 
smaller, then $\om_*$ increases. However the $y_{t*}$ obtained first 
decreases to a minima before rising again and becoming stable. We choose 
$\ep$ near this minima, so that we have more number of 
e-folds in the RG flows. It turns out that for each value of 
$f^2_*$ the position of occurrence of this minima will be different. 
For smaller $f^2_*$, the minima occurs at a smaller value of 
$\ep$. Thus for $f^2_*=10^{-6}$, $10^{-7}$ and $10^{-8}$,
the minima for $\ep$ occurs around 
$10^{-12}$, $10^{-15}$, and $10^{-16}$ respectively. 
We consider these cases in succession. 

The number of e-folds from the Planck's time to current 
galactic scale is $\sim130$. This stands then as another guiding 
principle to choose set of initial parameters. 
It is noticed that when $f^2_*$ is made more smaller
then the allowed upper value of $\xi_*$ (which is chosen 
so that we have $\gtrsim100$ e-folds) increases. 
This can be understood by considering the strength of 
vertices. For vertex containing one graviton leg and two 
scalar leg, the interaction strength $\sim\xi\sqrt{f^2}$.
Demanding perturbation theory to remain valid implies 
$\xi_*\sqrt{f^2_*} \lesssim1$, which explains the behaviour. 
We keep $\xi_*$ large so that there is sufficient communication 
between the matter and gravity sector. In table \ref{tab:inval} we 
tabulate our findings for $f^2_*=10^{-6}$, $10^{-7}$ and $10^{-8}$. 

\begin{table}[t]
\begin{tiny}
\begin{tabular}{| l | l | l | l | l | l | l | l | l | l | l | l |}
\hline 
\multicolumn{4}{|c|}{$f^2_*=10^{-6}$} &
\multicolumn{4}{c|}{$f^2_*=10^{-7}$} &
\multicolumn{4}{c|}{$f^2_*=10^{-8}$} \\
\hline \hline
$\xi_*$ & $\om_* \times10^{-4}$ & $y_{t*}$  & $T_r$ &
$\xi_*$ & $\om_* \times10^{-6}$ & $y_{t*}$  & $T_r$ &
$\xi_*$ & $\om_* \times10^{-6}$ & $y_{t*}$  & $T_r$   \\
\hline 
$1$ &$3.0536$ & $0.4522$ & $\sim215$ &
$10^2$ & $3.0536$ & $0.4536$  & $\gtrsim235$  &
$10^2$ & $3.0536$ & $0.4522$ & $\gtrsim230$\\
\hline
$10$ & $3.0536$ & $0.4532$ & $\sim215$ &
$5\times10^2$ & $3.0536$ & $0.4543$ & $\gtrsim267$ &
$10^3$ & $3.0536$ & $0.4532$ & $\gtrsim212$  \\
\hline
$10^2$ & $3.0549$ & $0.4543$ & $\gtrsim 217$ &
$10^3$ & $3.0537$ & $4546$ & $\sim437$ &
$10^4$ & $3.0549$ & $0.4542$ & $\gtrsim 235$  \\
\hline
$5\times10^2$ & $3.0869$ & $0.4562$ & $\gtrsim305$ &
$2\times10^3$ & $3.0542$ & $0.4551$ & $\sim250$ &
$5\times10^4$ & $3.0869$ & $0.4562$ & $\gtrsim305$  \\
\hline
$10^3$ & $3.1840$ & $0.4603$ & $\sim356$ &
$5\times10^3$ & $3.0569$ & $0.4561$ & $\sim120$ &
$10^5$ & $3.1840$ & $0.4603$ & $\sim356$  \\
\hline
$5\times10^3$ & $5.3737$ & $0.5355$ & $\sim60$ &
$10^4$ & $3.0669$ & $0.4587$ & $\sim72$ &
$5 \times10^5$ & $5.3736 $ & $0.5355$ & $\sim58$  \\
\hline
\end{tabular}
\end{tiny}
\caption{
Initial values for the various coupling parameters. These are for $f^2_*=10^{-6}$,
$10^{-7}$ and $10^{-8}$. We took $\lam_*=0.1$ and $\rho_*=1.5$. Here for three 
different values of $f^2_*$ we took $\ep = 10^{-12}$, $10^{-15}$ and 
$10^{-16}$. As $\xi_*$ increase, the value of $\om_*$ and $y_{t*}$ increase. 
The number of e-folds tend to decrease as $\xi_*$ increases. For smaller 
$\xi_*$ the reason we see less e-folds is due to numerical precision of machine. 
}
\label{tab:inval}
\end{table}

\begin{figure}
\centerline{
\vspace{0pt}
\centering
\includegraphics[width=3.0in]{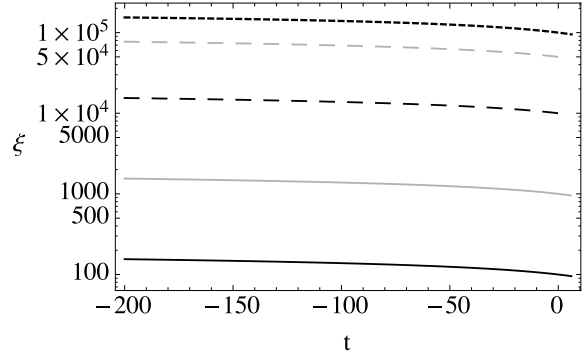}
}
 \caption[]{
Running of coupling $\xi$ for various values of initial conditions $\xi_*=
10^2$ (black solid line), $10^3$ (grey solid line), $10^4$ (big dashed black line), 
$5\times10^4$ (big dashed grey line) and $10^5$ (small dashed black line). For these 
flows we took $f^2_*=10^{-8}$, $\lam_*=0.1$, $\rho_*=1.5$ and 
$\ep = 10^{-16}$. 
}
\label{fig:xi1}
\end{figure}

We then plot the flows of various parameters for the 
choice of initial parameters written in table \ref{tab:inval}. 
Each flow is interesting to analyse. The flow of the 
coupling $\xi$ for various choices of the initial conditions 
is shown in figure \ref{fig:xi1}. The plot shown in 
figure \ref{fig:xi1} is for $f^2_*=10^{-8}$ only, for 
other values we observe similar qualitative behaviour
which will not be shown here.  
From the running of $\xi$ shown in figure \ref{fig:xi1} 
we notice that the parameter $\xi$ runs to smaller values in the UV regime. This might 
be an indication of the possible existence of an UV stable 
fixed point, but it is hard to give a robust answer in this paper. This however 
can be justified by looking at the beta-function of $\xi$ 
given in eq (\ref{eq:betas_Mat4}) whose {\it r.h.s.} can be 
seen to vanish for a certain choice of coupling parameters. 
%
\begin{figure}
\centerline{
\vspace{0pt}
\centering
\includegraphics[width=2.6in]{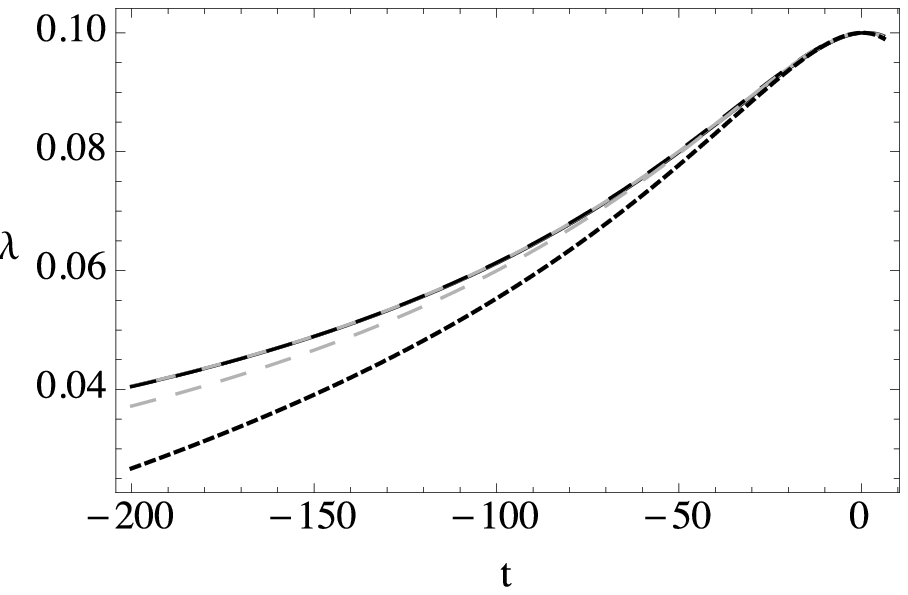}
\hspace{3mm}
\includegraphics[width=2.7in]{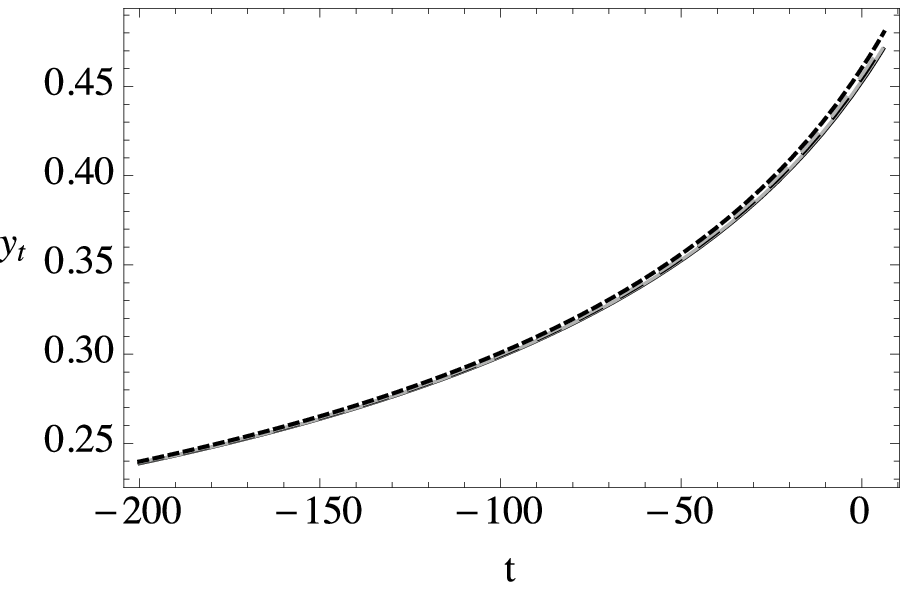}
}
 \caption[]{
Running for matter couplings $\lam$ and $y_t$ in left and right respectively. 
These are plotted for $f^2_*=10^{-8}$, $\lam_*=0.1$, $\rho_*=1.5$ and 
$\ep = 10^{-16}$. We considered the following initial conditions for 
$\xi_*=10^2$ (black solid line), $10^3$ (grey solid line), $10^4$ (big dashed black line), 
$5\times10^4$ (big dashed grey line) and $10^5$ (small dashed black line). 
}
\label{fig:lamyt1}
\end{figure}

The flow of matter couplings $\lam$ and $y_t$ for various initial conditions 
is shown in left and right of figure \ref{fig:lamyt1} respectively. 
For smaller values of $\xi_*$ the flow of 
these couplings remain almost same, while deviations are 
seen for large $\xi_*$. This is again plotted for $f^2_*=10^{-8}$, while 
for other values of $f^2_*$ qualitatively similar behaviour is seen. 
In the UV the flow of $\lam$ is seen to bend and run toward smaller 
values, which is caused by the yukawa coupling. 

\begin{figure}
\centerline{
\vspace{0pt}
\centering
\includegraphics[width=3.4in]{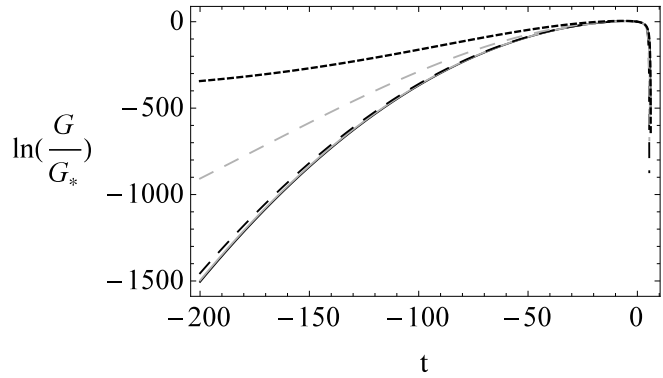}
}
 \caption[]{
The flow of the induced Newton's constant. Here 
we plot $\ln (G/G_*)$ for the case 
$f^2_*=10^{-8}$, $\lam_*=0.1$, $\rho_*=1.5$ and 
$\ep = 10^{-16}$. The flow is computed for five different 
values of $\xi_*=10^2$ (black solid line), $10^3$ (grey solid line), $10^4$ (big dashed black line), 
$5\times10^4$ (big dashed grey line) and $10^5$ (small dashed black line). 
Both in UV and IR the flow goes very close to zero. 
}
\label{fig:Gt1}
\end{figure}

The flow of the VeV induces a flow in the Newton's constant. The flow 
of the induced Newton's constant for various initial conditions 
is shown in figure \ref{fig:Gt1}. The induced Newton's 
constant goes to zero in the UV and in IR. 
In UV it is seen to go to zero at a finite energy scale. 
This is similar to the flow of 
Newton's constant observed in \cite{NarainA1, NarainA2,
NarainA3, NarainA4}, where the original action was not 
scale-invariant and contained Einstein-Hilbert piece 
in the higher-derivative action. This is somewhat interesting to
note. Again this is just a numerical observation and not 
a rigorous analytic argument. By varying the value of 
$f^2_*$ we notice that the qualitative features of the 
graph remains same.

\begin{figure}
\centerline{
\vspace{0pt}
\centering
\includegraphics[width=2.6in]{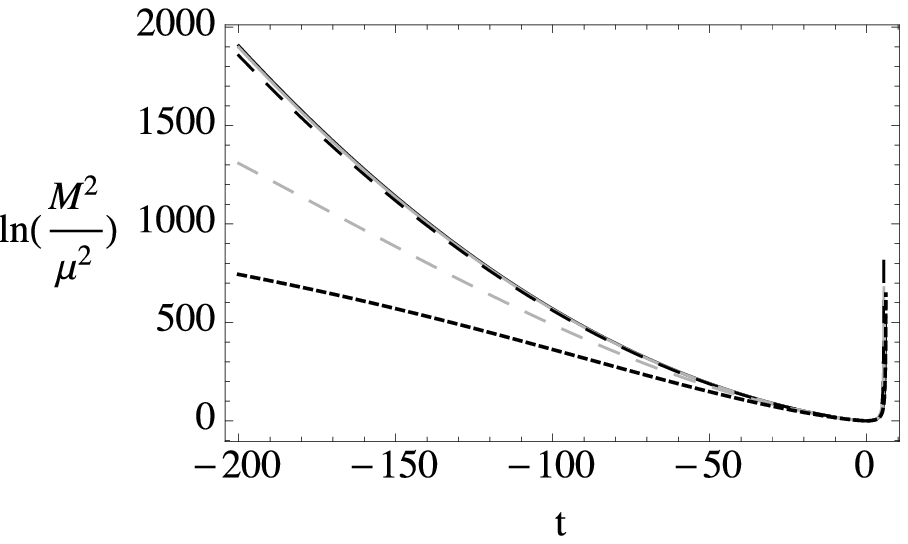}
\hspace{4mm}
\includegraphics[width=2.7in]{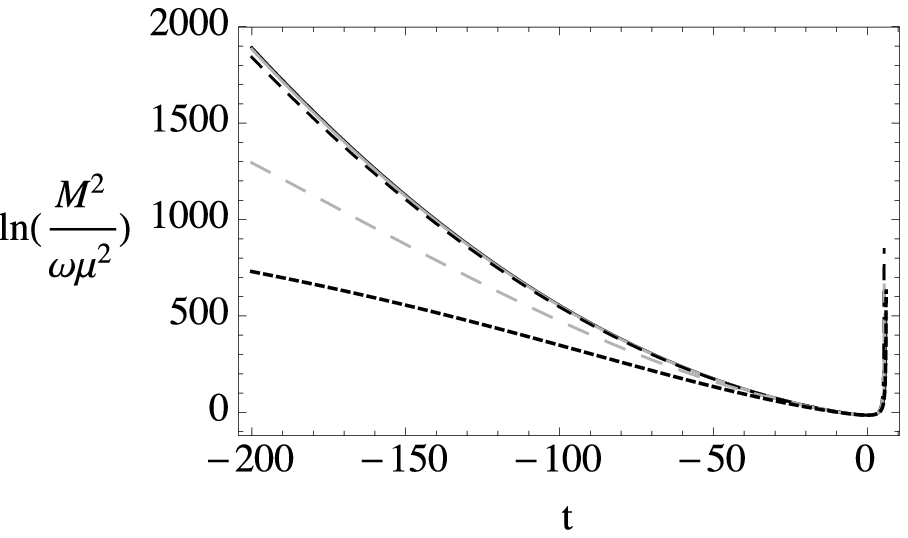}
}
 \caption[]{
The running of induced $\ln(M^2/\mu^2)$ and $\ln(M^2/\om\mu^2)$ for the case 
$f^2_*=10^{-8}$, $\lam_*=0.1$, $\rho_*=1.5$ and 
$\ep = 10^{-16}$. The flow is computed for five different 
values of $\xi_*=10^2$ (black solid line), $10^3$ (grey solid line), 
$10^4$ (big dashed black line), $5\times10^4$ (big dashed grey line) 
and $10^5$ (small dashed black line). 
In all cases it is seen that the flow has a minima. If the flow is such 
that we have $M^2/\mu^2>1$ throughout the flow, then this spin-2 ghost 
mode never goes on-shell, and theory satisfies unitarity.
}
\label{fig:UnitG1}
\end{figure}

Figure \ref{fig:UnitG1} shows the flow of parameters 
$M^2/\mu^2$ and $M^2/\om\mu^2$ in left and right 
respectively. The flow of $M^2/\mu^2$ is such that it is always 
above unity ($M^2/\mu^2>1$). This means that the 
propagator of metric fluctuation after symmetry breaking 
doesn't witness the ghost pole, as the problematic ghost 
mode is never realised and never goes on-shell. 
We observe this to happen for a large domain of 
coupling parameter space. A similar running of the parameter $M^2/\mu^2$ 
was also observed in \cite{NarainA1, NarainA2,
NarainA3, NarainA4}, and was used to establish 
unitarity criterion for the higher-derivative gravity. 
The flow of the parameter $M^2/\om\mu^2$ is 
different from the one observed in \cite{NarainA1, NarainA2,
NarainA3, NarainA4}, where a monotonic behaviour was seen.
In the present case we see a convex structure with a single 
minima in the flow. If we demand that $\left. M^2/\mu^2 \right|_*>1$, 
then it doesn't imply that $\left. M^2/\om\mu^2\right|_*>1$. 
However the reverse is always true {\it i.e.} $\left. M^2/\om\mu^2\right|_*>1$ implies 
$\left. M^2/\mu^2\right|_*>1$. By choosing $\rho_*$ to be large enough one can 
make the scalar mode also physically unrealisable.

\begin{figure}
\centerline{
\vspace{0pt}
\centering
\includegraphics[width=2.6in]{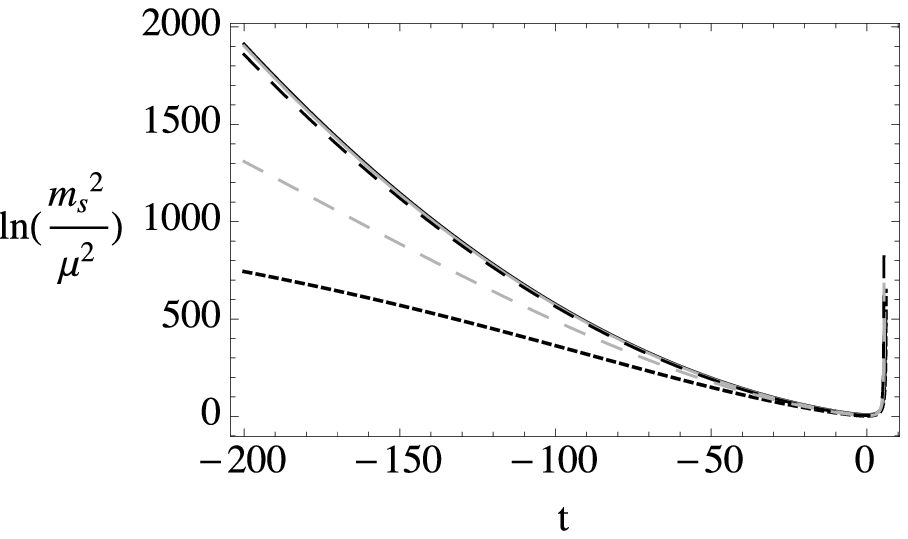}
\hspace{4mm}
\includegraphics[width=2.7in]{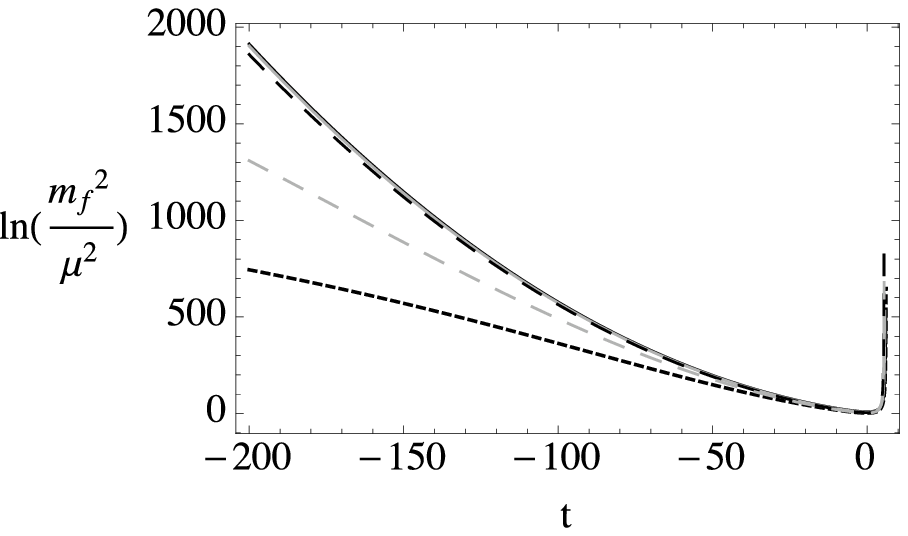}
}
 \caption[]{
The running of induced $\ln(m_s^2/\mu^2)$ and $\ln(m_f^2/\mu^2)$ for the case 
$f^2_*=10^{-8}$, $\lam_*=0.1$, $\rho_*=1.5$ and 
$\ep = 10^{-16}$. The flow is computed for five different 
values of $\xi_*=10^2$ (black solid line), $10^3$ (grey solid line), $10^4$ (big dashed black line), 
$5\times10^4$ (big dashed grey line) and $10^5$ (small dashed black line). 
In all cases it is seen that the flow has a minima. If throughout the flow 
both $m_s^2/\mu^2>1$ and $m_f^2/\mu^2>1$ then both of them 
never go on-shell. 
}
\label{fig:msmf1}
\end{figure}

We then consider the induced masses in the matter sector and 
consider the flow of combinations $m_s^2/\mu^2$ and $m_f^2/\mu^2$,
where $m_s$ and $m_f$ is given in eq. (\ref{eq:symbr_massG}). 
The induced RG running of them is shown in figure \ref{fig:msmf1}. 
The running of these is interesting in the sense that both of 
them has a minima. The value at the minima depends 
crucially on the initial parameters chosen to make the 
flow unitary. If we choose $\rho_*$ large enough then it is possible 
that flow of $m_s^2/\mu^2$ and $m_f^2/\mu^2$ will be such 
that the scalar and fermion will never be realised during the 
whole RG flow, and they never go on-shell. In that sense 
they affect the theory indirectly and only gravitationally but 
they never go on-shell.

\subsection{Fixed $f^2_*$}
\label{fixedM2G}

\begin{figure}
\centerline{
\vspace{0pt}
\centering
\includegraphics[width=1.8in, height=1.4in]{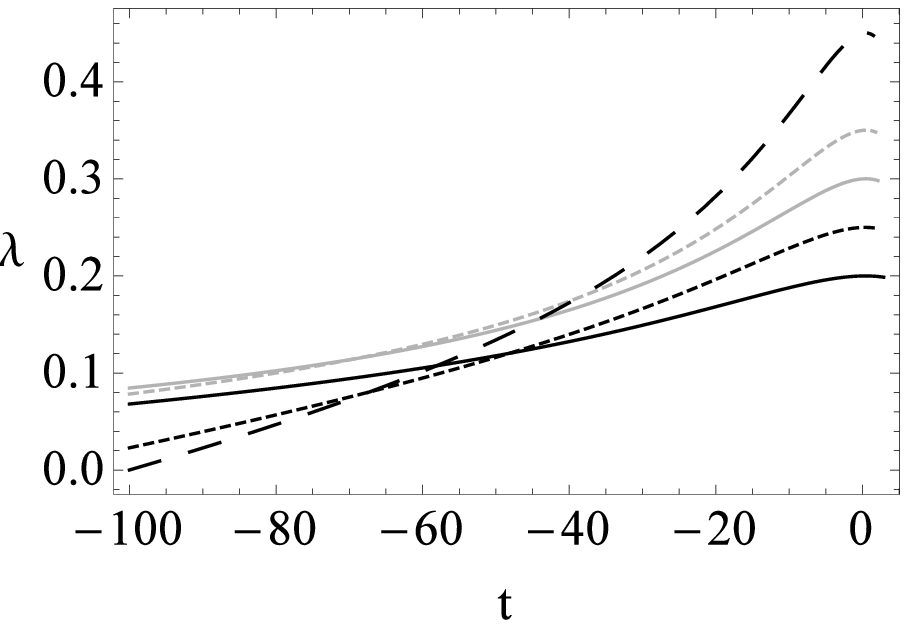}
\hspace{3mm}
\includegraphics[width=1.8in, height=1.4in]{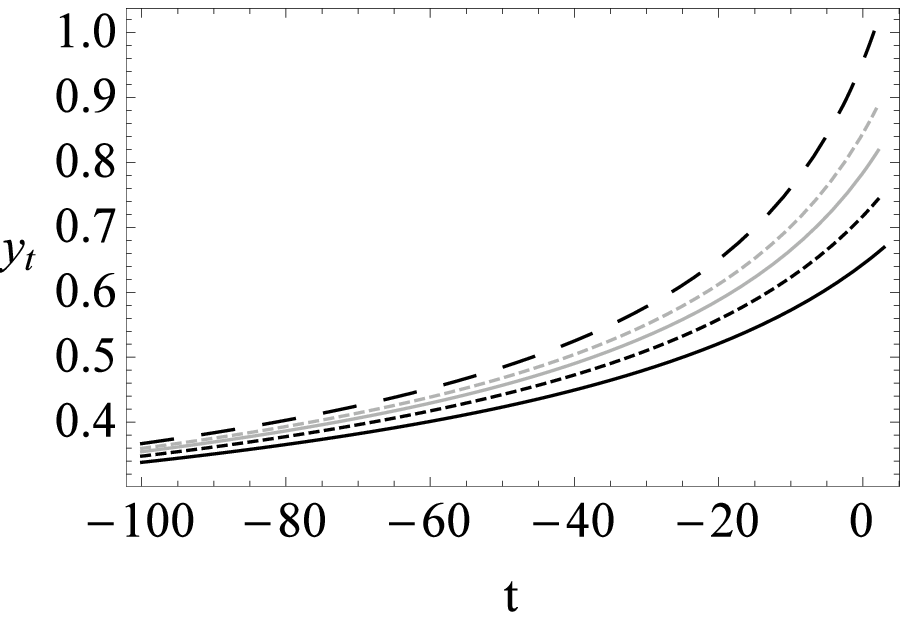}
\hspace{3mm}
\includegraphics[width=1.9in, height=1.4in]{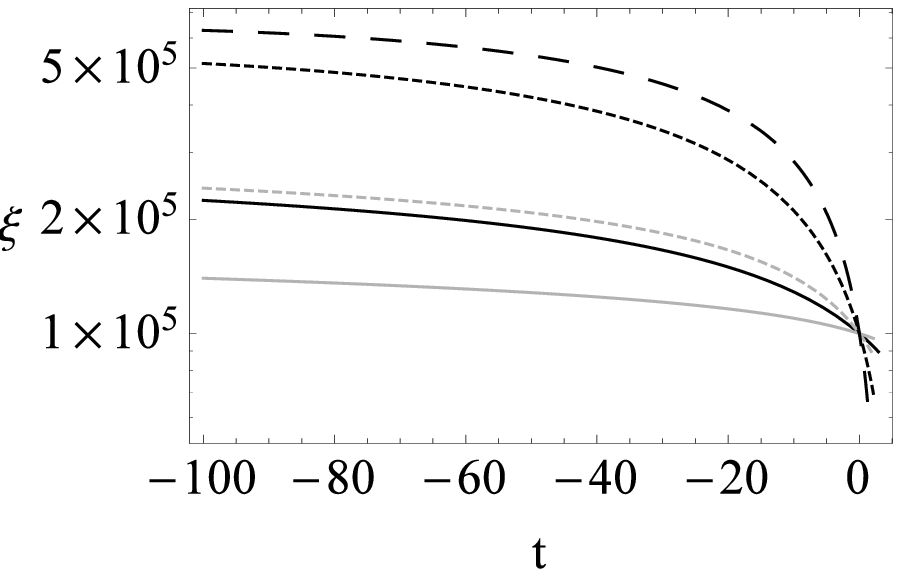}
}
 \caption[]{
Running for couplings $\lam$, $y_t$ and $\xi$ in left, centre and right respectively. 
These are plotted for $f^2_*=10^{-8}$, $\lam_*=0.2$ (black solid line), 
$0.25$ (black short-dashed line), 
$0.3$ (grey solid line), $0.35$ (grey short-dashed line) and 
$0.45$ (black long-dashed line). 
We took $\rho_*=1.5$ and $\xi_*=10^5$. 
}
\label{fig:lamyt2}
\end{figure}

In the previous subsection the case with fixed $\lam_*$ was investigated.
It is worth checking the robustness of the qualitative features 
when other parameters are varied. One particular important 
parameter is the $\lam_*$. It is important to see how the situation 
changes when $\lam_*$ is increased. For this we fix the value
of $f^2_*=10^{-8}$, $\xi_*=10^5$ $\rho_*=1.5$. We took a 
range of value of $\lam_*=0.2$, $0.25$, $0.3$, $0.35$ and $0.45$. 
Although the qualitative features are same but there are minor differences. 
In each case we always witness that the running $M^2/\mu^2$ has 
a minima, and the flow is always above unity. This further establishes
that there always exist a minima in the flow of induced $M^2/\mu^2$, and it also 
implies that by choosing right set of initial condition it is possible 
to make the massive tensor ghost innocuous. 

\begin{figure}
\centerline{
\vspace{0pt}
\centering
\includegraphics[width=1.8in, height=1.3in]{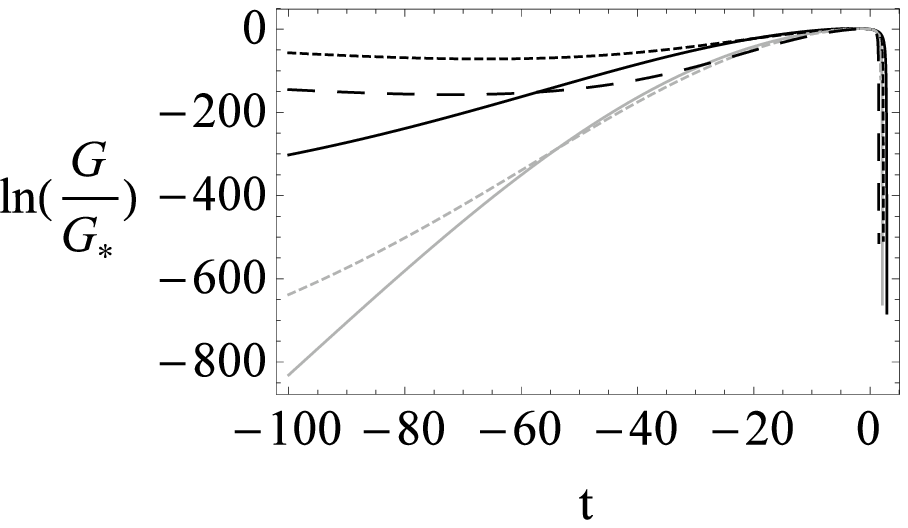}
\hspace{3mm}
\includegraphics[width=1.8in, height=1.3in]{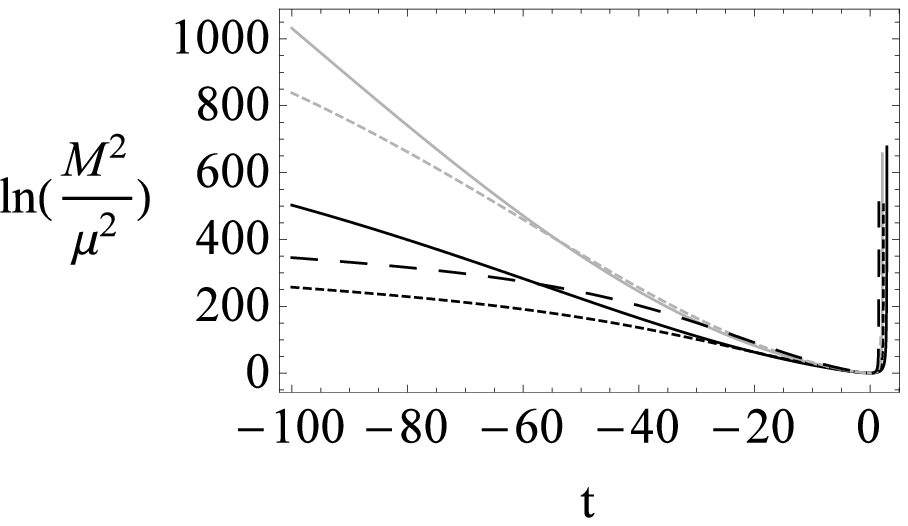}
\hspace{3mm}
\includegraphics[width=1.9in, height=1.3in]{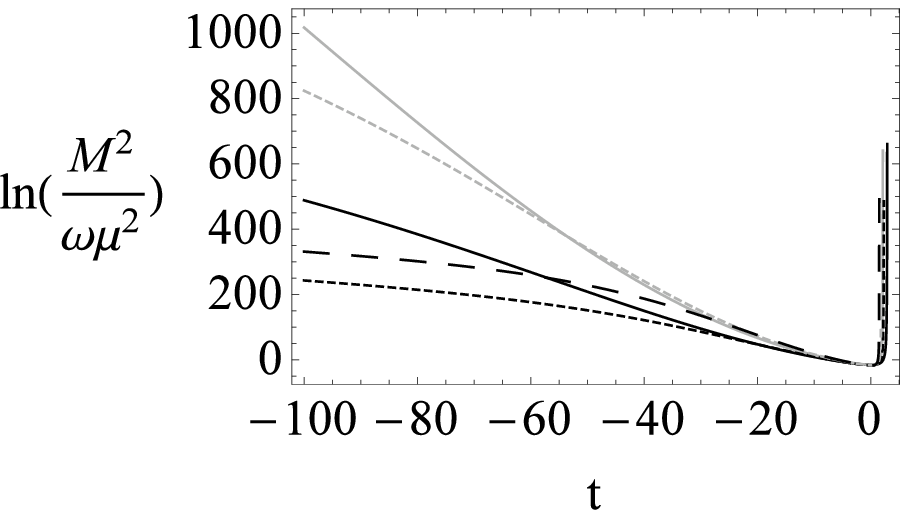}
}
 \caption[]{
The running of induced $\ln(G/G_*)$, $\ln(M^2/\mu^2)$ and $\ln(M^2/\om\mu^2)$ 
in left, centre and right respectively. These are plotted for 
$f^2_*=10^{-8}$, $\lam_*=0.2$ (black solid line), 
$0.25$ (black short-dashed line), 
$0.3$ (grey solid line), $0.35$ (grey short-dashed line) and 
$0.45$ (black long-dashed line). 
We took $\rho_*=1.5$ and $\xi_*=10^5$. 
}
\label{fig:UnitG2}
\end{figure}

As the system contain a mixture of several coupling which are 
all evolving in different manner, therefore the dynamics of 
system is rich and interesting. This becomes more apparent 
when we plot the running of various parameters. The flow of 
$\lam$, $y_t$ and $\xi$ is shown in figure \ref{fig:lamyt2}. 
These flows are very much similar to the ones shown 
for fixed $\lam_*$ in the previous sub-section. It is seen that as $\lam_*$ 
is increased the flow of $\xi$ decrease more sharply in the UV,
and in IR the flow goes to higher values, even though starting 
point is same. The running of yukawa coupling is simple,
in the sense that when $\lam_*$ increases, so does $y_{t*}$
and accordingly the whole RG trajectory for yukawa coupling. The flow of 
$\lam$ is interesting. For higher $\lam_*$, $\xi$ flows to higher values 
in the IR. This makes the $\lam$ to run faster toward zero in the 
IR. In the UV the RG trajectories for $\lam$ has self-similarity. 

\begin{figure}
\centerline{
\vspace{0pt}
\centering
\includegraphics[width=2.6in]{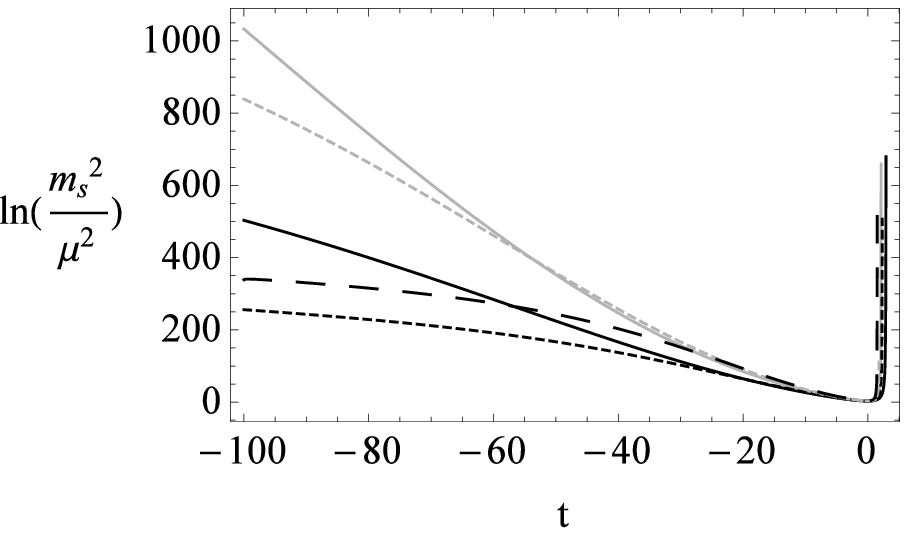}
\hspace{4mm}
\includegraphics[width=2.7in]{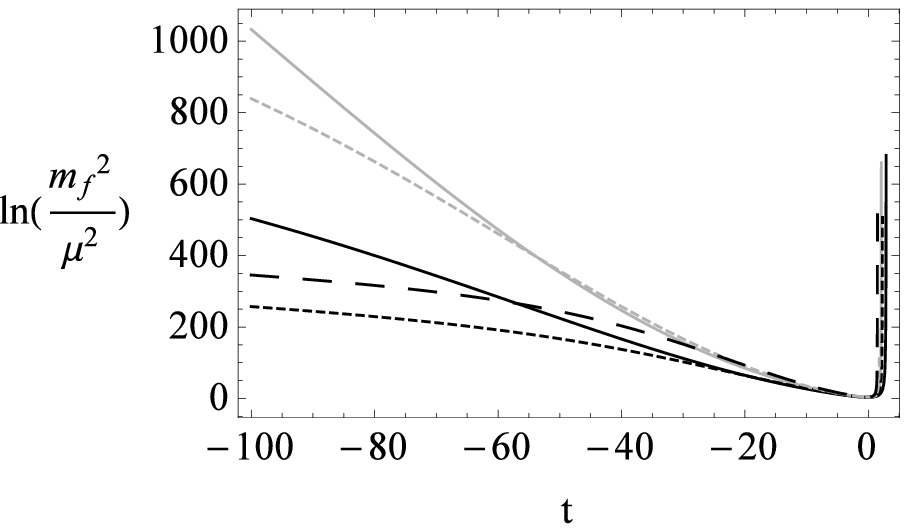}
}
 \caption[]{
The running of induced $\ln(m_s^2/\mu^2)$ and $\ln(m_f^2/\mu^2)$ in the left 
and right respectively. These are plotted for 
$f^2_*=10^{-8}$, $\lam_*=0.2$ (black solid line), 
$0.25$ (black short-dashed line), 
$0.3$ (grey solid line), $0.35$ (grey short-dashed line) and 
$0.45$ (black long-dashed line). 
We took $\rho_*=1.5$ and $\xi_*=10^5$. Choosing appropriate 
$\rho_*$ it is possible to make the both the scalar and fermion 
unrealisable. 
}
\label{fig:msmf2}
\end{figure}

The flow of $G$, $M^2/\mu^2$ and $M^2/\om\mu^2$ is shown in 
left, centre and right respectively in figure \ref{fig:UnitG2}. The 
qualitative behaviour is the same in the sense that the induced 
Newton's constant goes to zero in the UV at a finite energy scale.
It goes to zero in the IR. The RG flows for $M^2/\mu^2$ and $M^2/\om\mu^2$
have same qualitative features, and tensor ghost is physically 
unrealisable even when we increase $\lam_*$. Choosing appropriate 
$\rho_*$ will make the Riccion also physically unrealisable. 
The RG flow of Riccion mass is different from the 
one seen in \cite{NarainA1, NarainA2, NarainA3, NarainA4}. 
The flow of induced masses in the matter sector has similar 
qualitative features and is shown in figure \ref{fig:msmf2}.

\subsection{Fixing Planck's scale}
\label{planck}

\begin{figure}[ht]
\centerline{
\vspace{0pt}
\centering
\includegraphics[width=6in]{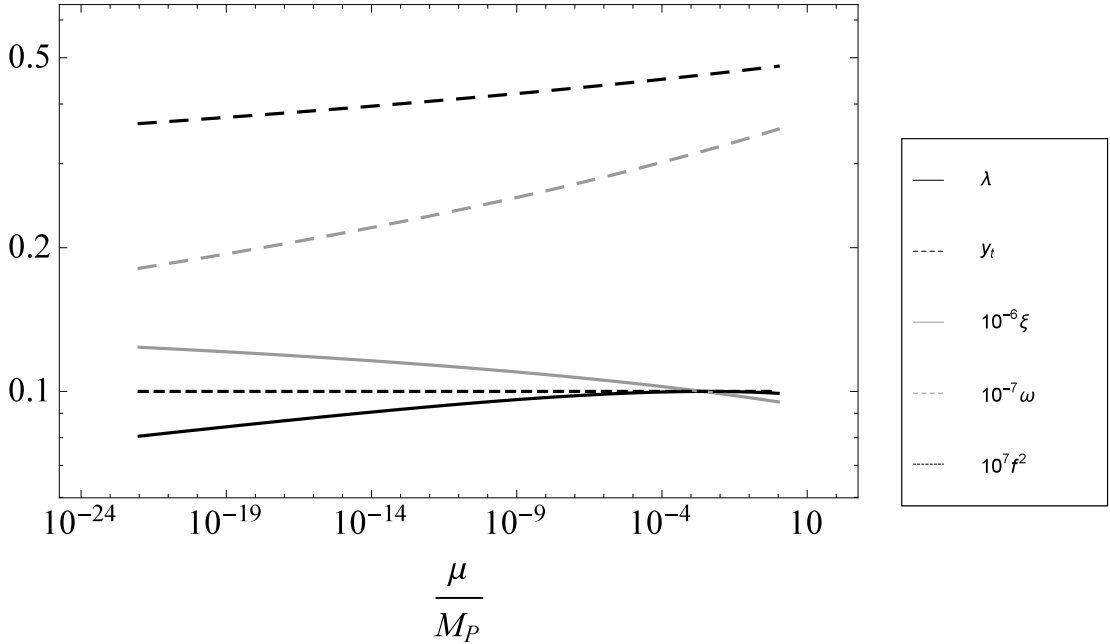}
}
 \caption[]{
The running of various dimensionless couplings. For this we took 
$f^2_*=10^{-8}$, $\lam_*=0.1$, $\xi_*=10^5$. 
We took $\rho_*=1.5$ and $\ep = 10^{-16}$, which 
gave $y_{t*}=0.46$ and $\om_*=3.18\times10^6$. 
Here $\mu$ is the running energy scale. 
}
\label{fig:coupall}
\end{figure}

\begin{figure}[h]
\centerline{
\vspace{0pt}
\centering
\includegraphics[width=6in]{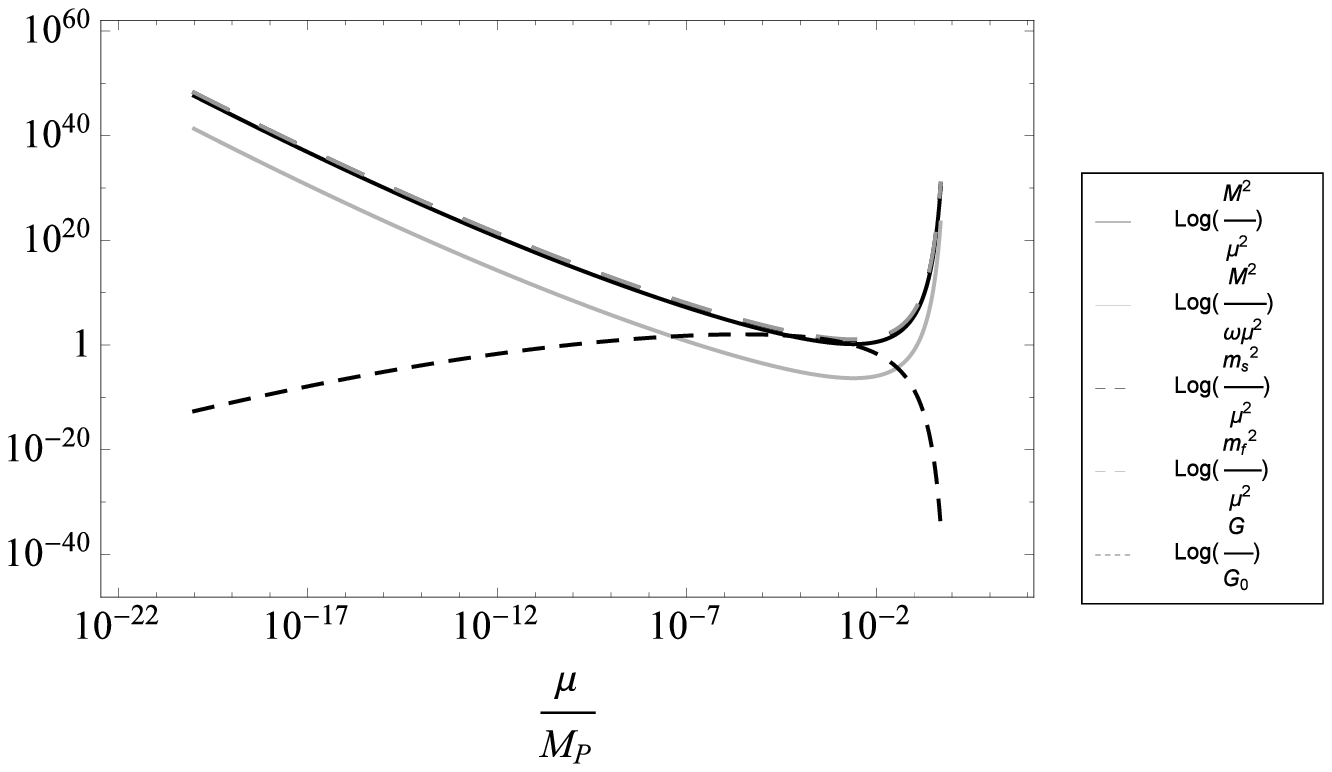}
}
 \caption[]{
The running of various induced masses and induced Newton's constant. 
For this we took $f^2_*=10^{-8}$, $\lam_*=0.1$, $\xi_*=10^5$. 
We took $\rho_*=1.5$ and $\ep = 10^{-16}$, which 
gave $y_{t*}=0.46$ and $\om_*=3.18\times10^6$. 
Here $\mu$ is the running energy scale. 
}
\label{fig:massall}
\end{figure}

The renormalisation group invariance insures that the flow of couplings doesn't 
depend on the reference point $\mu_0$. This gave us freedom to choose 
the reference point without any conditions. As a result for sake of convenience we choose it 
to be the point where the flow of $M^2/\mu^2$ has a minima, where the 
initial conditions for the flow are imposed. However an interesting thing to 
ask is to how to relate it with the phenomenology? In the sense how does 
the running of various parameter look like when compared to Planck's scale $M_{Pl}$,
whose value is around $1.22\times10^{19}$ GeV? This is interesting point to 
ponder on. For this we study the running of induced $G$, which runs 
strongly in the UV and goes to zero. From observations of astrophysics and 
cosmology we know $G_{Newton}$ remains constant for a large energy range. 
However it is usually expected that it will undergo strong running near the Planck's 
scale. For this reason we choose $M_{Pl}$ in the regime where induced $G$ 
witnesses a strong running {\it i.e.} near the point where induced $G$ goes to zero. 
Once this is fixed one can plot the flow of various coupling parameters and induced 
masses. These are presented in figure \ref{fig:coupall} and \ref{fig:massall}.

\section{Conclusion}
\label{conc}

Here in this paper the idea of gravity being induced 
from scale-invariant theory is considered. The fundamental theory 
is a coupled system of scale-invariant matter and higher-derivative gravity. 
The lorentzian path-integral of this fundamental theory incorporates 
quantum fluctuations from both matter and gravity. 
The matter sector is taken to be simple (a scalar and a dirac fermion). 

The effective action of the theory is computed in the $4-\ep$ 
dimensional regularisation procedure. The divergent part of which gives 
the RG flow of the various coupling parameters of the theory. 
These have been computed in past also. We did it again 
in order to verify the past results. We agree fully with the past 
literature \cite{Strumia1, Salvio2016}. We then compute the 
one-loop RG improved effective potential of the scalar field
on the flat space-time. This gets contribution from the both 
the gravitational and matter degrees of freedom. This effective potential 
however contains an instability which comes up as an imaginary 
piece in the effective potential. The straightforward interpretation 
of the appearance of an imaginary piece in the effective potential 
is an indication that the background (flat space-time and constant 
scalar background) is not stable, and will decay. 

The reason for the occurrence of this instability is probed. It is found that this is 
purely gravitational in nature, in the sense that it arose from 
occurrence of tachyonic modes in the spin-2 gravitational sector 
and the gravitational scalar sector. These kind of 
tachyonic instabilities have been investigated in past 
\cite{Gross1982, Srijit2014}, and is a characteristic feature of 
gravitational theories coupled with scalar in flat space-time. 
At finite temperature this instability (also known as Jeans 
instability) results in collapse of gas of gravitons. 
While the occurrence of this instability is a disturbing feature 
of the theory and is unavoidable, it is however an IR 
problem and has no effect on the UV physics. 
In this paper we considered a different feature of theory. 
We investigate the issue of ghost appearing due to the presence of higher-derivative 
terms in the theory, which affect also the UV physics. 
This is done by investigating only the real part of 
the effective potential and ignoring the instability 
caused by the tachyonic modes. 

The real part of effective potential develops a VeV. 
This breaks the scale symmetry and induces 
mass scale, which in turn generates Newton's constant 
and masses for matter fields. The propagator of the 
metric fluctuation field now has mass, and in the 
broken phase it is easy to see problematic 
massive tensor ghost of the theory which remains 
shrouded in the symmetric phase. The scalar mode (Riccion) of the 
metric fluctuation also picks a mass in the broken phase. The VeV has a running, 
which in turn induces a running in the various parameters that 
are generated in the broken phase. It therefore seems 
sensible to ask question on behaviour of massive tensor 
ghost under this running in the broken phase, which is the 
most important aim of the paper. 

The induced running in the various parameters generated 
in the broken phase allows to investigate the running of 
$M^2/\mu^2$ (where $M$ is the induced mass of tensor ghost). 
The crucial task of the paper was to see whether there exist a 
domain of coupling parameters space where it is possible to
make $M^2/\mu^2>1$ throughout the whole RG trajectory. 
Satisfactory arrangement of $M^2/\mu^2>1$ will imply that 
the massive tensor ghost is never physically realised and 
never goes on-shell. This issue is however studied 
numerically, as the complexity of the beta-functions and the
complicated running of the induced parameters in broken phase 
hinders to make analytic progress. 

The last part of paper 
is devoted to numerically studying this issue. A prescription 
to choose the set of initial conditions so that $M^2/\mu^2>1$
for whole RG trajectory is stated. This involves solving
certain constraints. It is realised that the flow of $M^2/\mu^2$
has a unique minima at a certain point along the RG trajectory. 
The existence of such a point was analytically proved in the 
context of higher-derivative gravity including Einstein-Hilbert 
term \cite{NarainA1, NarainA2}, as the RG equation were simpler. 
In the present paper however it is not possible to achieve this 
analytically and numerical support was taken to get evidence 
for the existence of such a minima. We do see that for a large 
domain of parameter space such a unique minima does exist, 
and by requiring that $M^2/\mu^2>1$ at this minima implies 
that $M^2/\mu^2>1$ for whole RG trajectory. This although is not 
a robust analytic proof but stands as a strong evidence. 
We considered different set of initial conditions by varying 
various parameters in a systematic fashion. In each case 
it was see that the minima in the flow of $M^2/\mu^2$ always 
exits and unitarity criterion can be met. 

In this domain of coupling parameter space where $M^2/\mu^2>1$ for the 
whole RG trajectory, the behaviour of other parameters are studied. 
The first important thing that is noticed is the existence of a 
finite UV cutoff in the theory, which was also noticed in 
\cite{NarainA1, NarainA2, NarainA3, NarainA4}. Even though we
do the analysis in dimensional regularisation, still 
a cutoff emerges dynamically from the RG equations. Beyond this 
point the flow cannot be continued and knowledge of 
higher-loop contributions are needed. 
In \cite{NarainA1, NarainA2, NarainA3, NarainA4}
we showed analytically that at this point the coupling $\om$ diverges. 
In the present context we noticed this numerically. 
The behaviour of matter coupling $\lam$ has an interesting flow. 
It is seen to increase monotonically, but in the UV this stops and 
starts to decrease. This is due to yukawa coupling. 
The flow of yukawa coupling $y_t$ increases monotonically, and stops when the 
cutoff is reached. In the case of $R\varphi^2$ coupling, the coupling starts 
to run only near the UV, where it is seen to go toward smaller values, hinting 
that there might perhaps exists a stable fixed point. 

The flow of the induced Newton's constant $G$ is interesting.
It approaches zero both in UV and IR. In UV it vanishes 
at finite energy scale. This was something which 
was also observed in \cite{NarainA1, NarainA2, 
NarainA3, NarainA4}, where Einstein-Hilbert was present 
in the bare action of the theory and was not induced. In that 
respect this is surprising that in the present picture of Einstein-Hilbert 
term being induced from scale-invariant theory, the flow of the 
induced gravitational coupling is qualitatively similar to the 
case where EH term is present in the theory to begin with. 
Such vanishing of induced $G$ is although unexpected 
but a welcome feature. This is opposite to the widely 
known feature in Einstein-Hilbert gravity (without 
higher-derivatives) where Newton's constant 
becomes very large in UV. However those results 
cannot be trusted as they appear in non-renormalizable theories.
In the presence of higher-derivatives the situation 
changes in UV. Such vanishing of Newton's constant 
means that in UV, gravity decouples 
from matter, although gravitational self-interactions continue 
to exists. Such a behaviour will have consequences in the 
early universe, and also justifies the use of flat 
background in the UV. This softening can also be 
used in addressing the Higgs naturalness problem \cite{Salvio2016}. 

The flow of combination $M^2/\om\mu^2$ is however a bit different 
than what has been witnessed in \cite{NarainA1, NarainA2, 
NarainA3, NarainA4}, in the sense that the function 
$M^2/\om\mu^2$ is no longer a monotonically 
decreasing function of RG time $t$. On the contrary 
it has a flow similar to $M^2/\mu^2$, having a single 
minima. But there is a region of RG time $t$ over which 
$M^2/\om\mu^2<1$. This is because of the choice of 
initial condition $M_*^2/\mu_*^2 = \rho_*$ we made. This will 
imply that there is a range of $t$ where this scalar 
mode will be realised and will go on-shell, and outside this 
region it will remain unrealised. This can play a role 
in early universe to drive inflation. On the other hand 
the parameter $\rho_*$ can be chosen appropriately large
in order to make this scalar mode ghost (never 
goes on-shell), while still having unitarity. 

The running of VeV also induces a running in the 
generated masses for the scalar and fermion. To analyse 
whether they are physically realised or not, we studied the 
behaviour of $m_s^2/\mu^2$ and $m_f^2/\mu^2$ respectively. 
It is seen that if we choose $\rho_*$ appropriately, then it is possible 
to make them not physically realisable. They never 
go on-shell but do effect the theory gravitationally. 

The RG flow equations for the dimensionless couplings are gauge 
independent at one-loop however at higher loop gauge dependence 
is expected to enter. The effective potential is gauge-dependent 
which is because the hessian carries gauge dependence. This gauge 
dependence then enters the VeV and any quantity which is related to 
VeV (induced Newton's constant and induced masses). 
In the paper we studied the problem in Landau gauge 
which is a physical gauge allowing propagation 
of only transverse modes and suppressing longitudinal ones. However 
such gauge dependence is expected not to change qualitative features. 
This was explicitly seen in the case of pure higher-derivative 
gravity without matter \cite{NarainA1, NarainA2}. 

In the action appearing in the integrand for the lorentzian path-integral,
the sign of coefficient of $C_{\mn\rho\sg}^2$ is taken 
to be negative while the sign of coefficient of $R^2$ is taken positive. This is 
done to avoid tachyons and make the poles (in broken phase) lie on real axis,
the inspiration of which comes from past study done in 
\cite{NarainA1, NarainA2, NarainA3, NarainA4}.
Such a choice further offers necessary convergence in the 
lorentzian path-integral in the feynman $+i\ep$-prescription.  
This sign choice then implies that the coupling $f^2$ 
is no longer asymptotically free different from what is seen in path-integral 
defining a positive-definite euclidean theory 
\cite{Fradkin1981,Fradkin1982, Avramidi1985, Einhorn2014, 
Jones2015, Jones1, Jones2} (which is an 
entirely different theory), but instead has a Landau pole. 
This Landau singularity however appears way beyond the point 
where the RG flow of all couplings stops. Also the occurrence of 
Landau singularity is very possibly a one-loop effect, as at higher loops 
the running of $f^2$ gets correction thereby hinting 
the occurrence of fixed point \cite{NarainA1,NarainA2}. 
Moreover in this theory the dimensionless perturbative parameters  
$f^2$, $f^2/\om$, $\xi \sqrt{f^2}$, $\xi \sqrt{f^2/\om}$, 
$\lam$ and $y_t$ remain small throughout the RG flow, 
thereby justifying the usage of perturbative 
approximation and we don't enter non-perturbative regimes
in our analysis. 

The analysis done in the paper is on a flat background. This is because
any generic background has locally flat regions allowed by 
(strong) equivalence principle. Also when one is probing
ultra-short distances, one can do the analysis on flat 
space-time, as the perturbative UV divergences are independent 
of the background. Moreover, the chances of tensor-ghost becoming 
physically realisable is more in UV (and nowhere else), 
therefore its avoidance is important in UV, which can be 
investigated on a flat space-time. 
However extrapolating flat space-time analysis in deep infrared can lead to 
erroneous conclusions. In performing this study cosmological 
constant was put to zero, and was argued that it can be maintained 
to be zero in a supersymmetric framework. However supersymmetry 
is broken in IR and this will generate cosmological constant back again. 
Moreover current observations also favours the existence of 
cosmological constant in order to drive the accelerated expansion 
of the universe. Therefore a proper treatment of IR physics in a
field theoretic language is needed. A possible direction would be 
to formulate the theory on a deSitter background
\cite{Einhorn2002, Polyakov2007, Marolf2012, Narain2014} 
(see also \cite{Akhmedov2013} and references therein). 
This will give more accurate description 
of the theory in the infrared. 

The existence of tachyonic instability is a further indication 
that the chosen background of flat space-time is unstable. 
While this is an IR effect and an unavoidable outcome of 
gravitational theories coupled with scalar on flat space-time, 
it signals the breakdown of flat space-time as the background. 
It is expected that performing 
the study on a curved background might address these issues. 
For this, one would require a more accurate description 
of the formulation of field theory on curved background, and 
obtain the results in low energy limit. 
Alternatively one may have to incorporate non-localities 
appropriately to deal with IR physics \cite{Maggiore2014, Maggiore2016}. 
This is a future direction and will be considered later. 

It is interesting to wonder whether the RG equations gets modified 
when the decoupling of massive spin-2 ghost mode occurs, in the 
same manner as in flat space-time non gravitational QFTs where 
decoupling theorem exists \cite{Appelquist1974}. In 
flat space-time QFTs a systematic order-by-order 
computation leads to decoupling of heavier modes in process 
occurring at energies less than the mass of heavy particle. 
This theorem has been suitably extended to the case of 
matter theories on curved background 
\cite{Gorbar2002,Gorbar2003,Gorbar2003yp}, where the 
beta-functions gets a correction in mass-dependent scheme. 
For spin-2 fields the situation is a bit more involved, 
as incorporating mass in a deffeomorphism 
invariant manner is a tricky task. A possible way to achieve is 
by including higher-derivative terms in the action, which 
immediately brings in ghosts. If the ghost mass however is always 
above the energy scale, then ghosts get avoided 
due to decoupling. But this occurs in the quantum theory where RG 
running of the ghost mass is always above the energy scale. 
This implies an effective decoupling 
in the sense this spin-2 ghost mode never goes on-shell 
and off-shell it doesnÕt contribute to imaginary part of amplitudes \cite{NarainA1,NarainA2}. 
But currently it is unclear how such a decoupling will modify the RG 
flow equations of various parameters. This in really worthy of investigations
and will be considered in future works.

\bigskip
\centerline{\bf Acknowledgements} 

I would like to thank Prof. Ramesh Anishetty for several useful discussions
and enlightening suggestions. I am thankful to Prof. Tianjun Li for 
support, encouragement and fruitful discussions. 
I would like to thank Nirmalya Kajuri, Nick Houston and Tuhin Mukherjee for useful 
discussions. 
I would like to thank IMSc for hosting my visit and providing hospitality, where 
initial stages of the work was done. I thank the referee for raising the point regarding the 
modification of RG equations under the decoupling of ghost modes. 

\appendix

\section{Expansions}

For the computation of the running of all couplings including 
wave-function renormalisation the background of flat space-time 
is sufficient. However for simplicity to compute the running of $R\phi^2$
coupling we employ heat kernel methods, for which we take
the background to be deSitter space-time. 

For the flat background we have $g_\mn = \eta_\mn + h_\mn$,
\beq
\label{eq:inverseM}
g^\mn = \eta^\mn - h^\mn + h^\mu{}_\al h^{\al\nu} + \cdots \, .
\eeq
The expansion of $\sqrt{-g}$ is,
\beq
\label{eq:sqrtmg}
\sqrt{-g} = 1 + \frac{1}{2} h + \frac{1}{8} h^2 - \frac{1}{4} h_\mn h^\mn + \cdots  \, .
\eeq
The tetrads ($e^a{}_\mu$) are related with the metric by the following,
\beq
\label{eq:tetmet}
\eta_{ab} e^a{}_\mu e^b{}_\nu = g_\mn \, .
\eeq
The inverse tetrads ($e_a{}^\mu$) are similarly related with the 
inverse metric. Using these relations one can work out the 
expansion of the tetrads and inverse-tetrads in terms of the 
metric fluctuation field $h_\mn$,
\beq
\label{eq:tetexp}
\begin{split}
e^a{}_\mu &= \bar{e}^a{}_\rho \left(
\de^\rho{}_\mu + \frac{1}{2} h^\rho{}_\mu - \frac{1}{8} h^\rho{}_\al h^\al{}_\mu + \cdots 
\right) \, , \\
e_a{}^\mu &= \bar{e}_a{}^\rho \left(
\de_\rho{}^\mu - \frac{1}{2} h_\rho{}^\mu + \frac{3}{8} h_\rho{}^\al h_\al{}^\mu + \cdots 
\right) \, .
\end{split}
\eeq
As the determinant of tetrad $e$ is just $\sqrt{-g}$ therefore its expansion 
is same as in eq. (\ref{eq:sqrtmg}). 

The expansion of the christoffel connection and spin connection can be 
performed by first writing them in terms of metric and tetrads respectively, 
and then using the expansion of metric and tetrad mentioned above 
to obtain their expansion.

The christoffel connection is given by,
\beq
\label{eq:chrisCon}
\G_\al{}^\mu{}_\bt = \frac{1}{2} g^{\mu\rho} 
[\pt_\al g_{\rho\bt} + \pt_\bt g_{\rho\al} - \pt_\rho g_{\al\bt}] \, ,
\eeq
Its expansion in terms of the $h_\mn$ around the flat background is,
\beq
\label{eq:Chrisexp}
\G_\al{}^\mu{}_\bt = \frac{1}{2} (\pt_\al h^\mu{}_\bt + \pt_\bt h^\mu{}_\al - \pt^\mu h_{\al\bt})
- \frac{1}{2} h^{\mu\rho} (\pt_\al h_{\rho\bt} + \pt_\bt h_{\rho\al} - \pt_\rho h_{\al\bt}) + \cdots \, .
\eeq
The spin-connection $\om_{\mu cd}$ for torsion-free space-time 
has a simple expression in terms of christoffel connection
\beq
\label{eq:spinC_Chris}
\om_\mu^{ad} = e^{d\nu} e^a{}_\lam \G_\mu{}^\lam{}_\nu - e^{d\nu}\pt_\mu e^a{}_\nu \, .
\eeq
The christoffel connection $\G_\al{}^\mu{}_\bt$ can be expressed 
in terms of tetrads and its inverse. This when plugged in 
eq. (\ref{eq:spinC_Chris}) gives spin-connection solely in terms 
of the tetrads, inverse tetrads and derivative of tetrad,
\beq
\label{eq:spinC_tet}
\om_\mu^{ad} = \frac{1}{2} \left[
e^{a\rho} (\pt_\mu e^d{}_\rho - \pt_\rho e^d{}_\mu) 
- e^{d\rho} (\pt_\mu e^a{}_\rho - \pt_\rho e^a{}_\mu)
+ e^b{}_\mu e^{a\rho}e^{d\nu} (\pt_\nu e_{b\rho} - \pt_\rho e_{b\nu})
\right] \, .
\eeq
By plugging the expansion of the tetrads and inverse tetrads one can 
obtain the expansion of the spin connection in terms of the 
metric fluctuation fields,
\beq
\label{eq:spinC_exp}
\om_\mu{}^{ad} = (\bar{e}^{a\ta} \bar{e}^{d\tau} - \bar{e}^{a\tau} \bar{e}^{d\ta})
\left[
-\pt_\ta h_{\tau\mu} - \frac{1}{4} h_\ta{}^\rho \pt_\mu h_{\tau\rho} 
+ \frac{1}{2} h_{\tau\rho} \pt_\ta h^\rho{}_\mu + \frac{1}{2} h_\ta{}^\rho \pt_\rho h_{\tau\mu}
+ \cdots
\right] \, .
\eeq
The Reimann curvature tensor is $R_{\mu\nu}{}^\rho{}_\sg = \pt_\mu \G_\nu{}^\rho{}_\sg
- \pt_\nu \G_\mu{}^\rho{}_\sg + \G_\mu{}^\rho{}_\lam \G_\nu{}^\lam_\sg
- \G_\nu{}^\rho{}_\lam \G_\mu{}^\lam_\sg$. We are interested only in the 
expansion of the Ricci tensor and Ricci scalar. 
\bea
\label{eq:RiccTen_exp1}
&&
R_\mn = \frac{1}{2} \left(\pt_\rho \pt_\mu h^\rho{}_\nu + \pt_\rho \pt_\nu h^\rho{}_\mu
- \Box h_\mn - \pt_\mu\pt_\nu h \right) + \pt_\rho \G^{(2)}_\mu{}^{\rho}{}_\nu 
- \pt_\mu \G^{(2)}_\rho{}^{\rho}{}_\nu 
\notag \\
&&
+ \frac{1}{4} \biggl[\pt^\lam h \left(\pt_\mu h_{\lam\nu} + \pt_\nu h_{\lam\mu} - \pt_\lam h_\mn \right)
- \pt_\mu h^{\rho\sg} \pt_\nu h_{\rho\sg} - 2\pt^\lam h^\rho{}_\mu \pt_\rho h_{\lam\nu}
+ 2 \pt^\lam h^\rho{}_\mu\pt_\lam h_{\rho\nu} \biggr] \, ,
\\
&&
\label{eq:RiccTen_exp2}
R = \pt_\mu\pt_\nu h^\mn - \Box h - \frac{1}{2} h^\mn \left(2\pt_\rho\pt_\mu h^\rho{}_\nu 
-\Box h_\mn - \pt_\mu\pt_\nu h\right) + \eta^\mn \left(
\pt_\rho \G^{(2)}_\mu{}^{\rho}{}_\nu - \pt_\mu \G^{(2)}_\rho{}^{\rho}{}_\nu \right)
\notag \\
&&
+ \frac{1}{4} \biggl[
\pt^\lam(2\pt^\sg h_{\lam\sg} - \pt_\lam h) + \pt^\rho h^{\nu\lam} \pt_\rho h_{\nu\lam}
-2 \pt^\lam h^{\rho\nu} \pt_\rho h_{\lam\nu}
\biggr] \, .
\eea

\section{Projectors}
\label{app_proj}

The metric fluctuation field $h_\mn$ around a general background can be decomposed
into various components by doing a transverse-traceless decomposition. This is equivalent
to doing decomposition of a vector into transverse and longitudinal components.
For the metric fluctuation field $h_\mn$ around a flat background, this decomposition can be
written in momentum space as,
\beq
\label{eq:TTdecomp}
h_{\mn} = 
h^T_{\mn} + \iota \left(q_{\mu} \xi_{\nu} + q_{\nu} \xi_{\mu} \right)
+ \left( \eta_\mn - \frac{q_\mu q_\nu}{q^2} \right) s
+ \frac{q_\mu q_\nu}{q^2} \, w \, .
\eeq
where the various components satisfies the following constraints,
\begin{gather}
\label{eq:constraints}
h^T_{\mu}{}^{\mu}=0 \, , \hspace{2mm} 
q^\mu h^T_\mn =0 \, , \hspace{2mm}
q^\mu \xi_\mu=0 \, .
\end{gather}
Here $h^T_\mn$ is a transverse-traceless symmetric tensor, $\xi_\mu$ is a
transverse vector and $s$ and $w$ are two scalars. 
This decomposition can be neatly written
by making use of flat space-time projectors, which projects various components of 
$h_\mn$ field into $h^T_\mn$, $\xi_\mu$, $s$ and $w$ respectively. 
These projectors are written in terms of the following two projectors,
\begin{gather}
\label{eq:LTproj}
L_{\mu\nu} = \frac{q_{\mu} \, q_{\nu}}{q^2} \, \ , \hspace{10mm}
T_{\mu\nu} = \eta_{\mn} - \frac{q_{\mu} \, q_{\nu}}{q^2} \, . 
\end{gather}
These are basically the projector for projecting out various components
of a vector field. They satisfy $q^\mu T_\mn=0$ and $q^\mu L_\mn=q_\nu$.
Using them the projectors for the rank-2 tensor field 
can be constructed. These are given by,
\bea
&&
\label{eq:spin2proj}
(P_2)_{\mu\nu}{}^{\alpha \beta}
 = \frac{1}{2} \left[ T_{\mu}{}^{\alpha} T_{\nu}{}^{\beta} + 
T_{\mu}{}^{\beta}T_{\nu}{}^{\alpha} \right] - \frac{1}{d-1} T_{\mu\nu}T^{\alpha\beta} \, ,
 \\
&&
\label{eq:spin1proj}
(P_1)_{\mu\nu}{}^{\alpha \beta} 
= \frac{1}{8} \left[ 
T_{\mu}{}^{\alpha} \, L_{\nu}{}^{\beta}
+T_{\mu}{}^{\beta} \, L_{\nu}{}^{\alpha} 
+ T_{\nu}{}^{\alpha} \, L_{\mu}{}^{\beta}
+T_{\nu}{}^{\beta} \, L_{\mu}{}^{\alpha} 
\right] \, ,
\\
&&
\label{eq:spinsproj}
(P_s)_{\mu\nu}{}^{\alpha \beta} 
= \frac{1}{d-1} T_{\mu\nu} \, T^{\alpha \beta} \, ,
\\
&&
\label{eq:spinwproj}
(P_w)_{\mu\nu}{}^{\alpha \beta} = L_{\mu\nu} \, L^{\alpha \beta} \, .
\eea
The projectors for spin-2, spin-1, spin-s and spin-w form an 
orthogonal set. In the scalar sector there are two more projectors 
(which are not projectors in the strict sense), which along with
spin-s and spin-w projectors form a complete set. They are given by,
\bea
&&
\label{eq:spinswproj}
(P_{sw})_{\mu\nu}{}^{\alpha \beta}  
= \frac{1}{\sqrt{d-1}} T_{\mu\nu} \, L^{\alpha \beta} \, ,
\\
&&
\label{eq:spinwsproj}
(P_{ws})_{\mu\nu}{}^{\alpha \beta}   
= \frac{1}{\sqrt{d-1}} L_{\mu\nu} \, T^{\alpha \beta}  \, .
\eea
The projectors in eqs. (\ref{eq:spin2proj}, \ref{eq:spin1proj}, \ref{eq:spinsproj} and \ref{eq:spinwproj})
forms a complete set in the sense that their sum is unity. 
\beq
\label{eq:projunity}
(P_2)_\mn{}^{\rho\sg} +(P_1)_\mn{}^{\rho\sg}
+ (P_s)_\mn{}^{\rho\sg} + (P_w)_\mn{}^{\rho\sg}
= \de_\mn^{\rho\sg} \, ,
\eeq
where $\de_\mn^{\rho\sg} = 1/2( \de_\mu^{\rho} \de_\nu^\sg 
+ \de^\rho_\nu \de^\sg_\mu)$.
Each of these projectors when act on $h_\mn$ projects out various 
spin components of the tensor field. 
\bea
\label{eq:projcomp}
&&
(P_2)_{\mn}{}^{\rho\sg} \, h_{\rho\sg} = h_{\mn}^T \, ,
\hspace{5mm}
(P_1)_{\mn}{}^{\rho\sg} \, h_{\rho\sg} =  \iota \left(
q_{\mu} \xi_{\nu} + q_{\nu} \xi_{\mu} \right)\, , 
\notag \\
&&
(P_s)_{\mn}{}^{\rho\sg} \, h_{\rho\sg} = (d-1) T_\mn s \, , \hspace{5mm}
(P_w)_{\mn}{}^{\rho\sg} \, h_{\rho\sg} = L_\mn w \, .
\eea
If the projectors $P_2$, $P_1$, $P_s$ and $P_w$ are written as 
$P_{22}$, $P_{11}$, $P_{ss}$ and $P_{ww}$ respectively, then 
all the projectors (including $P_{sw}$ and $P_{ws}$) satisfy the 
following algebra,
\beq
\label{eq:proj_algebra}
P_{ij} P_{mn} = \de_{jm} P_{in} \, ,
\eeq
where $i$, $j$, $m$ and $n$= \{$2$, $1$, $s$, $w$\}.

\section{Matter Propagator and Vertices}
\label{propver}

Here we write the propagator for matter fields and vertices of the action given in 
eq. (\ref{eq:Act}). These are obtained by doing the second variation 
of the action with respect to various fields. The first line of eq. (\ref{eq:Act})
gives the graviton propagator which is mentioned in eq. (\ref{eq:GR_prop}),
while the second line of eq. (\ref{eq:Act}) gives the propagator for the matter 
fields and various graviton-matter, matter-matter vertices. 
In the following we will be obtaining them one by one. 

\subsection{Propagators for Matter fields}
\label{invProp}

Here we write the inverse propagators for the various matter fields. 
These are obtained by doing the second variation of the action 
of the theory with respect to various fields.
The mixed terms in such kind of variation will be treated as interaction terms. 
From the second variation of the action given in eq. (\ref{eq:2ndvarActMat})
one can pick the terms corresponding to the scalar and fermion propagator. 
The operator whose inverse correspond to scalar propagator is,
\beq
\label{eq:scalar_prop}
\D_s = - \pt^2 \, .
\eeq
In the case of fermions the relevant inverse operator is given by,
\beq
\label{eq:fermi_prop}
(\D_F)_{ab} = i \g^\rho_{ab} \pt_\rho  \, .
\eeq

\subsection{Vertices}
\label{vert}

Here we specify the various vertices that are relevant for our one-loop 
computations. These can be categorised in 3 parts: (a) vertex with two 
internal graviton lines,  (b) vertex with one internal graviton line and one 
internal matter line, and (c) vertex with two internal matter lines. 

\subsubsection{Gravity-gravity}
\label{grgr}

\begin{figure}
\centerline{
\vspace{0pt}
\centering
\includegraphics[width=5in]{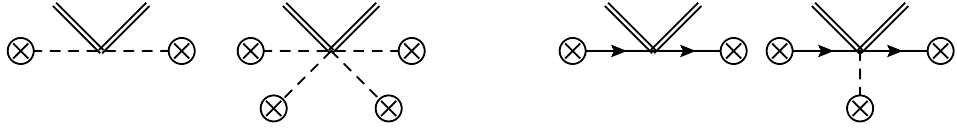}
}
 \caption[]{
Various vertices containing two internal graviton lines. Here the dashed line 
is scalar, solid line with arrow is fermion, while double line depicts graviton. 
The lines ending with circle containing cross are external legs. 
}
\label{fig:gragra}
\end{figure}

In these vertices there are two internal graviton lines. In the following the term 
$V^{\mn\rho\sg}$ comes from scalar field action, while the term $U^{\mn\rho\sg}$
comes from fermion field action. The vertices are depicted in 
figure \ref{fig:gragra}. 

\bea
&& \int {\rm d}^dx h_{\mn} (V^{\mn\rho\sg} + U^{\mn\rho\sg}) h_{\rho\sg}
\notag \\
&&
\label{eq:hhvert1}
V^{\mn\rho\sg} = \frac{1}{4} (\eta^{\mn}\eta^{\rho\sg} - \eta^{\mu\rho}\eta^{\nu\sg}
-\eta^{\mu\sg}\eta^{\nu\rho}) \left\{
\frac{1}{2} (\pt \varphi)^2 - \frac{\lam}{4} \varphi^4
\right\} - \frac{1}{2} \eta^{\rho\sg} \pt^\mu \varphi \pt^\nu \varphi 
\notag \\
&&
+ \eta^{\sg\nu} \pt^\mu \varphi \pt^\rho \varphi 
- \frac{1}{2} \xi \varphi^2 \left\{
- \eta^{\nu\sg} \pt^\mu \pt^\rho 
+ \frac{1}{4} (\eta^{\mu\rho} \eta^{\nu\sg} + \eta^{\mu\sg} \eta^{\nu\rho} 
-2 \eta^\mn \eta^{\rho\sg} )\Box + \eta^\mn \pt^\rho \pt^\sg
\right\} 
\notag \\
&&
- \xi \biggl\{
\varphi \pt^\mu \varphi (\eta^{\nu\rho} \pt^\sg
- 2 \eta^{\rho\sg} \pt^\nu) - \frac{1}{4} \varphi \pt^\al \varphi 
\left(3(\eta^{\mu\rho} \eta^{\nu\sg} + \eta^{\mu\sg} \eta^{\nu\rho})
-2 \eta^\mn \eta^{\rho\sg} \right) \pt_\al
\biggr\}
\notag \\
&&
+ \frac{1}{2} \xi (\pt^\nu \varphi  \pt^\bt \varphi 
+ \varphi \pt^\nu \pt^\bt \varphi )(\eta^{\mu\rho} \de^\sg_\bt
+ \eta^{\mu\sg} \de^\rho_\bt) \, , \\
&&
\label{eq:hhvert2}
U^{\mn\rho\sg} = \frac{i}{4} (\eta^\mn \eta^{\rho\sg} - \eta^{\mu\rho} \eta^{\nu\sg}
- \eta^{\mu\sg} \eta^{\nu\rho} ) \bar{\ta} \g^\tau \pt_\tau \ta 
- \frac{i}{2} \eta^\mn \bar{\ta} \g^\rho \pt^\sg \ta 
+ \frac{3i}{4} \eta^{\sg\nu} \bar{\ta} \g^\rho \pt^\mu \ta 
\notag \\
&&
- \frac{i}{8} \eta^{\sg\nu} \bar{\ta} \g^\tau [\g^\mu, \g^\rho] \ta \pt_\tau
+ \frac{i}{4} \bar{\ta} \g^\rho [\g^\mu, \g^\sg] \ta \pt^\nu
- \frac{i}{4} \eta^{\nu\sg} \pt_\al \left\{\bar{\ta}\g^\rho [\g^\al, \g^\mu] \ta\right\}
- \frac{i}{4} \eta^\mn \bar{\ta} \g^\rho [\g^\al, \g^\sg] \ta \pt_\al 
\notag \\
&&
- \frac{1}{4} y_t \varphi (\eta^\mn \eta^{\rho\sg} - \eta^{\mu\rho} \eta^{\nu\sg}
- \eta^{\mu\sg} \eta^{\nu\rho} ) \bar{\ta} \ta \, .
\eea
%

\subsubsection{Gravity-scalar}
\label{grsc}

\begin{figure}
\centerline{
\vspace{0pt}
\centering
\includegraphics[width=3.5in]{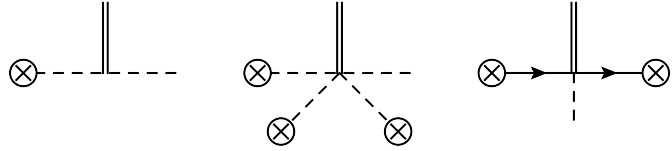}
}
 \caption[]{
Various vertices containing one internal graviton line and one internal scalar leg. 
Here the dashed line is scalar, solid line with arrow is fermion, while double line depicts graviton. 
The lines ending with circle containing cross are external legs. 
}
\label{fig:grasca}
\end{figure}

Here we write the vertex which has one internal graviton line and one internal 
scalar line. This vertex gets contribution from both scalar and fermion field actions. 
These vertices are depicted in figure \ref{fig:grasca}. 

\bea
&&
\int {\rm d}^dx \left[ h_{\rho\sg} (V_{h\phi})^{\rho\sg} \chi
+ \chi (V_{\phi h})^\mn h_\mn \right]
\notag \\
&&
\label{eq:Vhphi}
(V_{h\phi})_{\rho\sg} = - \frac{\lam}{2} \varphi^2 \eta_{\rho\sg} \varphi 
+ \frac{1}{2} \eta_{\rho\sg} \pt_\al \varphi \pt^\al - \pt_\rho \varphi \pt_\sg
- \xi \left\{
\pt_\rho\pt_\sg \varphi + \pt_\rho \varphi \pt_\sg 
+ \pt_\sg \varphi \pt_\rho + \varphi \pt_\rho \pt_\sg \right\} 
\notag \\
&&
+ \xi \eta_{\rho\sg} (\Box \varphi + 2 \pt_\al \varphi \pt^\al + \varphi \Box)
-\frac{1}{2} \eta_{\rho\sg} y_t \bar{\ta} \ta \, ,
\\
&&
\label{eq:Vphih}
(V_{\phi h})_\mn = - \frac{\lam}{2} \lam \varphi^2 \eta_\mn \varphi 
- \frac{1}{2} \eta_\mn \pt_\bt \varphi \pt^\bt - \frac{1}{2} \eta_\mn \Box \varphi
+ \pt_\mu \varphi \pt_\nu + \pt_\mu\pt_\nu \varphi 
\notag \\
&&
- \xi \varphi (\pt_\mu\pt_\nu - \eta_\mn \Box) 
-\frac{1}{2} \eta_\mn y_t \bar{\ta} \ta \, .
\eea

\subsubsection{Gravity-fermion}
\label{gravFerm}

\begin{figure}
\centerline{
\vspace{0pt}
\centering
\includegraphics[width=4.5in]{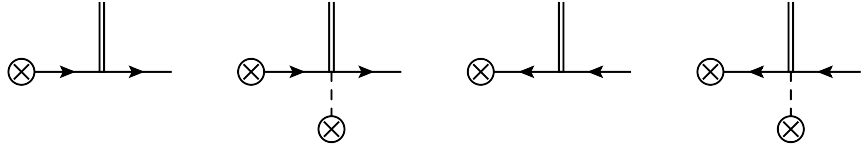}
}
 \caption[]{
Various vertices containing one internal graviton line and one internal fermion leg. 
Here the dashed line is scalar, solid line with arrow is fermion, while double line depicts graviton. 
The lines ending with circle containing cross are external legs. 
}
\label{fig:grafer}
\end{figure}

Here we write the vertex that contain one internal graviton line and 
one internal fermion line. These vertex comes only from the 
fermion field action. These vertices are depicted in figure \ref{fig:grafer}. 
\bea
&&
\int {\rm d}^dx \bigl[ \bar{\eta}_d (V_{\bar{\psi}h})_d^{\rho\sg} h_{\rho\sg}
+ h_\mn (V_{h\bar{\psi}}^T)_c^\mn \bar{\eta}^T_c \bigr] \, ,
\notag \\
&&
\label{eq:VbarpsiH1}
(V_{\bar{\psi}h})_{d\rho\sg} = \frac{i}{2} (\eta_{\rho\sg} \g^\tau_{de} \pt_\tau \ta_e 
- (\g_\rho)_{de} \pt_\nu \ta_e) -\frac{i}{4} (\g_\rho[\g^\al, \g_\sg])_{de} \ta_e \pt_\al 
-\frac{y_t}{2} \varphi \eta_{\rho\sg} \ta_d \, , 
\\
&&
\label{eq:VbarpsiH2}
(V_{h\bar{\psi}}^T)_c^\mn = - \frac{i}{2} \left\{
\eta^\mn \pt_\tau \ta_e (\g^{T\tau})_{ec} - (\pt^\nu \ta^T_e) \g^{T\mu}_{ec}
\right\} - \frac{i}{4} (\pt_\al \ta^T_e) \{[\g^{\mu T}, \g^{\al T}]\g^{\nu T}\}_{ec}
\notag \\
&&
- \frac{i}{4} (\ta^T_e) \{[\g^{\mu T}, \g^{\al T}]\g^{\nu T}\}_{ec} \pt_\al 
+ \frac{1}{4} \eta^\mn y_t \varphi \ta^T_c
\, .
\eea
\bea
&&
\int {\rm d}^dx \bigl[ h_\mn (V_{h\psi})^\mn_a \eta_a 
+ \eta^T_a (V_{\psi h}^T)^{\rho\sg} h_{\rho\sg} \bigr] \, , 
\notag \\
&&
\label{eq:Vhpsi1}
(V_{h\psi})^{\mn}_c = \frac{i}{2} \left\{
\eta^\mn \bar{\ta}_e \g^\tau_{ec} \pt_\tau - \bar{\ta}_e \g^\nu_{ec} \pt^\mu
\right\} + \frac{i}{4} \left\{
(\pt_\al \bar{\ta}_e) (\g^\nu[\g^\al, \g^\mu])_{ec} 
+ \bar{\ta}_e (\g^\nu[\g^\al, \g^\mu])_{ec} \pt_\al \right\} 
\notag \\
&&
- \frac{y_t}{4} \eta^\mn \varphi \bar{\ta}_c \, , 
\\
&&
\label{eq:Vhpsi2}
(V^T_{\psi h})^{\rho\sg}_b = \frac{i}{2} \left\{
\eta^{\rho\sg} \g^{T\tau}_{be} (\bar{\ta}^T_e \pt_\tau 
+ \pt_\tau \bar{\ta}^T_e )
-  \g^{T\sg}_{be} (\bar{\ta}^T_e \pt^\rho + \pt^\rho \bar{\ta}^T_e) \right\}
+ \frac{i}{4} ([\g^{T\rho}, \g^{T\al}]\g^{T\sg})_{be} \bar{\ta}^T_e \pt_\al
\notag \\
&&
+ \frac{y_t}{2} \eta^{\rho\sg} \bar{\ta}^T_b \varphi \, .
\eea
%

\subsubsection{Matter-matter}
\label{matmat}

\begin{figure}
\centerline{
\vspace{0pt}
\centering
\includegraphics[width=6in]{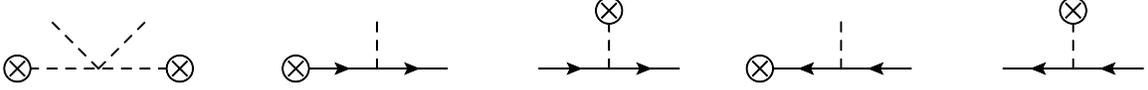}
}
 \caption[]{
Various vertices containing two internal matter legs (scalar-scalar, scalar-fermion, fermion-fermion). 
Here the dashed line is scalar while solid line with arrow is fermion. 
The lines ending with circle containing cross are external legs. 
}
\label{fig:matmat}
\end{figure}

Here we write the vertices which has two internal matter lines. These 
will be either both scalar lines, one scalar and one fermion line or both 
fermion lines. These vertices are depicted in figure \ref{fig:matmat}. 
\bea
&&
\int {\rm d}^dx \bigl[
- \chi (V_s) \chi + \bar{\eta}_a (V_{\bar{\psi}\psi})^{ab} \eta_b 
+ \eta^T_a (V_{\bar{\psi}\psi}^T)^{ab} \bar{\eta}^T_b 
\notag \\
&&
+ \bar{\eta}_a (V_{\bar{\psi}\phi})^a \chi 
+ \chi (V_{\bar{\psi}\phi}^T)^b \bar{\eta}^T_b 
+ \chi (V_{\phi\psi})^b \eta_b + \eta^T_a (V_{\phi\psi}^T)^a \chi 
\bigr] 
\notag \\
&&
\label{eq:Vs}
V_s = 3 \lam \varphi^2 \, , \\
&&
(V_{\bar{\psi}\psi})^{ab} = - y_t \varphi \de^{ab} \, , 
\hspace{5mm}
(V^T_{\bar{\psi}\psi})^{ab} =  y_t \varphi \de^{ab} \, ,
\\
&&
(V_{\bar{\psi}\phi})^a = - y_t \ta^a \, , \hspace{5mm}
(V^T_{\bar{\psi}\phi})^a = y_t \ta^{Ta} \\
&&
(V_{\phi\psi})^b = -y_t \bar{\ta}^b \, , 
\hspace{5mm}
(V^T_{\phi\psi})^b = y_t \bar{\ta}^{Tb} \, .
\eea

\section{Cubic Equation}
\label{cubic}

Here we will consider the roots of generic cubic equation with real 
coefficients. Such an equation emerges in section. \ref{effpot} while 
computing the contribution to the scalar effective potential 
from the scalar sector of theory. In order to compute the contribution 
of the scalar to the effective potential we need analyse a cubic 
equation in $-\Box$ operator written in eq. (\ref{eq:sca_op_veff}). Here 
in this section we will consider a generic cubic equation of the form
\beq
\label{eq:cubgen}
a z^3 + b z^2 + c z + d =0 \, .
\eeq
By a change of variable $z=u-b/3a$, this equation becomes a depressed cubic
\bea
\label{eq:depcub}
&&
u^3 - \frac{\D_0}{3a^2} u + \frac{\D_1}{27a^3} =0 \, ,\\
&&
\D_0 = b^2 - 3ac \, , \hspace{3mm}
\D_1=2b^3 - 9abc +27d a^2 \, .
\eea
By choosing $u=v + \D_0/9a^2v$, this equation can be converted 
in to a quadratic equation in $v^3$ 
\beq
\label{eq:quadv}
v^6 + \frac{\D_1}{27a^3} v^3 + \frac{\D_0^3}{729a^6} = 0 \, .
\eeq
This quadratic equation can be solved by known algebraic 
methods and has two roots. The nature of roots can be 
determined by the sign of the discriminant of this quadratic 
equation.
\beq
\label{eq:quaddis}
-27a^2 \D = \D_1^2 - 4 \D_0^3 \, ,
\eeq
where $\D$ was also defined in eq. (\ref{eq:disc}). The two roots of the 
quadratic equation will be given by,
\beq
\label{eq:rootquad}
v^3 = \frac{-\D_1 \pm \sqrt{\D_1^2 - 4\D_0^3}}{54 a^3} 
= \frac{1}{27a^3} l^3_{1,2} \, .
\eeq
This can be solved easily by taking cube-root. Here there will be 
three roots for $v$. Corresponding to each cube-root we have a 
root for the eq. (\ref{eq:cubgen}), which is obtained by plugging 
$v$ back in to $u$ and $z$. The three roots will be given in terms of 
$l_1$ and $l_2$
\bea
\label{eq:cuberoot}
&&
z_1 = \frac{1}{3a} \left(
-b + l_1 + l_2
\right)\, ,
\notag\\
&&
z_2 = \frac{1}{3a} \left(
-b + e^{2\pi i/3} l_1+ e^{4\pi i/3}l_2
\right) \, ,
\notag\\
&&
z_3 = \frac{1}{3a} \left(
-b + e^{4\pi i/3} l_1 + e^{2\pi i/3}l_2
\right) \, .
\eea
In the case when the discriminant $\D_1^2 - 4\D_0^3>0$, 
we have one real root and a complex conjugate pair.
In this case we can write the exponentials in terms of sine and cosine functions. One
can then write the real and imaginary part of these roots. 
This complex conjugate pair can be written in polar form also with
\bea
\label{eq:ccpolar}
&&
r = \frac{1}{3a} \sqrt{
b^2 + l_1^2 + l_2^2 -2 \cos(\frac{2\pi}{3}) (b l_1 + b l_2 + l_1 l_2) }
\, , \notag \\
&&
\tan \ta = \frac{\sin(2\pi/3) (l_1 - l_2)}{-b + \cos(2\pi/3)(l_1 + l_2)} \, ,
\hspace{4mm}
{\rm with} \,\, -\frac{\pi}{2} \leq \ta \leq \frac{\pi}{2} \, .
\eea
In the case when the discriminant $\D_1^2 - 4\D_0^3<0$, we have 
a complex conjugate pair of roots for $v^3$, meaning $l_1$ and 
$l_2$ will be complex conjugate. This will imply that expression 
for $z_1$, $z_2$ and $z_3$ are real. However now $\ta$ appearing 
in eq. (\ref{eq:ccpolar}) will be imaginary.



\end{document}